\documentclass[longauth, onecolumn]{aa}
\onecolumn
\usepackage[T1]{fontenc}
\usepackage{amstext}
\usepackage{graphicx}
\usepackage[varg]{txfonts}  
\bibpunct{(}{)}{;}{a}{}{,}  
\usepackage[colorlinks, citecolor=blue, linkcolor=blue]{hyperref}

\newcommand{\toprule}{
     \hline\hline
     \noalign{\smallskip}
}
\newcommand{\midrule}{
      \noalign{\smallskip}
      \hline
      \noalign{\smallskip}
}
\newcommand{\bottomrule}{
        \noalign{\smallskip}
        \hline
}

\newcommand*{\fnref}[1]{\textsuperscript{\ref{#1}}}

\makeatletter

\iftrue
    \usepackage{twoopt}
    \usepackage{xstring}
    \newcommand{\adshref}[2]{\StrRight{#1}{19}[\adsid]\href{http://adsabs.harvard.edu/abs/\adsid}{#2}}
    
    \let\orgcitep\citep
    \let\orgcitet\citet
    \let\orgcitealt\citealt
    \let\orgcitealp\citealp
    \let\orgciteauthor\citeauthor

    \renewcommandtwoopt{\cite}[3][][]{\adshref{#3}
        {\def\hyper@linkstart##1##2{}
        \let\hyper@linkend\@empty\orgcitet[#1][#2]{#3}}}
    \renewcommandtwoopt{\citep}[3][][]{\adshref{#3}
        {\def\hyper@linkstart##1##2{}
        \let\hyper@linkend\@empty\orgcitep[#1][#2]{#3}}}
    \renewcommandtwoopt{\citet}[3][][]{\adshref{#3}
        {\def\hyper@linkstart##1##2{}
        \let\hyper@linkend\@empty\orgcitet[#1][#2]{#3}}}
    \renewcommandtwoopt{\citealt}[3][][]{\adshref{#3}
        {\def\hyper@linkstart##1##2{}
        \let\hyper@linkend\@empty\orgcitealt[#1][#2]{#3}}}
    \renewcommandtwoopt{\citealp}[3][][]{\adshref{#3}
        {\def\hyper@linkstart##1##2{}
        \let\hyper@linkend\@empty\orgcitealp[#1][#2]{#3}}}
    \renewcommandtwoopt{\citeauthor}[3][][]{\adshref{#3}
        {\def\hyper@linkstart##1##2{}
        \let\hyper@linkend\@empty\orgciteauthor[#1][#2]{#3}}}
    \newcommandtwoopt{\citeyearads}[3][][]
        {\href{http://adsabs.harvard.edu/abs/#3}
        {\def\hyper@linkstart##1##2{}
        \let\hyper@linkend\@empty\citeyear[#1][#2]{#3}}}
   \iftrue
        \renewcommandtwoopt{\cite}[3][][]{\adshref{#3}{\orgcitet[#1][#2]{#3}}}
        \renewcommandtwoopt{\citep}[3][][]{\adshref{#3}{\orgcitep[#1][#2]{#3}}}
        \renewcommandtwoopt{\citet}[3][][]{\adshref{#3}{\orgcitet[#1][#2]{#3}}}
        \renewcommandtwoopt{\citealt}[3][][]{\adshref{#3}{\orgcitealt[#1][#2]{#3}}}
        \renewcommandtwoopt{\citealp}[3][][]{\adshref{#3}{\orgcitealp[#1][#2]{#3}}}
    \fi
\fi

\providecommand{\tabularnewline}{\\}
\providecommand{\tabnewline}{\\[0.3em]}

\def\instrefs#1{{\def\scsep{\def\scsep{,}}\@for\w:=#1\do{\scsep\ref{inst:\w}}}}
\renewcommand{\inst}[1]{\unskip$^{\instrefs{#1}}$}

\usepackage{xspace}
\newcommand{\MEarth}{M_\oplus}
\newcommand{\teegarden}{\object{Teegarden's Star}\xspace}
\newcommand{\gaia}{{\em Gaia}\xspace}

\renewcommand*\aa@pageof{, page \thepage{} of \pageref*{LastPage}} 

\makeatother
\onecolumn

\begin{document}
\lefthyphenmin=3

\title{The CARMENES search for exoplanets around M dwarfs}
\subtitle{Two temperate Earth-mass planet candidates around Teegarden's Star\thanks{Tables \ref{tab:RV_VIS} and \ref{tab:RV_NIR} are only available at the CDS via anonymous ftp to \url{cdsarc.u-strasbg.fr} (\url{130.79.128.5}) or via \url{http://cdsweb.u-strasbg.fr/cgi-bin/qcat?J/A+A/vol/page}}}

\author{
    M.~Zechmeister\inst{iag}
    \and S.~Dreizler\inst{iag}
    \and I.~Ribas\inst{ice,ieec}
    \and A.~Reiners\inst{iag}
    \and J.~A.~Caballero\inst{cabesac}
    \and F.~F.~Bauer\inst{iaa}
    \and V.~J.~S.~B\'ejar\inst{iac,ull}
    \and L.~Gonz\'alez-Cuesta\inst{iac,ull}
    \and E.~Herrero\inst{ice,ieec}
    \and S.~Lalitha\inst{iag}
% ...
}

\institute{
    \label{inst:iag}Institut f\"ur Astrophysik, Georg-August-Universit\"at, Friedrich-Hund-Platz 1, 37077 G\"ottingen, Germany\\
    \email{zechmeister@astro.physik.uni-goettingen.de}
    \and \label{inst:ice}Institut de Ci\`encies de l'Espai (ICE, CSIC), Campus UAB, C/Can Magrans s/n, 08193 Bellaterra, Spain
    \and \label{inst:ieec}Institut d'Estudis Espacials de Catalunya (IEEC), 08034 Barcelona, Spain
    \and \label{inst:cabesac}Centro de Astrobiolog\'ia (CSIC-INTA), ESAC campus, Camino bajo del castillo s/n, 28692 Villanueva de la Ca\~nada, Madrid, Spain
    \and \label{inst:iaa}Instituto de Astrof\'isica de Andaluc\'ia (IAA-CSIC), Glorieta de la Astronom\'ia s/n, 18008 Granada, Spain
    \and \label{inst:iac}Instituto de Astrof\'isica de Canarias (IAC), V\'ia L\'actea s/n, 38205 La Laguna, Tenerife, Spain
    \and \label{inst:ull}Departamento de Astrof\'isica, Universidad de La Laguna (ULL), 38206 La Laguna, Tenerife, Spain
% ...
}

\abstract{Teegarden's Star is the brightest and one of the nearest ultra-cool dwarfs in the solar neighbourhood. For its late spectral type (M7.0\,V), the star shows relatively little activity and is a prime target for near-infrared radial velocity surveys such as CARMENES.}
{As part of the CARMENES search for exoplanets around M dwarfs, we obtained more than 200 radial-velocity measurements of Teegarden's Star and analysed them for planetary signals.}
{We find periodic variability in the radial velocities of Teegarden's Star. We also studied photometric measurements to rule out stellar brightness variations mimicking planetary signals.}
{We find evidence for two planet candidates, each with 1.1\,$\MEarth$ minimum mass, orbiting at periods of 4.91 and 11.4\,d, respectively. No evidence for planetary transits could be found in archival and follow-up photometry. Small photometric variability is suggestive of slow rotation and old age.}
{The two planets are among the lowest-mass planets discovered so far, and they are the first Earth-mass planets around an ultra-cool dwarf for which the masses have been determined using radial velocities.
}


\date{Received 13 March 2019 / Accepted 14 May 2019}

\maketitle
\twocolumn  
\sloppy

\section{Introduction}

Since the first exoplanet discoveries in the mid-1990s, more than 800 exoplanets have been detected with the radial-velocity (RV) method, while thousands of them have been detected with dedicated transit searches. Despite this great success, only very few planets have been found so far around dwarf stars later than M4.5\,V, although these very late stellar types are very numerous. We know\footnote{\url{http://exoplanets.org}, accessed on 2019-04-10.} only two planet hosts with effective temperatures cooler than 3000\,K, but these are remarkable. One of them is \object{Proxima Centauri} (M5.5\,V), which is the closest star to the Sun and hosts an Earth-mass planet in its habitable zone \citep{Anglada2016Natur.536..437A}. The other one is TRAPPIST-1 (\object{2MUCD~12171}; M8.0\,V), which hosts seven transiting planets and is located at a distance of $12.0\pm0.4$\,pc \citep{Gillon2016Natur.533..221G, Gillon2017Natur.542..456G}.

The paucity of planet detections around very late-type stars is mainly an observational bias due to the faintness in the visible wavelength range and red colours of these stars. The CARMENES\footnote{Calar Alto high-Resolution search for M dwarfs with Exoearths with Near-infrared and optical \'Echelle Spectrographs.} project \citep{Quirrenbach2018SPIE10702E..0WQ}, as well as a number of other projects (e.g. \citealp[SPIRou\footnote{SPectropolarim\`etre InfraROUge.},][]{Donati2018haex.bookE.107D}; \citealp[IRD\footnote{InfraRed Doppler spectrograph.},][]{Kotani2018SPIE10702E..11K}; \citealp[HPF\footnote{Habitable Zone Planet Finder.},][]{Mahadevan2014SPIE.9147E..1GM}; \citealp[NIRPS\footnote{Near Infra-Red Planet Searcher.},][]{Wildi2017SPIE10400E..18W}), aim to address this bias.
Here we report the discovery of two planet candidates orbiting \object{Teegarden's Star}. First, we introduce the host star in Sect.~\ref{sec:Tgard}. The RV measurements and the intensive photometric monitoring are presented in Sect.~\ref{sec:Obs}, and they are analysed in Sects.~\ref{sec:Analysis} and \ref{sec:Photo}. The results are discussed in Sect.~\ref{sec:Discussion}.

\section{\label{sec:Tgard}Teegarden's Star}

\subsection{Basic stellar parameters}

\teegarden was discovered in this century by \cite{Teegarden2003ApJ...589L..51T}. It is the 24th nearest star to the Sun\footnote{\url{http://www.astro.gsu.edu/RECONS/TOP100.posted.htm}} with a distance of 3.831\,pc. The spectral type is M7.0\,V \citep{Alonso-Floriano2015A&A...577A.128A}, making it the brightest representative of this and later spectral classes with a $J$ magnitude of 8.39\,mag ($V{\,=\,}15.08$\,mag).
\cite{Schweitzer2019A&A...625A..68S} derived the effective temperature $T_\mathrm{eff}$, metallicity [Fe/H], and surface gravity $\log g$ from fitting PHOENIX synthetic spectra \citep{Husser2013A&A...553A...6H} to CARMENES spectra following the method of \cite{Passegger2018A&A...615A...6P}. They obtained the luminosity~$L$ with Gaia DR2 parallax and integrated broad-band photometry as described in \citet[in prep.]{Cifuentespre038}. \cite{Schweitzer2019A&A...625A..68S} then estimated the stellar radius $R$ via Stefan-Boltzmann's law and finally a stellar mass $M$ of $0.089\,M_\odot$ by using their own linear mass-radius relation.
Table~\ref{tab:stellar-params} summarises a number of basic parameters mostly compiled from our M-dwarf database Carmencita \citep{Caballero2016csss.confE.148C}. For comparison, we also provide the $T_\mathrm{eff}$ from \citet{Rojas2012ApJ...748...93R} and [Fe/H] from \citet{Dieterich2014AJ....147...94D}, pointing to disagreement of stellar parameter determination at the very cool end.

\begin{figure}
    \centering
    \includegraphics[width=1\linewidth]{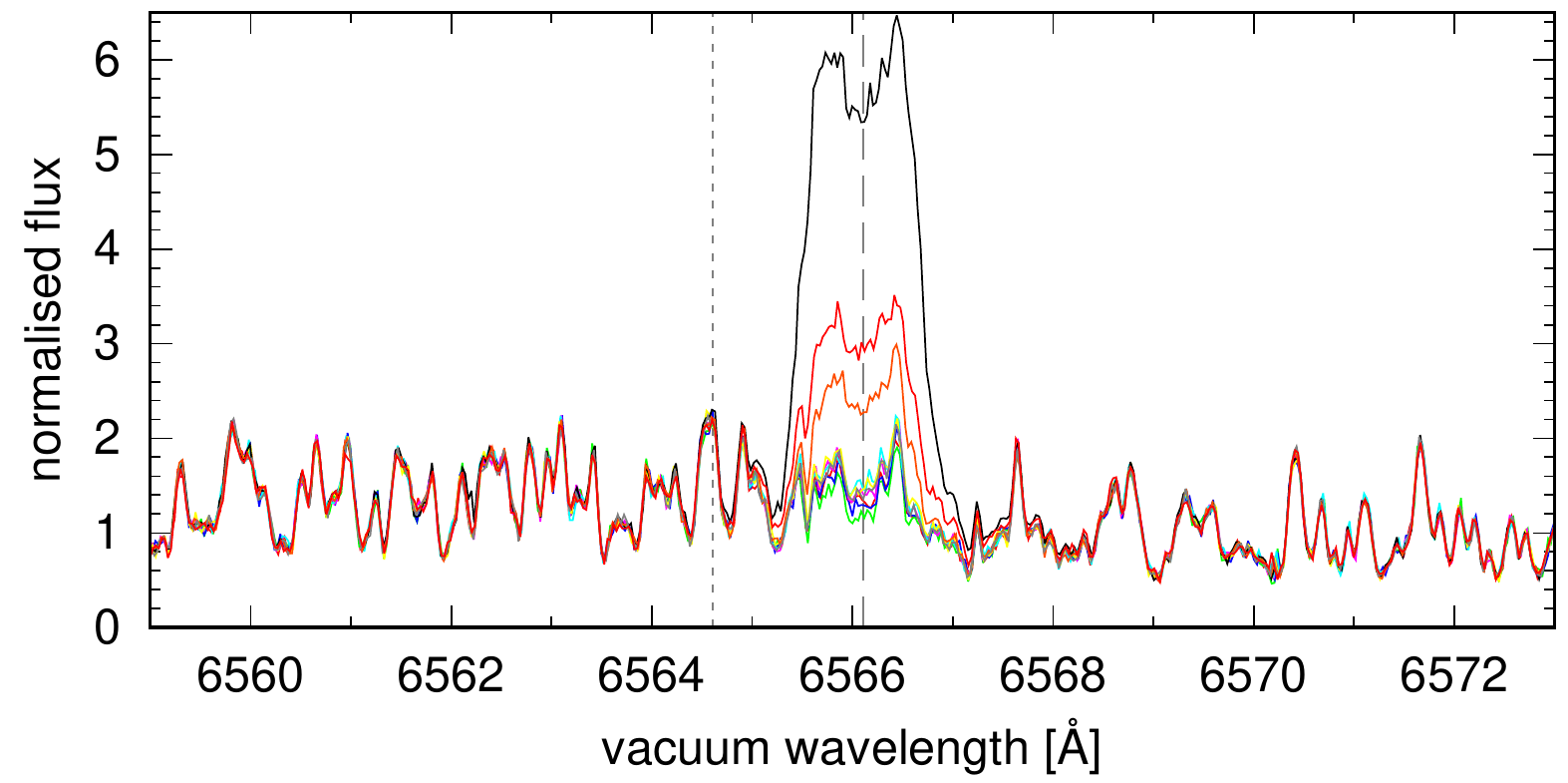}
    \caption{\label{fig:Halpha}Largest H$\alpha$ flare (black) in our CARMENES data. These ten spectra were taken between 2017-10-26 and 2017-11-01. The vertical line marks H$\alpha$ at rest wavelength ($\lambda_\mathrm{vac}=6564.61$\,\AA, grey dotted) and at a shift of $\gamma= +68.375$\,km/s (grey dashed).}
\end{figure}

\begin{table}
    \caption{\label{tab:stellar-params}Stellar parameters for \teegarden.}
    \centering
    \begin{tabular}{@{}lcc}
        \toprule
        Parameter & Value & Ref. \tabularnewline
        \hline
        \noalign{\medskip}
        Discovery name    & \teegarden                & Tee03\tabularnewline
        Alias name        & GAT 1370                  & Gat09\tabularnewline
        Karmn\tablefootmark{a} & J02530+168           & Cab16\tabularnewline
        $\alpha$          & ~~02 53 00.89             & \gaia\tabularnewline
        $\delta$          &  +16 52 52.6~~            & \gaia\tabularnewline
        $\mu_\alpha \cos{\delta}$ [mas/yr]& $~~3429.53 \pm 0.33$ & \gaia\tabularnewline
        $\mu_\delta$ [mas/yr] & $ -3806.16 \pm 0.31$  & \gaia\tabularnewline
        $l$ [deg]\tablefootmark{b} & 45.82636650288   & \gaia\tabularnewline
        $b$ [deg]\tablefootmark{b} & +0.30111982472   & \gaia\tabularnewline
        $\pi$ [mas]       & $261.01  \pm 0.27$        & \gaia\tabularnewline
                          & $259.6~~ \pm 0.6~~$       & Wei16\tabularnewline
                          & $259.3~~ \pm 0.9~~$       & Gat09\tabularnewline
        $d$ [pc]          & $3.831 \pm 0.004$         & \gaia\tabularnewline
        $V$ [mag]         & $15.08 \pm 0.12 $         & Hen15\tabularnewline
        $J$ [mag]         &  $8.39 \pm 0.03 $         & 2MASS\tabularnewline

        Sp. type          & M7.0\,V                   & Alo15\tabularnewline
        $T_{\rm eff}$ [K] & $2904\pm51$               & Sch19\tabularnewline
                          & $2637\pm30$               & Roj12\tabularnewline
        $L$ [$L_{\odot}$] & $0.000\,73 \pm 0.000\,01$ & Sch19\tabularnewline
                          & $0.000\,73 \pm 0.000\,02$ & Die14\tabularnewline
        $R$ [$R_{\odot}$] & $0.107 \pm 0.004$         & Sch19\tabularnewline
                          & $0.127 \pm 0.004$         & Die14\tabularnewline
        $M$ [$M_{\odot}$] & $0.089 \pm 0.009$         & Sch19\tabularnewline
        {[Fe/H]} [dex]    & $-0.19 \pm 0.16$          & Sch19\tabularnewline
                          & $-0.55 \pm 0.17$          & Roj12\tabularnewline

        $U$ [km/s]        & $-69.46 \pm 0.31$         & Cor16\tabularnewline
        $V$ [km/s]        & $-71.17 \pm 0.15$         & Cor16\tabularnewline
        $W$ [km/s]        & $-58.68 \pm 0.25$         & Cor16\tabularnewline
        $\gamma$ [km/s]   & +68.375                   & Rei18\tabularnewline
        $v\sin i$ [km/s]  & $< 2$                       & Rei18\tabularnewline
        $\log L_{\mathrm{H}\alpha}/L_\mathrm{bol}$ & $-5.37$ & This work\tabularnewline
        $\log L_\mathrm{X}/L_\mathrm{bol}$ & $< -4.23$ & Ste14, this work\tabularnewline
        Age [Gyr]         & $>8$                      & This work\tabularnewline
        \bottomrule
    \end{tabular}
    \tablefoot{
        \tablefoottext{a}{Carmencita identifier.}
        \tablefoottext{b}{Heliocentric ecliptic longitude and latitude.}
    }
    \tablebib{
        2MASS: \citet{Skrutskie2006AJ....131.1163S};
        Alo15: \citet{Alonso-Floriano2015A&A...577A.128A};
        Cab16: \citet{Caballero2016csss.confE.148C};
        Cor16: \citet{Cortes2016UCM-PhD};
        Die14: \citet{Dieterich2014AJ....147...94D};
        \gaia{}: \citet{GAIA2018A&A...616A...1G};
        Gat09: \citet{Gatewood2009AJ....137..402G};
        Hen15: \citet{Henden2015AAS...22533616H};
        Rei18: \citet{Reiners2018A&A...612A..49R};
        Roj12: \citet{Rojas2012ApJ...748...93R};
        Sch19: \citet{Schweitzer2019A&A...625A..68S};
        Ste14: \citet{Stelzer2014MNRAS.442..343S};
        Tee03: \citet{Teegarden2003ApJ...589L..51T};
        Wei16: \citet{Weinberger2016AJ....152...24W}.
    }
\end{table}

\subsection{\label{sec:age}Activity, rotation period, and age estimates}

During their lifetime, low-mass stars lose angular momentum, leading to slower rotation and lower magnetic activity at old ages \citep[e.g.][]{Barnes2007ApJ...669.1167B, Irwin2009IAUS..258..363I}. Two useful measures of magnetic activity are non-thermal emissions in H$\alpha$ and in X-rays. \citet{Reiners2018A&A...612A..49R} reported for \teegarden a normalised H$\alpha$ luminosity of $\log L_{\mathrm{H}\alpha}/L_\mathrm{bol} = -5.18$ based on one CARMENES spectrum. The average over all CARMENES spectra is $\log L_{\mathrm{H}\alpha}/L_\mathrm{bol} = -5.25$. 
This value can be compared to other ultra-cool dwarfs, for example those measured in \citet{Reiners2010ApJ...710..924R}. Out of their sample of 24 stars of spectral types M7.0\,V and M7.5\,V, only one exhibits H$\alpha$ emission as low as \teegarden (\object{LP 321-222}: $\log L_{\mathrm{H}\alpha}/L_\mathrm{bol} < -5.27$). In \cite{Jeffers2018A&A...614A..76J}, it also belonged to the least active stars in this spectral bin.
However, \teegarden exhibits occasional flares (Fig.~\ref{fig:Halpha}), and $\log L_{\mathrm{H}\alpha}/L_\mathrm{bol}$ varies from $-5.58$ up to $-4.26$ during the CARMENES observations (see also Fig.~\ref{fig:ActInd}).

The stellar rotation period of \teegarden has not been measured. \citet{Newton2016ApJ...821...93N} listed a photometric period of 18.460\,d with an amplitude of $0.0013 \pm 0.0021$\,mag based on MEarth data. Because the uncertainty is larger than the amplitude itself, a detection cannot be claimed.
The line profiles of \teegarden do not exhibit any significant rotational broadening in the CARMENES spectra \citep[$v\sin i<2$\,km/s,][]{Reiners2018A&A...612A..49R}. When a stellar radius of 0.107\,$R_\odot$ is assumed, this implies a minimum stellar rotation longer than $P_\mathrm{rot}>2.7$\,d.
To estimate the rotation period of \teegarden, we can use H$\alpha$-$P_\mathrm{rot}$ relations. \cite{Newton2017ApJ...834...85N} showed that stars with masses of about 0.1\,$M_{\odot}$ and no significant H$\alpha$ emission rotate at periods $P_\mathrm{rot} \approx 100$\,d or longer (see their Fig.~5). Similar estimates can be obtained from the works of \citet[][Fig.~8]{West2015ApJ...812....3W} or \citet[][Fig.~2]{Jeffers2018A&A...614A..76J}.

The age of \teegarden can be estimated in various ways. Again, the normalised H$\alpha$ luminosity can be used. \citet{West2008AJ....135..785W} showed the distribution of H$\alpha$ activity from an analysis of 38\,000 M dwarfs. Their low-resolution Sloan Digital Sky Survey (SDSS) spectra are sensitive to H$\alpha$ equivalent widths of 1\,\AA{} in emission, which translates into $\log L_{\mathrm{H}\alpha}/L_\mathrm{bol}$ well above $-5$ at this spectral type. From their data they concluded that stars of spectral type M7\,V show emission in excess of this threshold for as long as $8.0_{-1.0}^{+0.5}$\,Gyr. Our average measurements of H$\alpha$ emission are significantly below this threshold, so that \teegarden would be classified as inactive according to the work of \citet{West2008AJ....135..785W}.

An X-ray measurement of \teegarden was attempted by \citet{Stelzer2014MNRAS.442..343S}, who reported an upper limit of $\log F_\mathrm{X}/\mathrm{mW\,m^2} < -13.03$ from the non-detection. Normalised with an updated bolometric flux of $\log F_\mathrm{bol}/\mathrm{mW\,m^2} < -8.79$ (calculated with distances and luminosity from Table~\ref{tab:stellar-params}), this upper limit of $\log L_\mathrm{X}/L_\mathrm{bol} < -4.23$ does not provide strong constraints on age or rotation rate, but it is consistent with our conclusions from the H$\alpha$ measurement \citep[see, e.g.,][]{Wright2011ApJ...743...48W, Reiners2014ApJ...794..144R}.

The age has been also estimated from a comparison to theoretical stellar evolutionary tracks. We have derived an age of $7\pm3$\,Gyr from a Bayesian approach \citep{delBurgo2018MNRAS.479.1953D} using the stellar evolution library PARSEC v1.2S, taking as input the iron-to-hydrogen ratio ($[\mathrm{Fe/H}] = -0.55 \pm 0.17$), the colour $r-J$ (determined from SDSS $r$ and 2MASS $J$ bands), and absolute magnitude $M_J$ (calculated from 2MASS $J$ and \gaia DR2 parallax), and their uncertainties.

According to our kinematic analysis \citep{Cortes2016UCM-PhD} with $U$, $V$, and $W$ velocities revised with the latest \gaia DR2 astrometry and our radial velocity, we conclude that \teegarden belongs to an evolved population of the Galaxy, probably the thick disc. This is consistent with our age estimates.

In summary, \teegarden is relatively magnetically quiet for its spectral type M7.0\,V. Most late-M~dwarfs known show higher levels of magnetic activity. Different approaches (H$\alpha$, X-ray, $v\sin i$, evolutionary track, and space motions) are compatible with the conclusion that \teegarden is most likely a very old (8--10\,Gyr) and very slowly rotating star ($P_{\rm rot} \gtrsim 100$\,d).

\subsection{Transit visibility zones}

\begin{figure}
    \centering
    \includegraphics[width=1\linewidth]{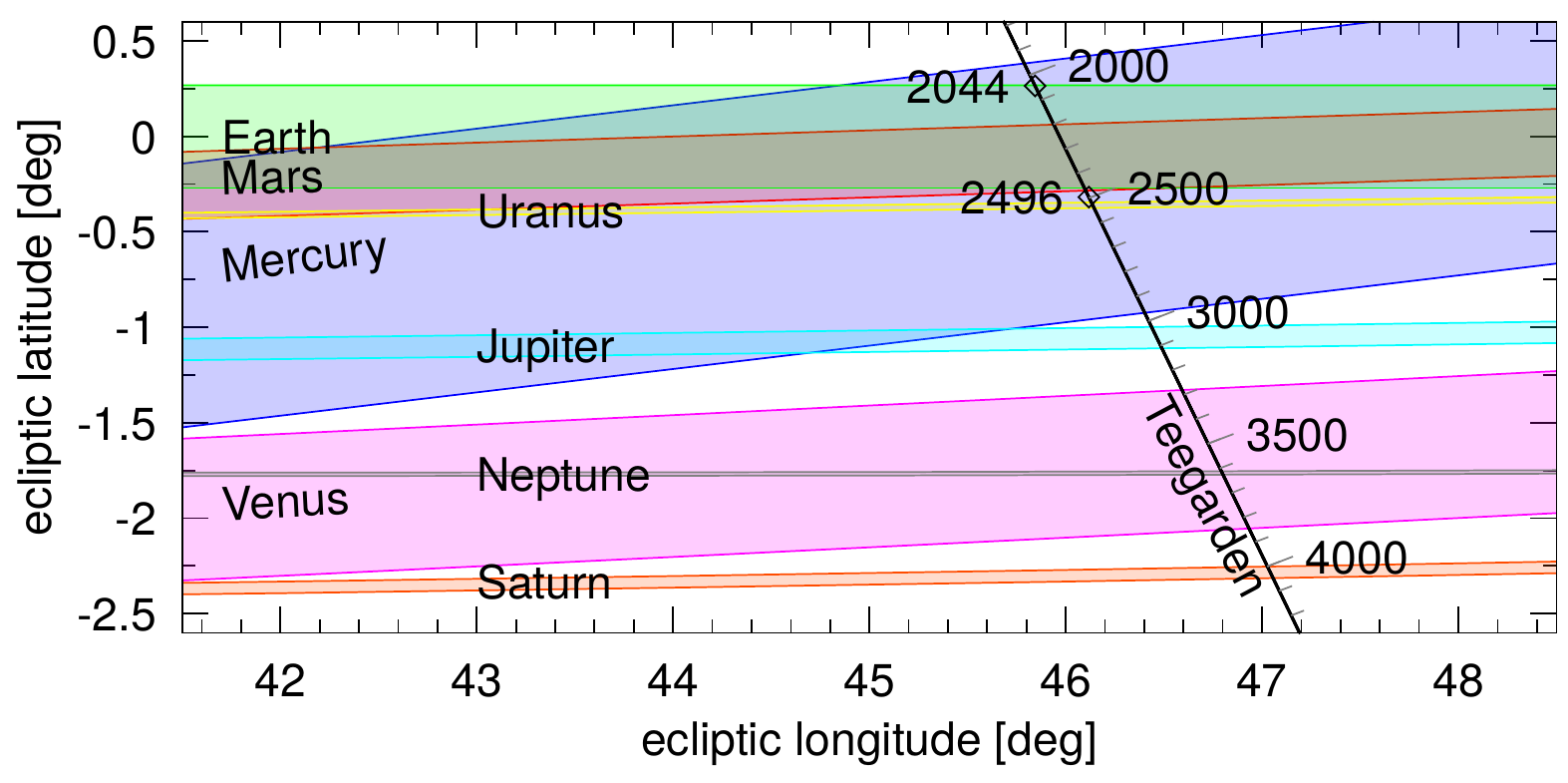}
    \caption{\label{fig:ecliptic}Transits of solar system objects as seen from \teegarden. Current location of transit regions for Earth (green), Mars (red), and Mercury (blue). The trace of \teegarden (black) due to its proper motion is marked with first Earth transit (2044) and last Earth transit (2496).}
\end{figure}

The ecliptic latitude of \teegarden is only 0.30\,deg, which means that the star is very close to the plane of Earth's orbit around the Sun. An observer within a band of $\pm0.25$\,deg could see the Earth transiting in front of our Sun \citep{Heller2016AsBio..16..259H, Wells2018MNRAS.473..345W}. Currently, \teegarden is outside the transit bands of Earth and Mars, while Mercury has already been visible in transit since 1956 (Fig.~\ref{fig:ecliptic}). However, accounting for the fast stellar proper motion (it has the 15th largest proper motion, \citealt{Wenger2000A&AS..143....9W}), 
 we derive that \teegarden will move into the transit band of Earth in 2044 and into that of Mars in 2190.
For more than 200 years, these three planets will exhibit transits as seen from \teegarden. This is also the maximum number of solar system planets that can be simultaneously observed in transit. In about 2438, the Mars transits will stop, and slightly later, in 2496, there will be the last transit of Earth, while Mercury will stay in total for a millennium until 2872 in the transit visibility zone. Over the next 1000\,yr, the transits of Venus, Jupiter, Saturn, and Neptune will be observable from \teegarden (Fig.~\ref{fig:ecliptic}).

According to \cite{Wells2018MNRAS.473..345W}, the probability of falling within the overlapping transit zones of Mercury, Earth, and Mars is $2.1\cdot10^{-4}$. So far, there is only one known exoplanet (K2-101~b, \citealt{Barros2016A&A...594A.100B}; $R = 2R_\oplus$, \citealt{Mann2017AJ....153...64M}) within three overlapping transit zones (Jupiter, Saturn, and Uranus).

\section{\label{sec:Obs}Observations}

\subsection{CARMENES spectra}

\begin{figure}
    \centering
    \includegraphics[width=1\linewidth]{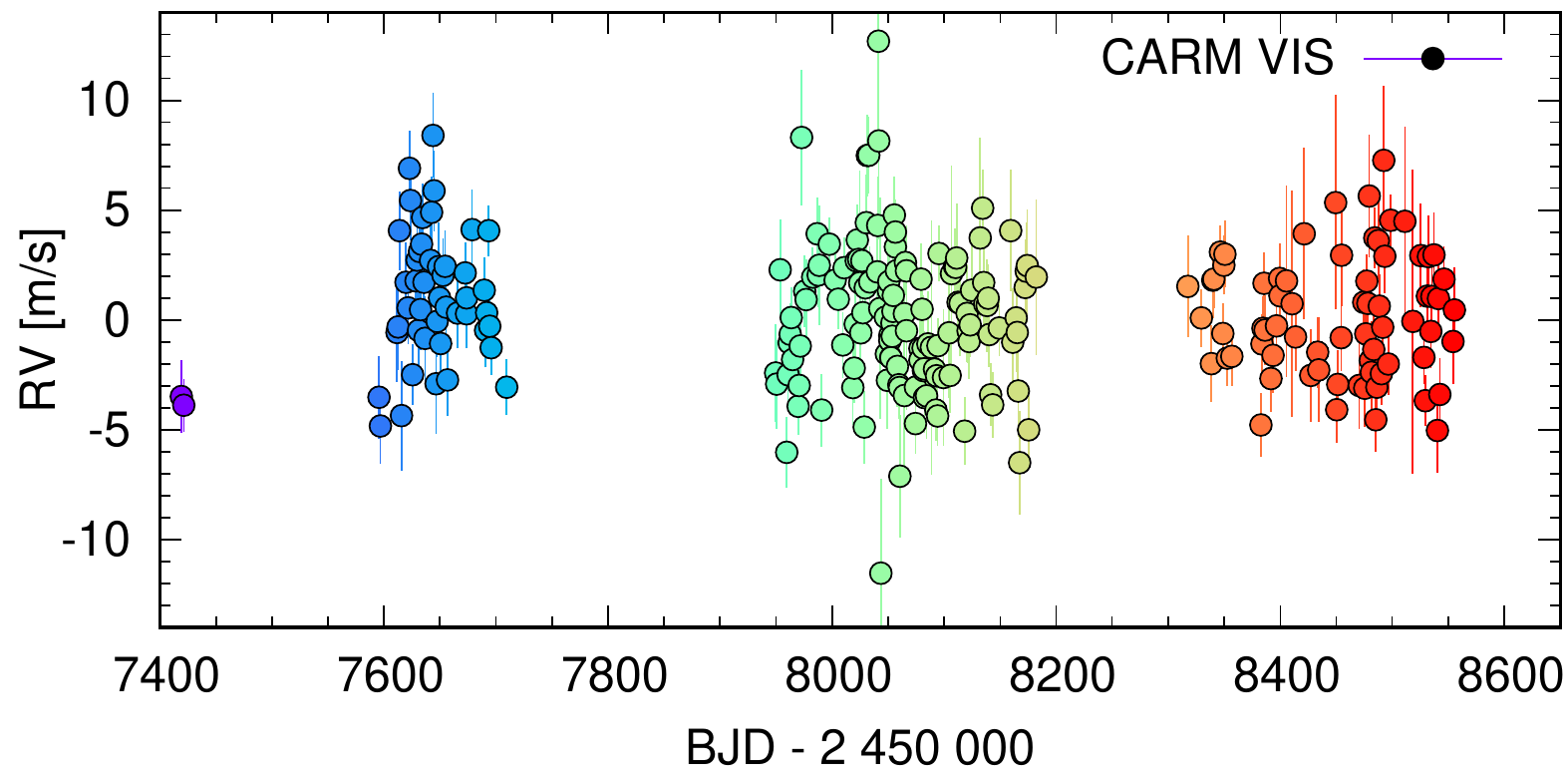}
    \includegraphics[width=1\linewidth]{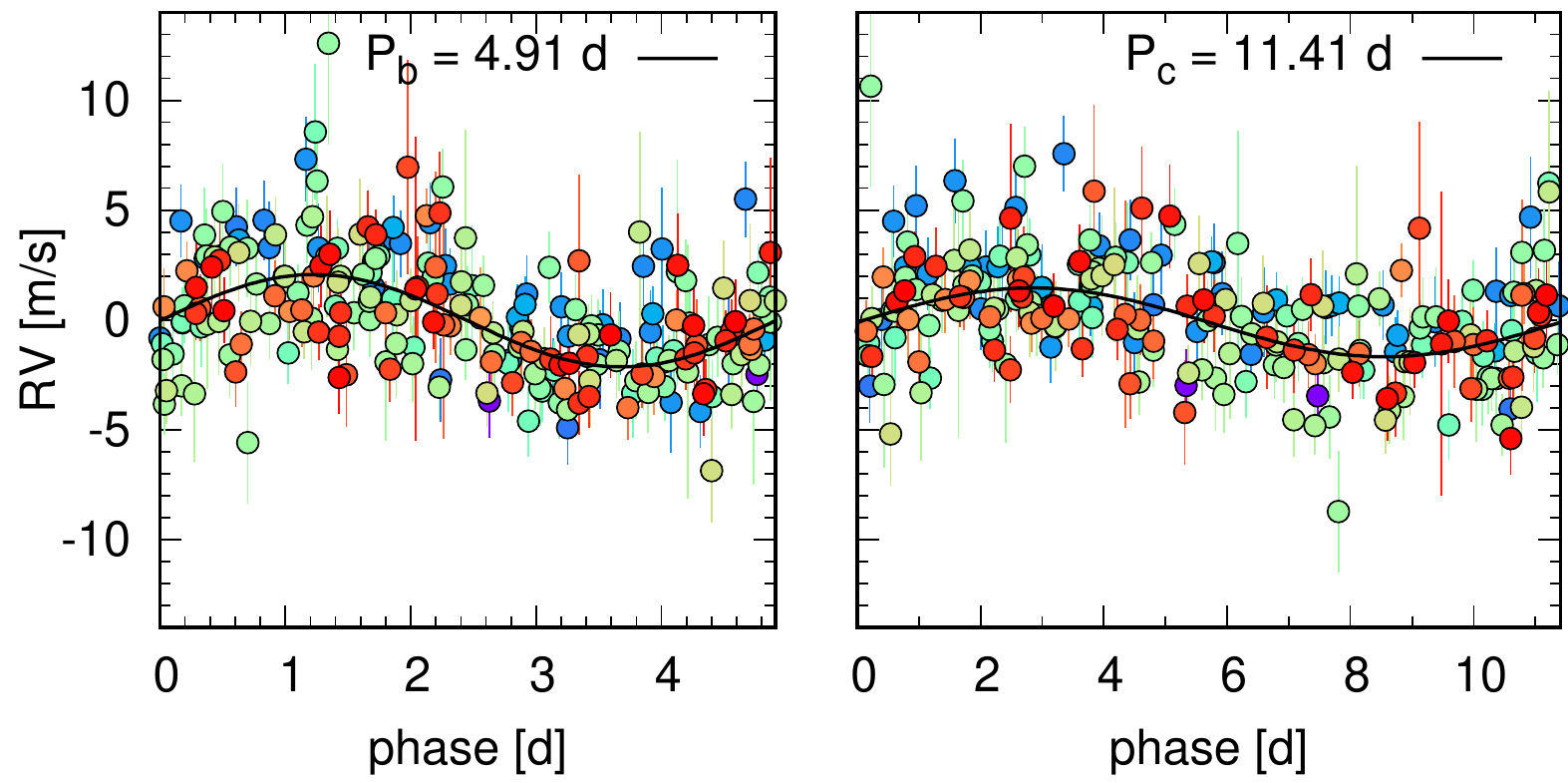}
    \caption{\label{fig:RV} {\em Top panel}: CARMENES RV time series for \teegarden colour-coded with time.
       {\em Bottom panel}: RVs phase-folded to the periods of planets b ({\em left}) and c ({\em right}). In each case the contribution of the other planet was subtracted.
    }
\end{figure}

The CARMENES instrument consists of a visual (VIS) and a near-infrared (NIR) spectrograph covering 520--960\,nm and 960--1710\,nm with a spectral resolution of 94\,600 and 80\,400, respectively \citep{Quirrenbach2014SPIE.9147E..1FQ,Quirrenbach2018SPIE10702E..0WQ}. It is located at the 3.5\,m Zeiss telescope at the Centro Astron\'omico Hispano-Alem\'an (Almer\'ia, Spain). Since the start of CARMENES in January 2016, we obtained 245 spectra of \teegarden within the CARMENES guaranteed time observations (GTO)  survey for exoplanets \citep{Reiners2018A&A...612A..49R}. The spectra have typical signal-to-noise ratios of 58 per pixel around 746\,nm  (see also Fig.~\ref{fig:Halpha}) and exposure times of 30\,min.
We extracted the spectra with the CARACAL pipeline, based on flat-relative optimal extraction \citep{Zechmeister2014A&A...561A..59Z} and wavelength calibration that was performed by combining hollow cathode lamps (U-Ar, U-Ne, and Th-Ne) and Fabry-P\'erot etalons \citep{Bauer2015A&A...581A.117B, Schaefer2018SPIE10702E..76S}. The instrument drift during the nights is tracked with the Fabry-P\'erot in the simultaneous calibration fibre.

We derived the radial velocities from the spectra with the SERVAL\footnote{\url{https://github.com/mzechmeister/serval}} code \citep{Zechmeister2018A&A...609A..12Z}.
The data were corrected for nightly zero-point offsets, which were determined using a large set of RV standard stars from the GTO sample as described in \cite{Trifonov2018A&A...609A.117T} and \cite{Tal-Or2019MNRAS.484L...8T}. These corrections were about 3\,m/s at the beginning of the survey and 1\,m/s after instrumental improvements (e.g. fibre coupling).
Figure~\ref{fig:RV} (top) shows the RV time series.

\subsection{\label{sec:photo_inst}Photometry}

\begin{figure}
    \centering
    \includegraphics[width=1\linewidth, trim={0 0.8cm 0 0}, clip]{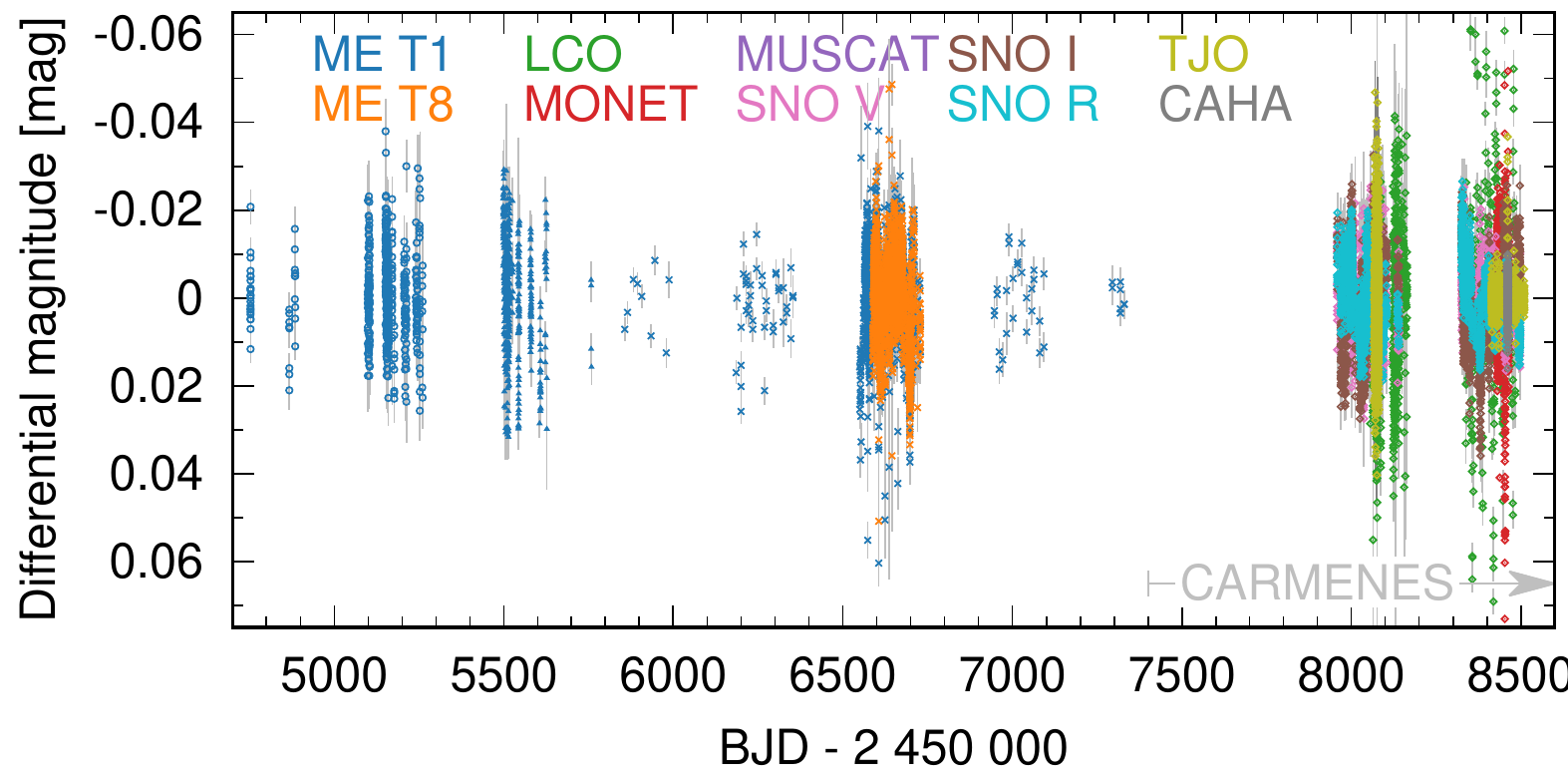}
    \includegraphics[width=1\linewidth, trim={0 0.8cm 0 0}, clip]{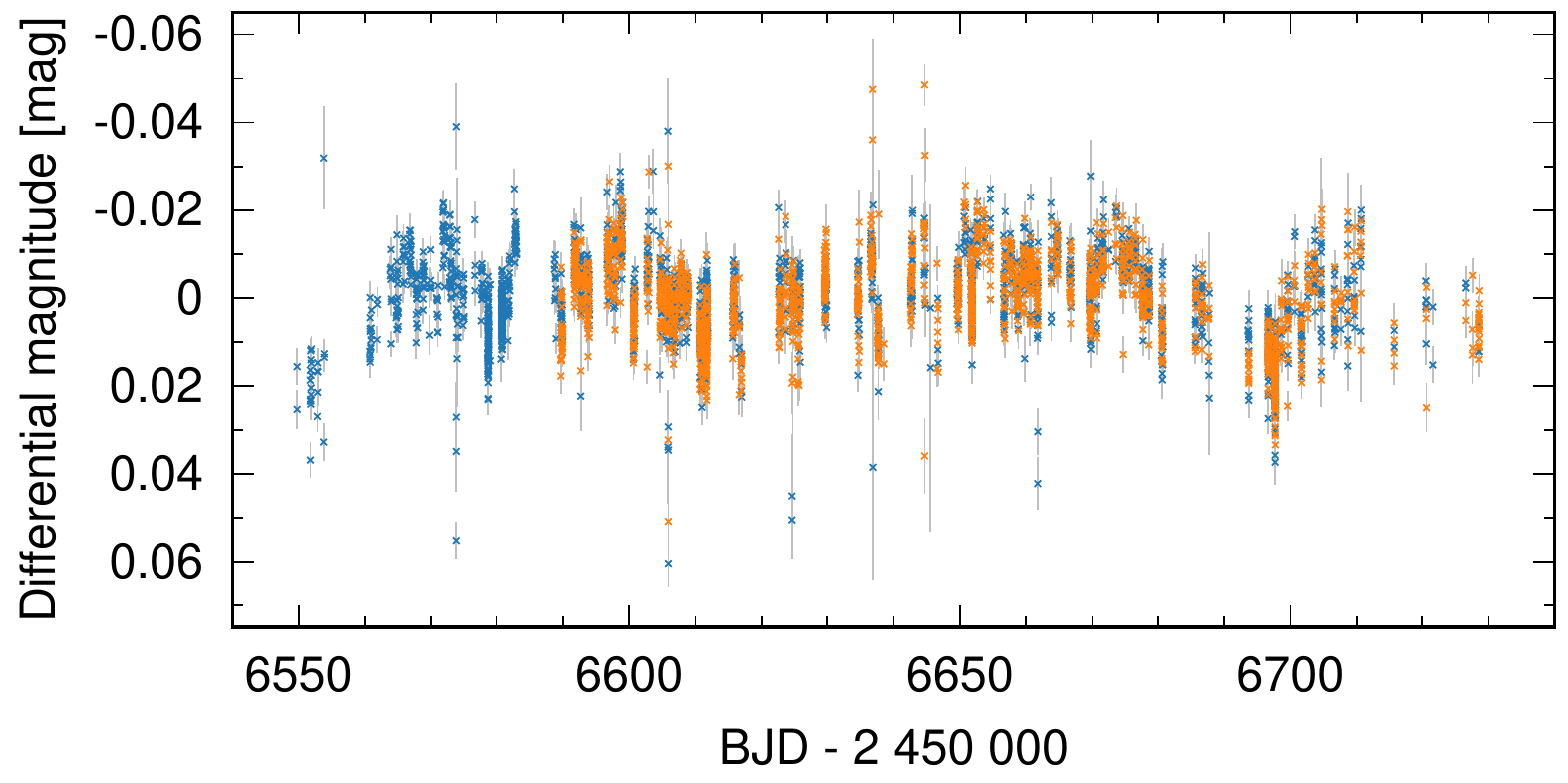}
    \includegraphics[width=1\linewidth]{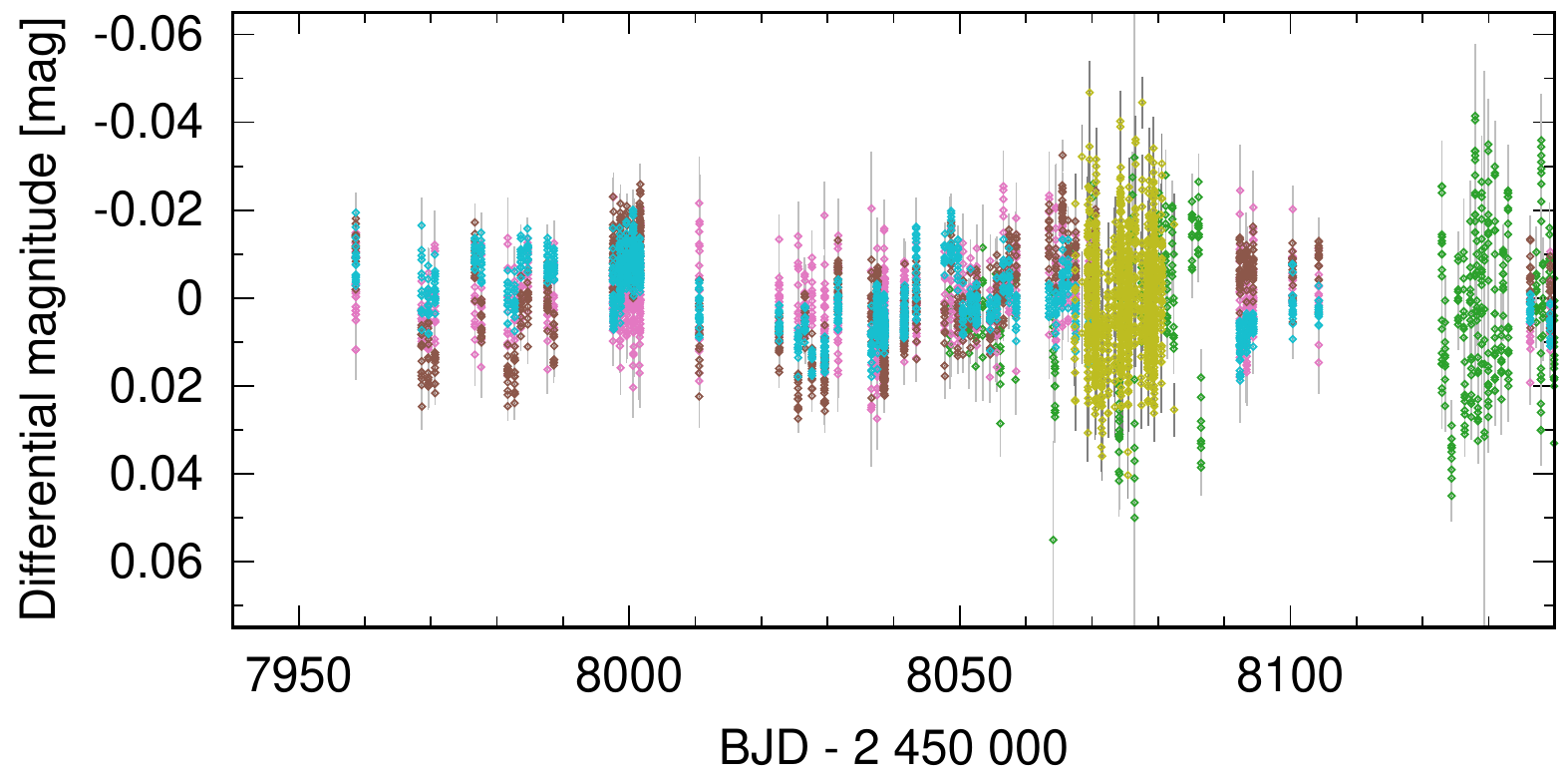}
    \caption{\label{fig:LC}Photometry from different instruments and bands ({\em top}) with 200\,d zooms to MEarth data ({\em middle}) and LCO, SNO, and TJO data from 2017 ({\em bottom}).}
\end{figure}

Complementary to the spectroscopic data, we also obtained our own and archival photometric data from various instruments. On one hand, the photometric data were used to estimate the rotation period of \teegarden from (quasi-)periodic brightness variations (Sect.~\ref{sec:phot_rot}). On the other hand, we searched for transits (Sect.~\ref{sec:phot_transit}) guided by the orbital solutions from the RV analysis. In Appendix~\ref{sec:phot_inst} we briefly describe the seven instruments whose basic properties are summarised in Table~\ref{tab:photometry}. Figure~\ref{fig:LC} and Table~\ref{tab:photometry_sets} provide an overview of the available photometric data, temporal sampling, and precision. As explained in Sect.~\ref{sec:phot_rot}, we removed the strongest signal in each data set after the non-detection of a common periodicity for all data sets. This assumes that the strongest peak is due to remaining systematic effects.

All CCD measurements were obtained by the method of synthetic aperture photometry using typically 2$\times$2 binning. Each CCD frame was corrected in a standard way for bias, dark, and flat field by instrument-specific pipelines. From a number of nearby and relatively bright stars within the frames, the best sets were selected as reference stars.

\section{\label{sec:Analysis}Spectroscopic data analysis}

\subsection{Radial velocities and orbit parameters\label{sec:RVanalysis}}

\begin{table}
    \centering
    \caption{\label{tab:orbit}Orbital planet and statistical parameters of \teegarden b and c.}
    \begin{tabular}{@{}lcc}
        \toprule
        Keplerian parameters & Planet b & Planet c\\
        \midrule
        $P$ [d]        & $    4.9100^{+0.0014}_{-0.0014} $ & $    11.409^{ +0.009}_{ -0.009} $\tabnewline
        $K$ [m/s]      & $      2.02^{  +0.19}_{  -0.20} $ & $      1.61^{  +0.19}_{  -0.19} $\tabnewline
        $e$\tablefootmark{a}     & $      0.00^{  +0.16} $ & $      0.00^{  +0.16}           $\tabnewline
        $\omega$ [deg] & $        77^{    +52}_{    -79} $ & $       286^{    +101}_{    -74} $\tabnewline
        $t_{\rm p} - 2\,458\,000$ [d]      & $    52.3^{  +0.7}_{  -1.1} $ & $     53.2^{   +3.2}_{   -2.3} $\tabnewline
        \midrule
        Derived parameters\\
        \midrule
        $a$ [au]                 & $0.0252^{+0.0008}_{-0.0009}$ & $ 0.0443^{+0.0014}_{-0.0015} $\tabnewline
        $m \sin i$ [$M_\oplus$]  & $ 1.05^{   +0.13}_{  -0.12}$ & $    1.11^{   +0.16}_{  -0.15} $\tabnewline
        $m$ [$M_\oplus$]\tablefootmark{b}   & $ 1.25^{   +0.68}_{  -0.22}$ & $    1.33^{   +0.71}_{  -0.25} $\tabnewline
        $\sin i$\tablefootmark{b}   & \multicolumn{2}{c}{$      0.87^{   +0.12}_{  -0.31}$}\tabnewline
        $F$ [$S_\oplus$]\tablefootmark{c}  & $1.15^{ +0.08}_{-0.08}$  & $0.37^{+0.03}_{-0.03}$  \tabnewline
        \hline
        \noalign{\smallskip}
        Statistical parameters\\
        \midrule
        $N$          & \multicolumn{2}{c}{238}\tabularnewline
        $T$ [d]      & \multicolumn{2}{c}{1136}\tabularnewline
        Internal uncertainties [m/s] & \multicolumn{2}{c}{1.67}\tabularnewline
        Jitter [m/s] & \multicolumn{2}{c}{$1.21^{   +0.16}_{  -0.16}$}\tabnewline
        wrms [m/s]   & \multicolumn{2}{c}{2.06}\tabularnewline
        $\ln L$ & \multicolumn{2}{c}{$-520.98$}\tabularnewline
        $\Delta\ln L$ & \multicolumn{2}{c}{69.16}\tabularnewline
        \bottomrule
    \end{tabular}
    \tablefoot{
        \tablefoottext{a}{One-side distributed; 68\,\% interval.}
        \tablefoottext{b}{For geometrically randomly oriented orbits (uniform distribution in $\cos{i}$).}
        \tablefoottext{c}{Insolation with stellar parameters adopted from \cite{Schweitzer2019A&A...625A..68S}.}
    }
\end{table}

We secured 245 spectra with CARMENES. Seven spectra without simultaneous Fabry-P\'erot drift measurements are excluded from the RV analysis. This leaves 238 RVs that cover a time span of 1136\,d (Fig.~\ref{fig:RV}, top panel). The effective data uncertainty $(\sum \sigma^{-2}_i/N)^{-0.5}$ is 1.67\,m/s and the weighted root-mean-square (rms) is 2.82\,m/s, leaving an unexplained noise (stellar jitter, calibration noise,  or planetary signals) of 2.19\,m/s.

\begin{figure}
    \centering
    \includegraphics[width=1\linewidth]{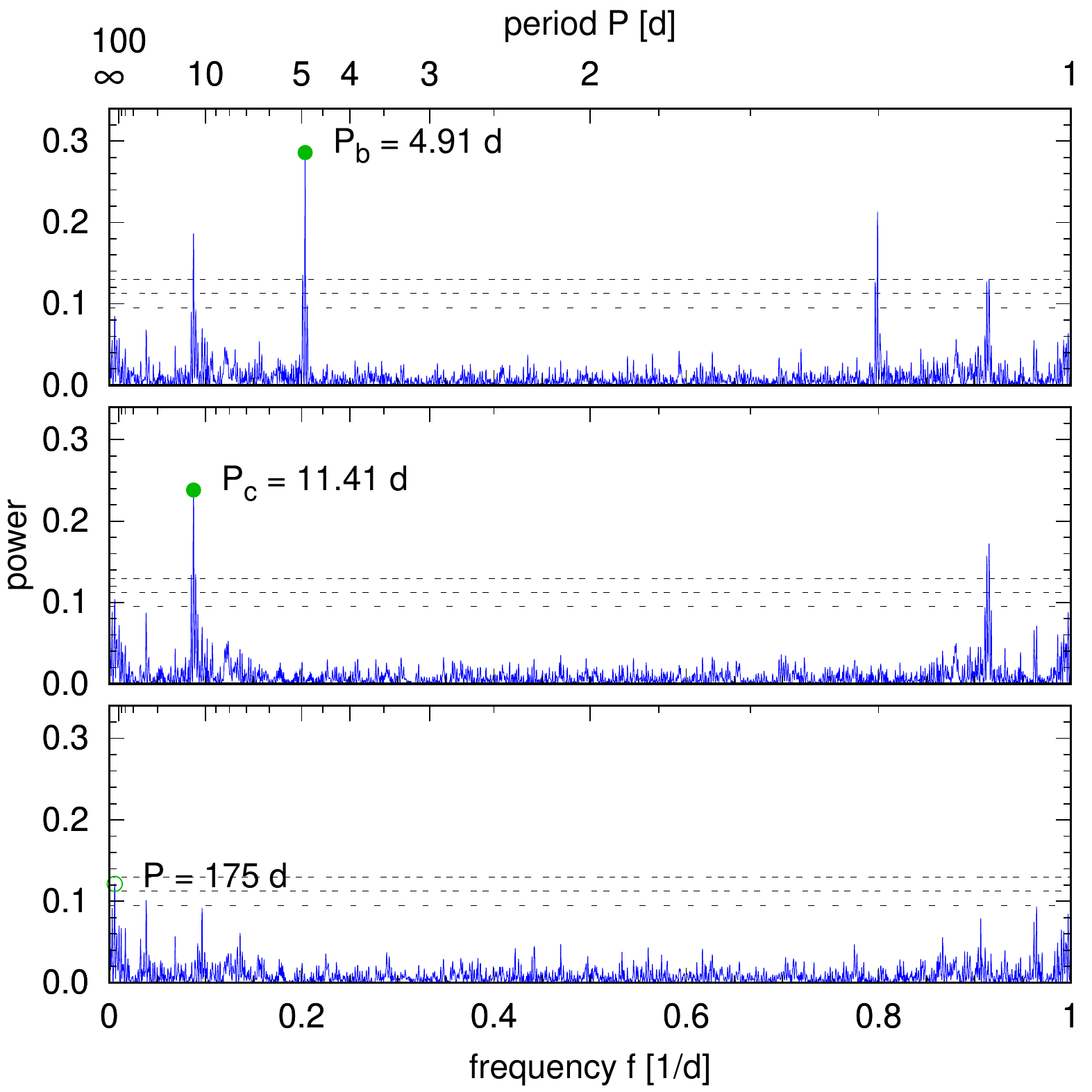}
    \caption{\label{fig:GLS}GLS periodogram of the RVs ({\em top}), after subtraction of the first signal $P_\mathrm{b}$ ({\em middle}), and after subtraction of the second signal $P_\mathrm{c}$ (residuals, {\em bottom}). Horizontal lines indicate the 1, 0.1, and 0.01\,\% FAP levels.}
\end{figure}

The GLS periodogram\footnote{\url{https://github.com/mzechmeister/GLS}} \citep{Zechmeister2009A&A...496..577Z} of the RVs (Fig.~\ref{fig:GLS}, top) shows two strong peaks at $P_\mathrm{b}=4.910\,\mathrm{d}$ and $P'_\mathrm{b} = 1.25\,\mathrm{d}$, which are related to each other by the typical one-day aliasing ($1/P'_\mathrm{b} = 1/\mathrm{d} - 1/P_\mathrm{b} $). The 4.910\,d period is clearly preferred (analytical false-alarm probability $\mathrm{FAP}_\mathrm{b} = 8.1\cdot 10^{-15}$ versus $\mathrm{FAP}'_\mathrm{b} = 5.9\cdot 10^{-10}$). In Fig.~\ref{fig:RV} (bottom left) the RVs are phase-folded to period $P_\mathrm{b}$ and corrected for the contribution of a second signal $P_\mathrm{c}$ (see below).

The residuals after subtraction of the signal $P_\mathrm{b}$ shows a further peak at $P_\mathrm{c} = 11.41$\,d (Fig.~\ref{fig:GLS}, middle panel) with an $\mathrm{FAP}=1.6\cdot 10^{-11}$. The RVs are folded to this period in Fig.~\ref{fig:RV} (bottom right).
After subtracting the signal $P_\mathrm{c}$, the periodogram has the highest peak at a (probably spurious) 175\,d period (Fig.~\ref{fig:GLS}, bottom). The signal at $P_{\mathrm d}=25.94$\,d ($K_{\mathrm d}=0.9$\,m/s) deserves further attention. It reaches an $\mathrm{FAP}=0.5\,\%$ and requires more data for confirmation.

We modelled the RVs with two Keplerian signals. We chose an additive noise term (jitter $\sigma_\mathrm{jit}$) to encompass further unmodelled sources (e.g. stellar activity, further planets, and additional instrumental or observational errors) that could be present in our data. The jitter was added in quadrature to the data uncertainties, which thereby re-weighted the data and balanced between a weighted and an unweighted fit (in case of $\sigma_\mathrm{jit}=0$ and $\sigma_\mathrm{jit}\rightarrow\infty$, respectively). The ten Keplerian parameters, the RV zero-point, and the jitter term were optimised by maximising a Gaussian likelihood. We used uniform priors for all parameters and explored the posteriors with a Markov chain Monte Carlo (MCMC) method \citep[{\tt emcee},][]{Foreman-Mackey2013PASP..125..306F} as well as with the curvature matrix.

Figure~\ref{fig:MCMC_bc} shows the posterior distribution for the Keplerian parameters from 500\,000 samples. Their medians and $1\sigma$ uncertainties (estimated from equal-tailed 50th--16th and 84th--50th percentile spreads) are summarised in Table~\ref{tab:orbit}. The covariance matrix, which was obtained by fitting a 12-dimensional paraboloid\footnote{\url{https://github.com/mzechmeister/serval/blob/master/src/paraboloid.py}} to the 150 best samples with the highest likelihood values, provides similar estimates as indicated by the error ellipses. The covariance matrix can handle only linear correlations.
The amplitudes are $K_\mathrm{b} = 2.0$\,m/s and $K_\mathrm{c} = 1.6$\,m/s; the weighted rms is 2.06\,m/s (best sample), and the jitter is 1.21\,m/s.

Both orbits are circular within the eccentricity uncertainties. The eccentricity posteriors cumulate near zero and have a one-sided distribution. Therefore, we give a 68\,\% percentile in Table~\ref{tab:orbit}.
In case of low eccentricities, the time of periastron passage $t_\mathrm{p}$ and periastron longitude $\omega$ are degenerate. Therefore, both parameters are poorly constrained, as indicated by their broad distribution, in particular for planet c. To preserve the phase information, which is encoded in their strong correlation (coefficients of 0.9987 and 0.985 in the covariance matrix), we provide a time of inferior conjunction $t_{\rm ic}$ (i.e. a potential transit time $t_{\rm tr}$) derived from the samples (Fig.~\ref{fig:MCMC_tr}). These transit times have smaller uncertainties than does the time of periastron passage.

From the Keplerian parameters, we derived planetary minimum masses using a Gaussian prior for the stellar mass (Fig.~\ref{fig:MCMC_mass}; because the stellar mass is not constrained by the likelihood, the posterior matches the prior). The minimum masses are $m_\mathrm{b}\sin i=1.05\MEarth$ and $m_\mathrm{c}\sin i=1.11\MEarth$. To estimate true masses, we further drew for each sample an inclination from a uniform distribution of $\cos i$, which corresponds to geometrically random orientations \citep{Kuerster2008A&A...483..869K}. The median values of the true masses are around 16\,\% higher than the minimum masses ($\cos i = 0.5 \rightarrow 1/\sin{i}=1.155$).

In the following sections we analyse the RV data in more detail. We also investigate further activity indicators to determine whether we can attribute both signals to a planetary origin.

\subsection{Temporal coherence of the RV signal}

\begin{figure}
    \centering
    \includegraphics[width=1\linewidth]{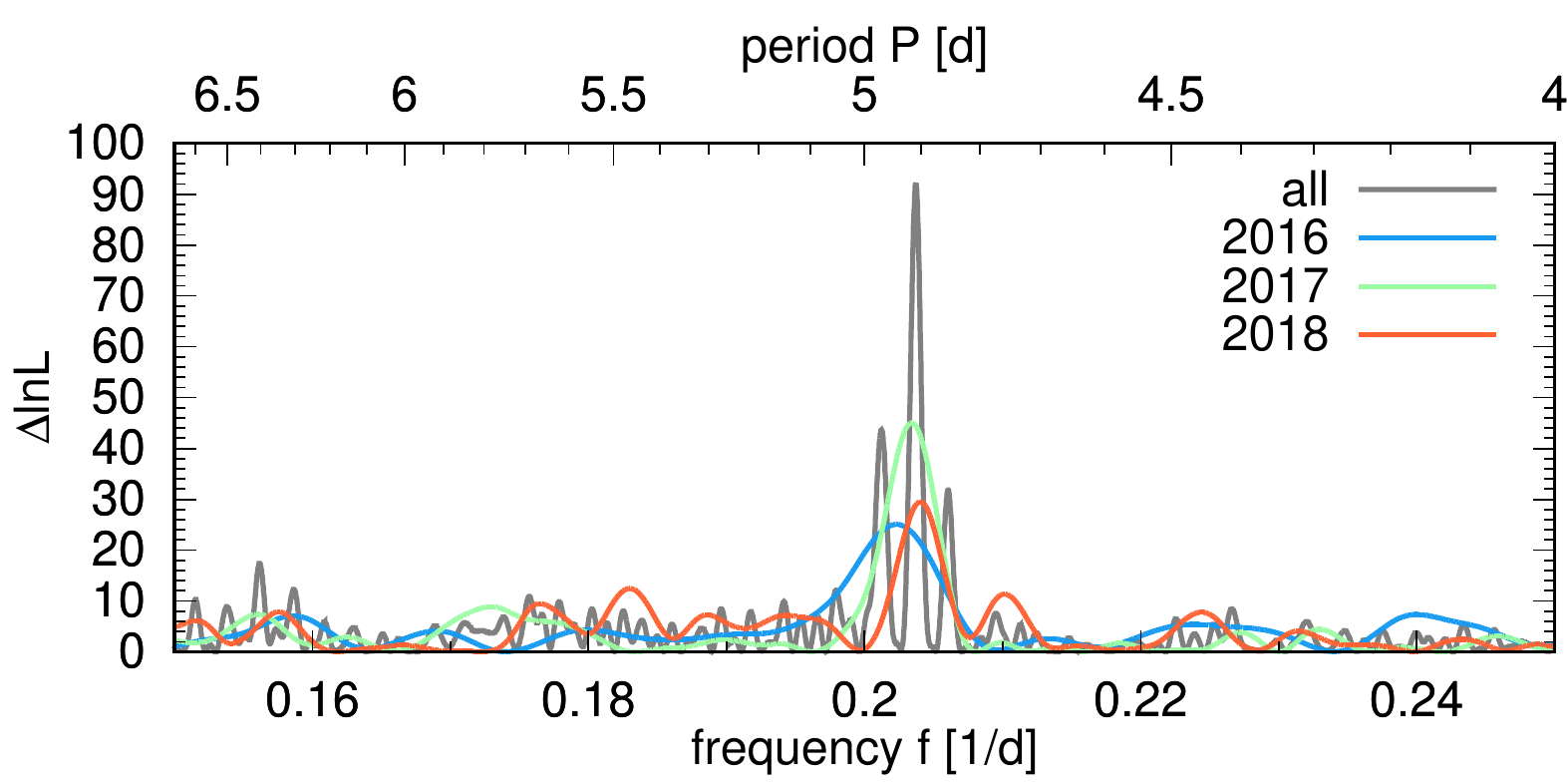}
    \includegraphics[width=1\linewidth]{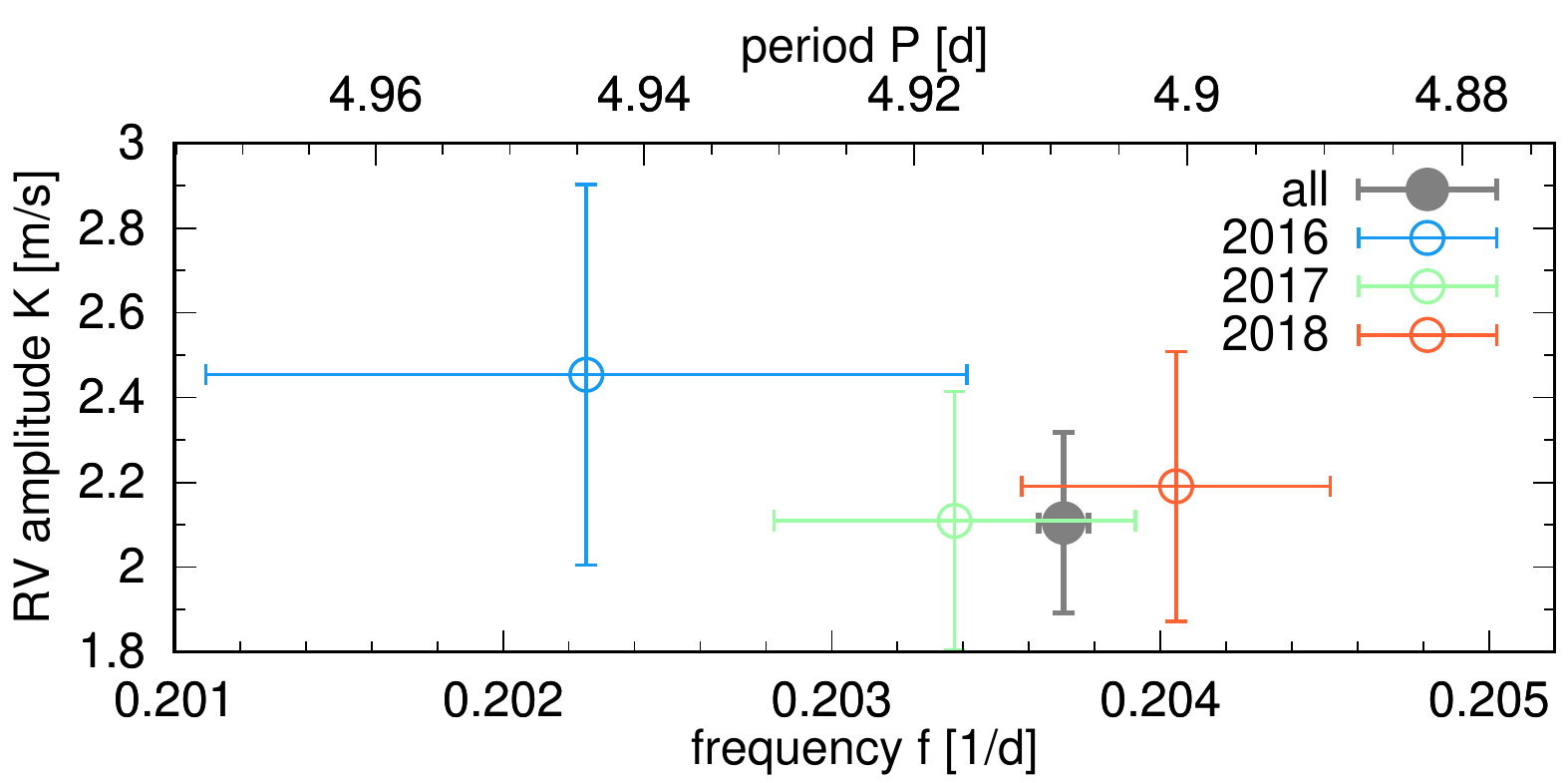}
    \caption{\label{fig:GLSseasons} {\em Top panel}: GLS periodograms around $P_\mathrm{b}$ for three CARMENES seasons (same colour-coding as in Fig.~\ref{fig:RV}). {\em Bottom panel}: RV amplitude and period of the highest peaks along with uncertainty estimates. The grey curve and (filled) symbol corresponds to the joint data set.}
\end{figure}

We subdivided the RV data into the three seasons to analyse the stability of the RV signal. We computed for each season the best-fitting period and amplitude (Fig.~\ref{fig:GLSseasons}). As expected, the parameter uncertainties increase due to the smaller number of data points (2016: 44, 2017: 125, and 2018: 67). However, the amplitudes and periods between the seasons are all consistent within their $1\sigma$ uncertainties. This result indicates that the 4.91\,d period is stable over at least 1000\,d.

\subsection{\label{sec:EchelleMLP}Chromatic coherence of the RV signal}

\begin{figure}
    \centering
    \includegraphics[width=1\linewidth]{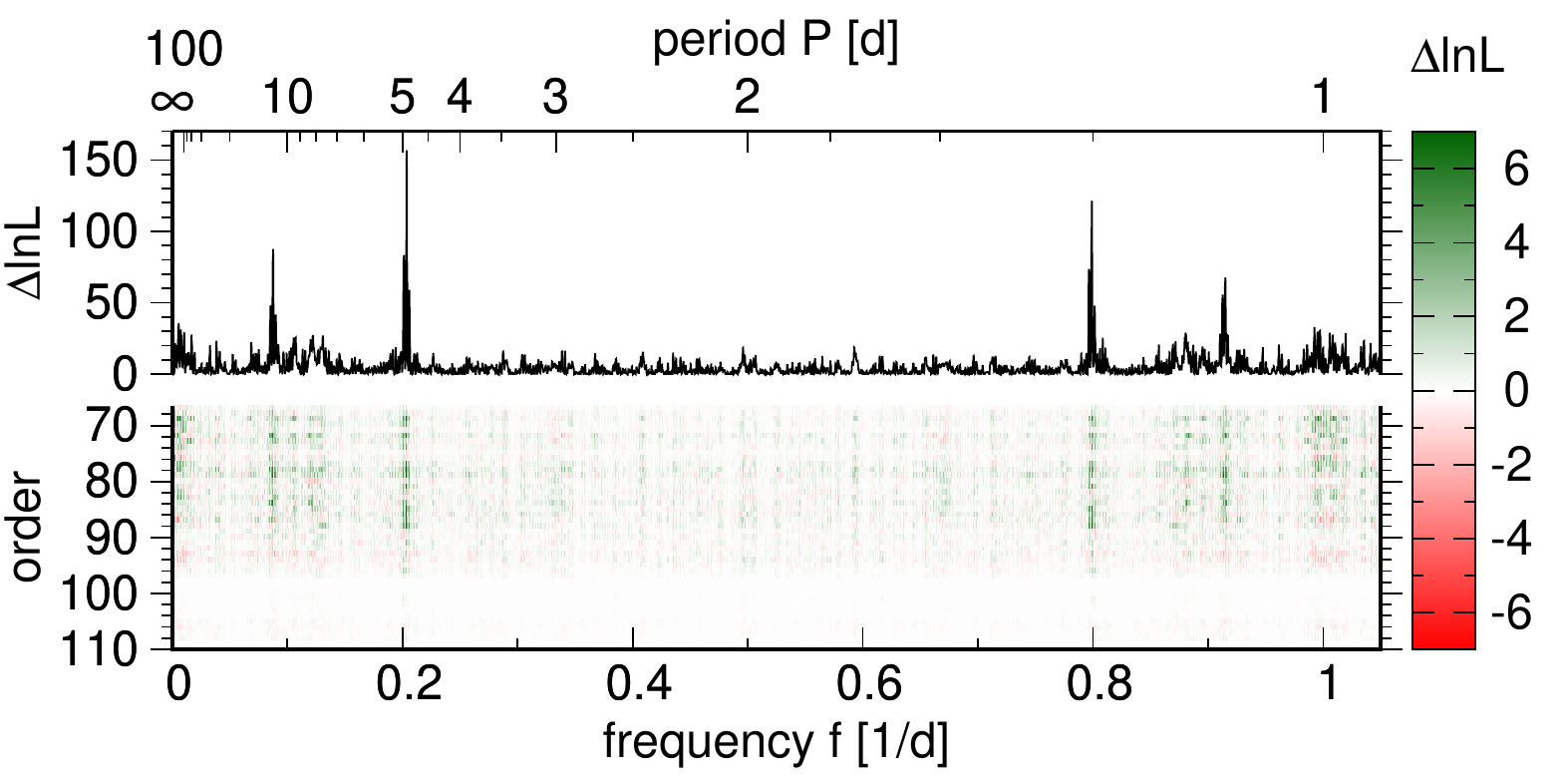}
    \includegraphics[width=1\linewidth]{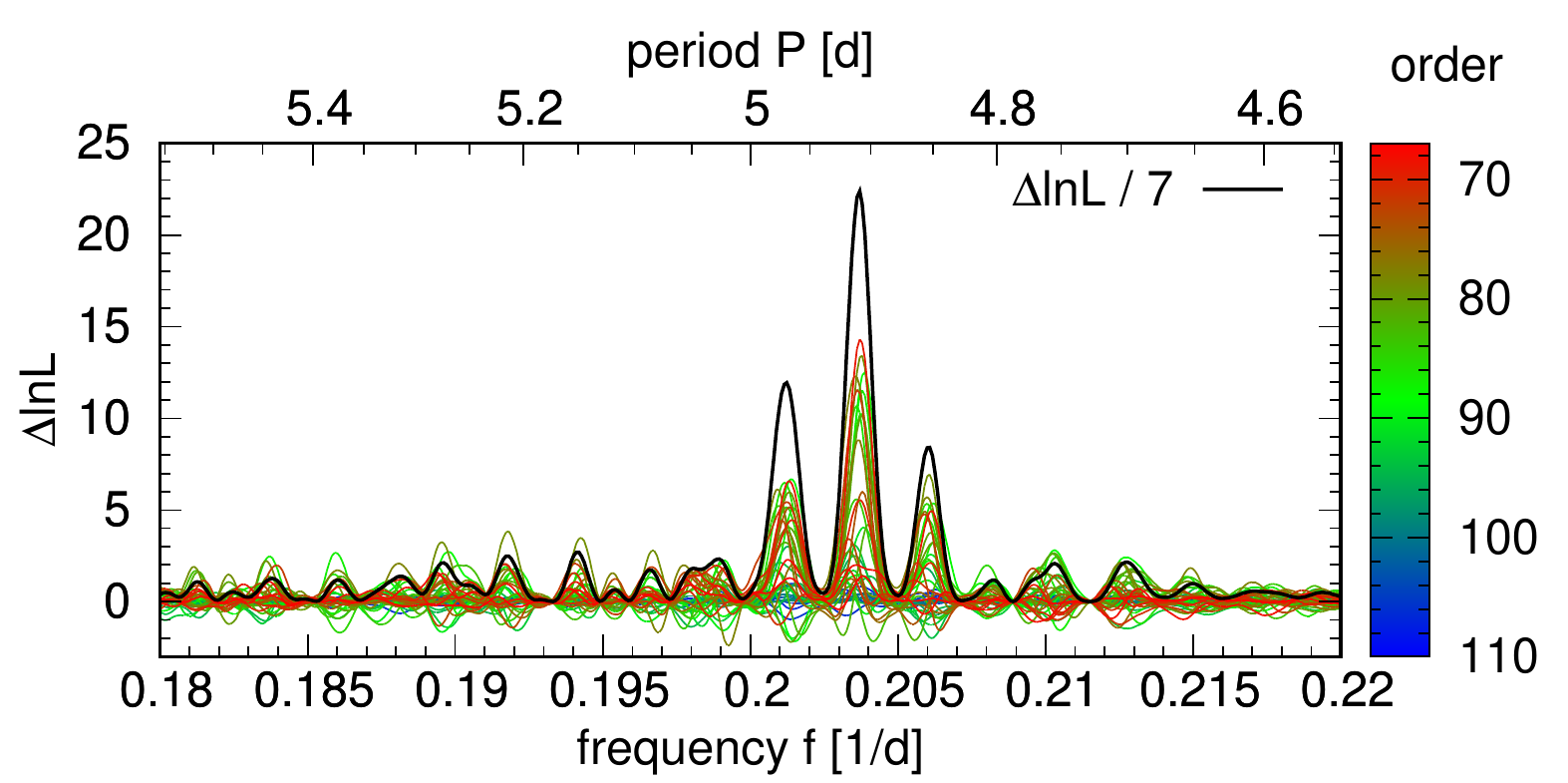}
    \caption{\label{fig:MLP-RV} {\em Top panel}: total maximum likelihood periodogram for the order-wise RVs (black curve) and an order-frequency map colour-coded by the likelihood contribution. 
    {\em Bottom panel}: Zoom-in to the period $P_\mathrm{b}$ and colour-coding with the echelle order. For the sake of clarity, the total likelihood (black curve) is downscaled by a factor~7.}
\end{figure}

The RV of each epoch is the average of individual RVs measured in 42 echelle orders. During this averaging, information about a possible wavelength dependence of the RV signal is lost.
To check the contributions of the orders to a periodic signal, we here treated each order as an independent instrument.
Therefore, we fit for a fixed trial frequency $f$ a sine simultaneously to all orders with one common amplitude and phase. Each order $o$ has its own offset $\mu_o$ and jitter term $\sigma_o$. We optimised the parameters by maximising the total likelihood
\begin{align}
   \ln L = \sum_o \ln L_o \label{eq:lnL},
\end{align}
where the likelihood for one order summed over all epochs $n$ is
\begin{align}
    \ln L_o = -\frac 1 2 \sum_{n} \ln 2\pi(\sigma^2_{n,o}+\sigma^2_o) + \frac{[RV_{n,o} - RV(t_{n,o})-\mu_o]^2}{\sigma^2_{n,o}+\sigma^2_o},
\end{align}
with the model $RV(t) = a\cos{\tau ft} + b\sin{\tau ft}$.
By looping over a sequence of frequencies $f_k$, we computed a maximum likelihood periodogram. Figure~\ref{fig:MLP-RV} shows the likelihood improvement with respect to the likelihood $L_0$ of a constant model ($a=b=0$),
\begin{align}
    \Delta\ln L = \ln L - \ln L_0 = \sum_o \ln L_o - \sum_o \ln L_{0,o} = \sum_o \Delta \ln L_o \label{eq:dlnL}.
\end{align}

The evaluation of this maximum likelihood periodogram takes much longer than the GLS periodogram because there are many more parameters (including non-linear jitter terms).

The contributions of the orders can of course be of different strengths and depend on their RV information content, as well as their signal-to-noise ratio. The log-likelihood change of individual orders can be even negative when the period in such an order has an amplitude or phase that is significantly different from the rest. However, the total log-likelihood change cannot be negative in this simultaneous approach ($\Delta\ln L \geq 0$). In Fig.~\ref{fig:MLP-RV} no order significantly contradicts ($\Delta\ln L_o > -2$) at $P_\mathrm{b}$, so the orders contribute in a positive way, as expected for a planetary signal.

\subsection{Near-infrared radial velocities}

\begin{figure}
    \centering
    \includegraphics[width=1\linewidth]{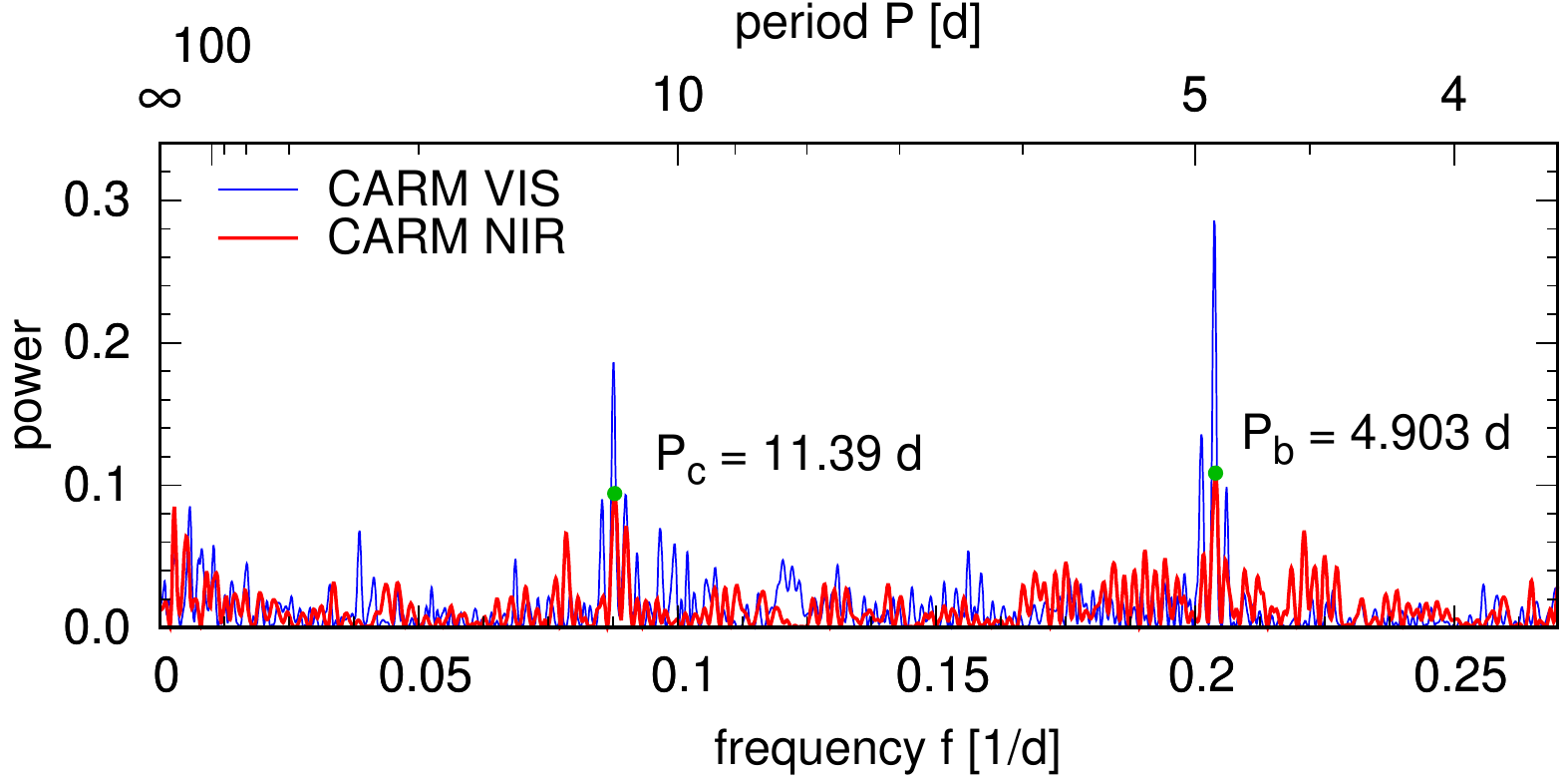}
    \caption{\label{fig:GLS_NIR} GLS periodogram of NIR RVs compared to VIS RVs.}
\end{figure}

In \cite{Reiners2018A&A...612A..49R} we showed with CARMENES data that M dwarfs have a lower RV content in the NIR than expected from previous predictions \citep[e.g.][]{Reiners2010ApJ...710..432R,Rodler2011A&A...532A..31R}. Coupled with the current instrument and pipeline performance, the NIR RVs of \teegarden have a weighted rms of 5.75\,m/s, making the detection of 1--2\,m/s signals difficult. Still, signals with $P_{\rm b}$ and $P_{\rm c}$ are the highest peaks in the GLS periodogram of the NIR RVs (Fig.~\ref{fig:GLS_NIR}). The formal FAP of $P_{\rm b}$ is 0.1\,\%, that is, significant. The small signal at $P_{\rm d}$ in the VIS RVs is not seen in the NIR RVs, however.

\subsection{Activity indicators}

\begin{figure}
    \centering
    \includegraphics[width=1\linewidth]{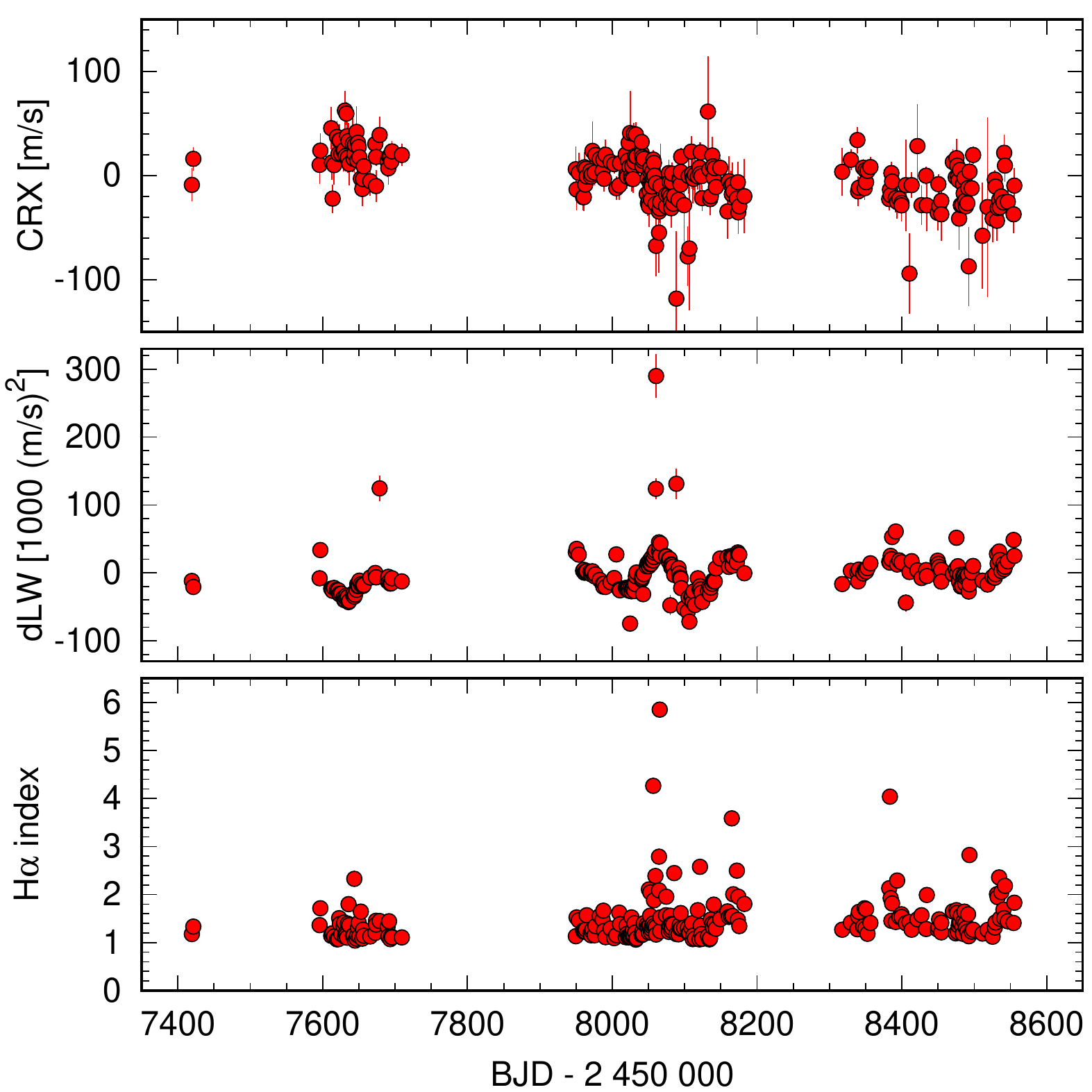}
    \caption{\label{fig:ActInd}Time series of the activity indicators: chromatic index CRX ({\em top}), differential line width dLW ({\em middle}), and H$\alpha$ index ({\em bottom}).}
\end{figure}

\begin{figure*}
    \centering
    \includegraphics[width=1.\linewidth]{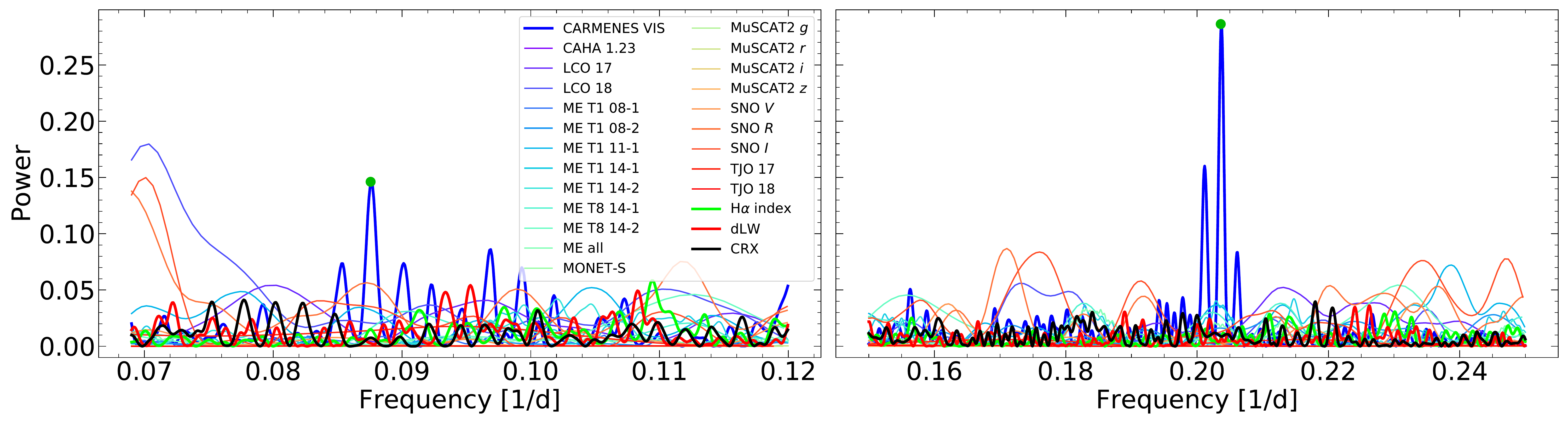}
    \caption{\label{fig:ActivityAll}GLS periodograms of photometry and various activity indicators compared to CARMENES VIS RVs (thick blue line) around $P_{\rm c}$ ({\em left panel}) and $P_{\rm b}$ ({\em right panel}).}
\end{figure*}

We extracted three activity indicators from the CARMENES spectra as described in \cite{Zechmeister2018A&A...609A..12Z}. The time series of the chromatic index (CRX), the differential line width (dLW), and the H$\alpha$ index are shown in Fig.~\ref{fig:ActInd} and their periodograms in Fig.~\ref{fig:GLS_ActInd}.

The CRX time series has some negative excursions around ${\rm BJD}-2\,450\,000 = 8050 ... 8100$, which coincides with dLW excess and also increased H$\alpha$ activity. During this time, the largest H$\alpha$ emission occurred (Fig.~\ref{fig:Halpha}). We see no clear periodicity for CRX (Fig.~\ref{fig:ActivityAll}). The highest periodogram peak is at 1\,000\,d, which captures the slight trend. After its removal, the highest peak is at 120\,d, which captures the previous feature of increased H$\alpha$ activity. No power is found at the planet periods during this pre-whitening. This result agrees with the analysis from Sect.~\ref{sec:EchelleMLP}, which exploits the wavelength-dependent information in a different way.

The differential line width dLW as well as the full width at half-maximum (FWHM) or contrast are also used as activity indicators.
However, they can be also affected by instrumental and observational effects, such as focus change of the spectrograph, sky-background, or broadening due to differential barycentric motion. We realised that dLW outliers are indeed caused by a close Moon separation during cloudy nights, which leads to an increased sky background. Ignoring these points, dLW is dominated by long-term variations, but it is not clear whether their origin is related to stellar rotation. The dLW has no power at the two RV periods (Fig.~\ref{fig:ActivityAll}). 

The H$\alpha$ time series shows a few flare events. Even after excluding them, we see no power at the RV periods (Fig.~\ref{fig:ActivityAll}).

\section{\label{sec:Photo}Photometry analysis}

As mentioned in Sect.~\ref{sec:photo_inst}, we collected photometric data for two purposes. First, we wished to rule out that the RV signals are due to stellar rotation. Second, the small stellar radius allows a ground-based transit search down to Earth-size planets, which is the expected radius range of the two planets.

Unfortunately, \teegarden was not observed by the \emph{Kepler} space telescope \citep{Borucki2010Sci...327..977B}, although it broadly covered the ecliptic during its \emph{K2} mission. Moreover, \teegarden is not included in the scheduling plan of the Transiting Exoplanet Survey Satellite mission \citep[TESS,][]{Ricker2015JATIS...1a4003R}, which excludes ecliptic latitudes below 6\,deg.

Because \teegarden is very red, ground-based photometry suffers from quite severe colour effects: all comparison stars are bluer. Prior to an analysis, we therefore detrended all the photometric data. MEarth data offer the possibility for detrending using additional information, such as CCD position, airmass, or the so-called common mode (a kind of detrended differential magnitude corrected for main known systematic effects)\footnote{\label{fn:DR7}\url{https://www.cfa.harvard.edu/MEarth/DataDR7.html}}. We investigated possible detrending options and decided for a final data set using the position of the star on the CCD and the common mode. For a consistency check, we also used the original data. Our own data were detrended using either airmass or night-by-night polynomial fits. 

Despite these corrections, we analysed the various data sets individually because the different filters, CCD sensitivities, and comparison stars very likely lead to systematic differences in possible activity-induced photometric signals. Only the MEarth data were also combined into a seven-year-long light curve (excluding the 2010--2011 data using the $I_{715-895}$ interference filter). 

\subsection{\label{sec:phot_rot}Rotation period}

Because a measured rotation period is not available (Sect.~\ref{sec:age}), we re-analysed all available data (Table~\ref{tab:photometry_sets}, Fig.~\ref{fig:LC}) using GLS periodograms. There is no common periodicity detectable that were present in all data or even in a sub-set. The combined original MEarth data indicate a period at 5.11\,d. The TJO data periodogram also shows a peak at this period, but not as the strongest one. In the de-trended MEarth data, this period is not detectable, however. Two data sets (MEarth from 2014 and LCO) show a 28\,d periodicity that we attribute to uncorrected Moon contamination. A common property of most data sets is a periodicity between 50\,d and 80\,d. This might be an indication of a rotation period in that range, but it is below our detection threshold, probably due to a combination of aliasing and short lifetimes of active regions (shorter than this possible rotation period).

Assuming that the various periodicities in our photometric data are due to uncorrected systematics, we removed the largest signal from each data set using a sinusoidal fit. In the pre-whitened data, a weak rotation period may become detectable. However, a common period in the photometric data is again lacking. As demonstrated in Fig.~\ref{fig:ActivityAll}, there is no significant period at one of the two planet orbital periods.

We conclude that we cannot detect a rotation period in \teegarden. The photometric non-detection could indicate a long stellar rotation period resulting in small amplitudes and relatively short lifetimes of active regions. Neither activity indicators nor photometry contradict the interpretation of the two RV signals as being due to planets.

\subsection{\label{sec:phot_transit}Transit search}

\begin{figure}
    \centering
    \includegraphics[width=1\linewidth, trim=0cm 0.8cm 0cm 0cm, clip]{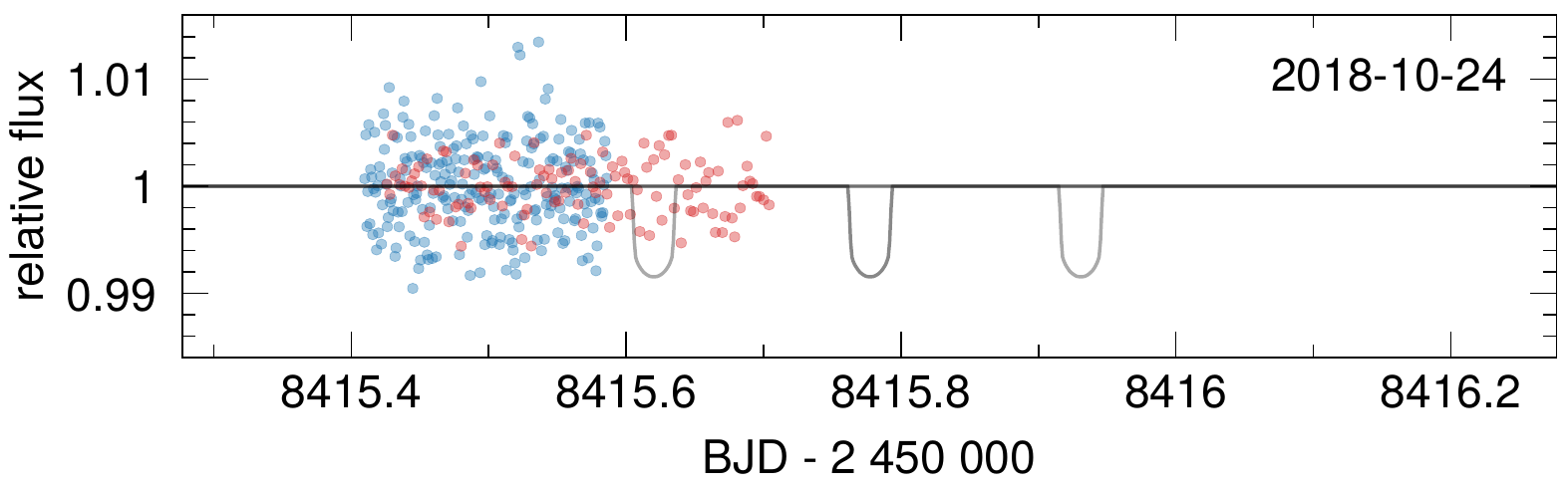}
    \includegraphics[width=1\linewidth, trim=0cm 0.8cm 0cm 0cm, clip]{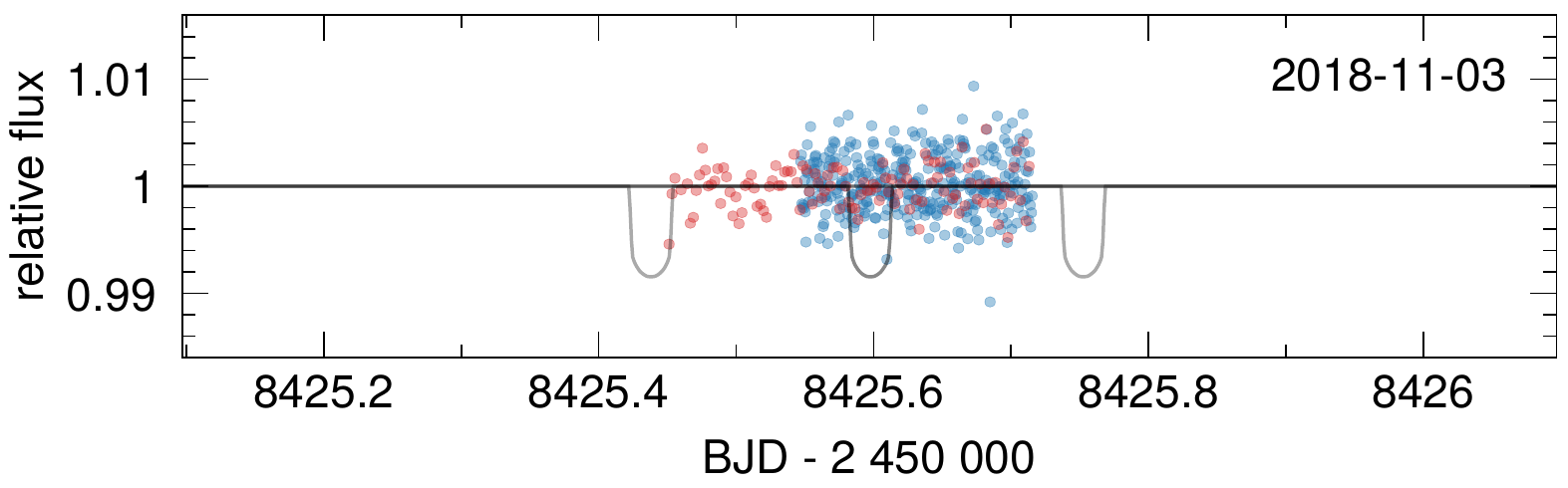}
    \includegraphics[width=1\linewidth, trim=0cm 0.8cm 0cm 0cm, clip]{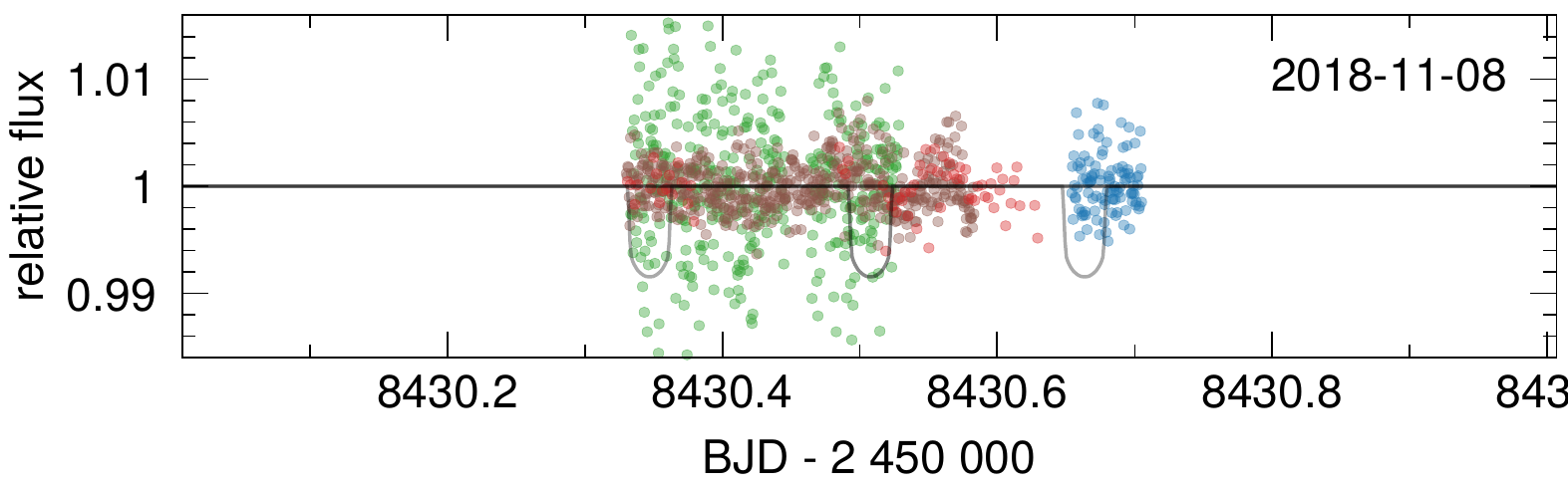}
    \includegraphics[width=1\linewidth]{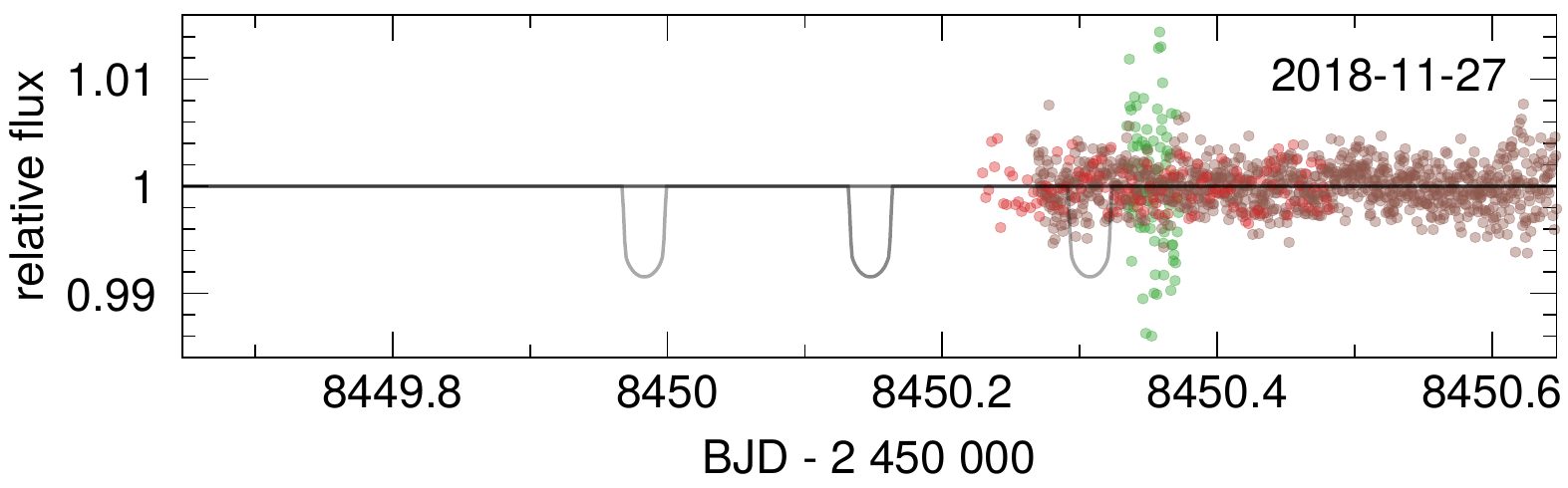}
    \includegraphics[width=1\linewidth]{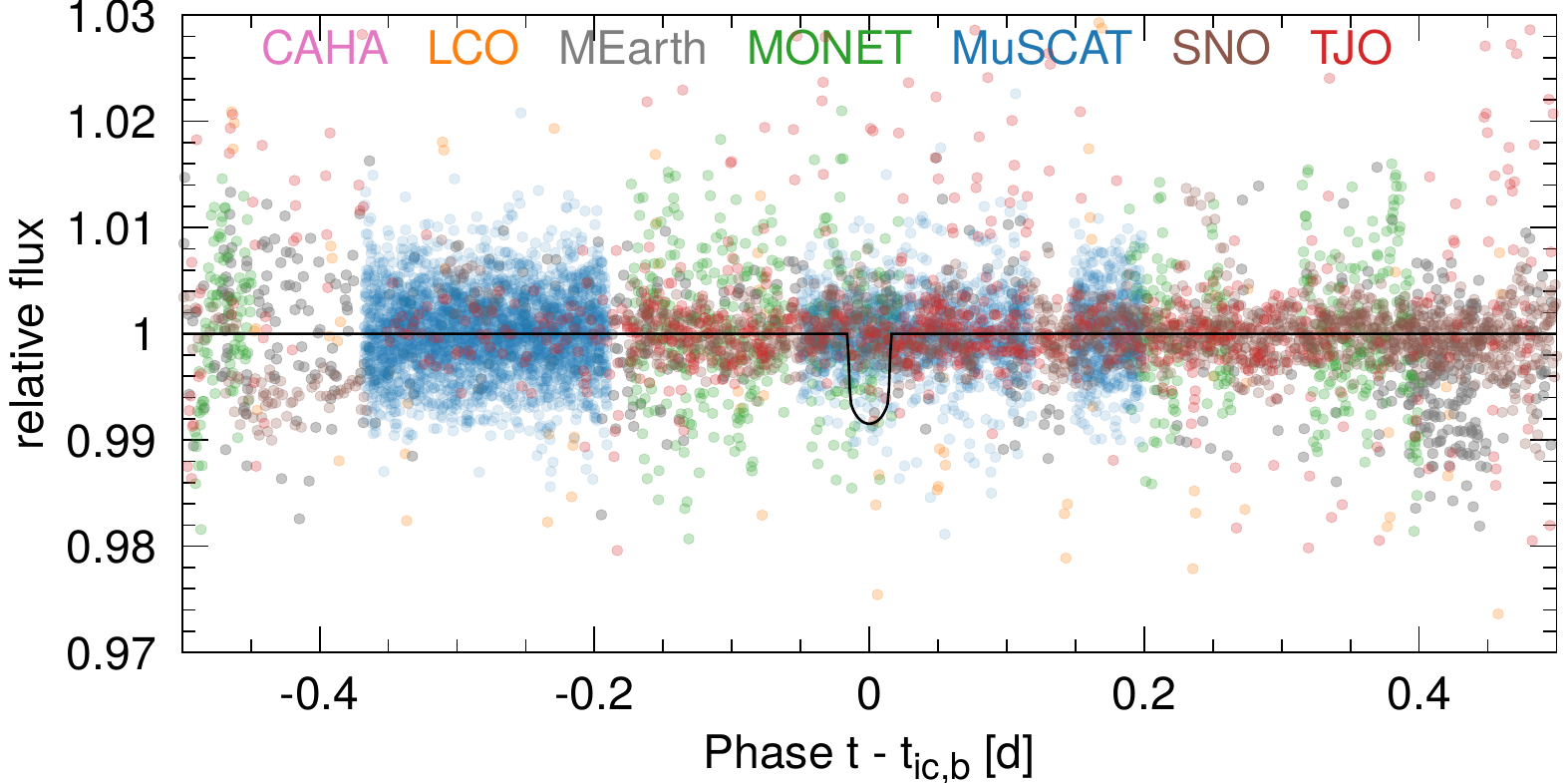}
    \caption{\label{fig:transit_b}Transit search with various de-trended photometric data sets. {\em Top panels:} four high-cadence nights from the transit search. The transit model (black curve) shows the expected signal for planet b and for maximum transit depth. The left and right models (grey) indicate the predicted $1\sigma$ uncertainty for the transit window (0.15\,d, 3.6\,h). {\em Bottom panel:} all photometric data phase-folded with $P_\mathrm{b}$ and $t_\mathrm{ic,b}$.}
\end{figure}

The transit probability of planet b is $p_\mathrm{tr,b} \approx R_\star/a_\mathrm{b} \approx 2.1\,\%,$ 
and its maximum transit depth would be up to $R_\mathrm{b}^2 / R_\star^2 \approx 0.7\,\%$ with a transit duration of $\Delta t_\mathrm{b} \lessapprox 40$\,min.
We focussed our follow-up on planet b because the transit probability of planet c is lower ($p_\mathrm{tr,c}=1.1\%$), the transit duration is longer ($\Delta t_\mathrm{c} \lessapprox 60$\,min), and it has fewer and more uncertain transit windows, which makes ground-based transit searches more challenging.

Our photometric data cover the potential transit phase of planet b densely. Figure~\ref{fig:transit_b} shows the photometric data phase-folded to $P_\mathrm{b}$. An example of an expected transit signal is overplotted. A combined fit of RV data and photometric data (including an offset and a jitter term for each photometric data set) does not indicate a transit signal, and neither does a box-least-squares search \citep{Kovacs2002A&A...391..369K} in the photometry. Because not all individual data sets cover the full orbital phase space and they have different photometric precision, an estimate of our detection limit is difficult to compute and would not be very reliable. A conservative estimate is a detection limit of 2\,mmag for the depth of a transit signal over the full transit window of 1\,d width.

\section{\label{sec:Discussion}Discussion and conclusions}

\begin{figure}
    \centering
    \includegraphics[width=1\linewidth]{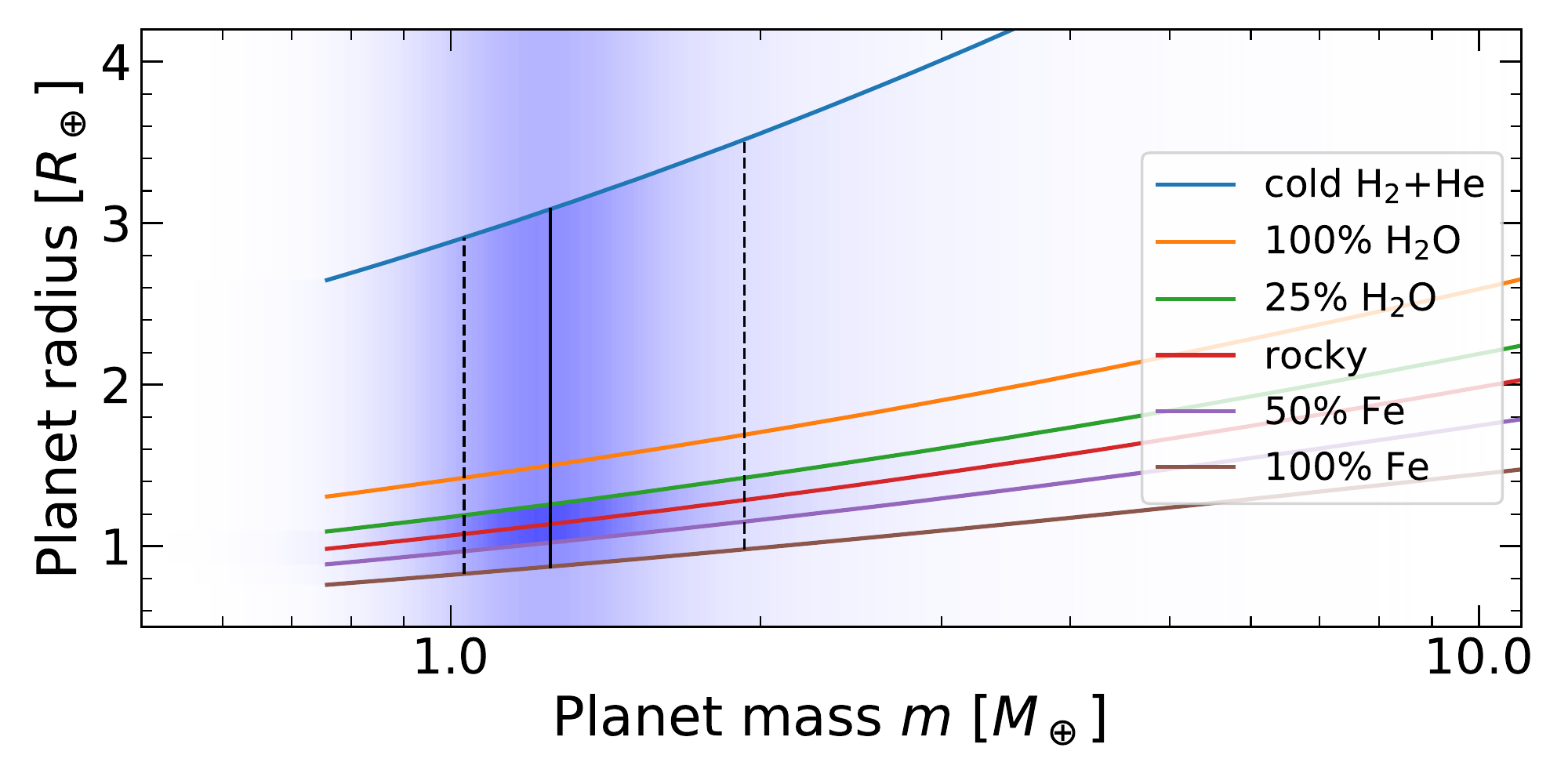}
    \caption{\label{fig:MR} Mass-radius relation for various compositions \citep{Zeng2016ApJ...819..127Z}. The probability for the true mass of planet b is horizontally shaded by the blue gradient; the vertical lines indicate the median (solid black) and 17 and 83\% percentiles (dashed black). The region of less extreme compositions between 25\,\% H$_2$O and 50\,\% Fe is additionally highlighted.}
\end{figure}

\begin{figure}
    \centering
    \includegraphics[width=1\linewidth]{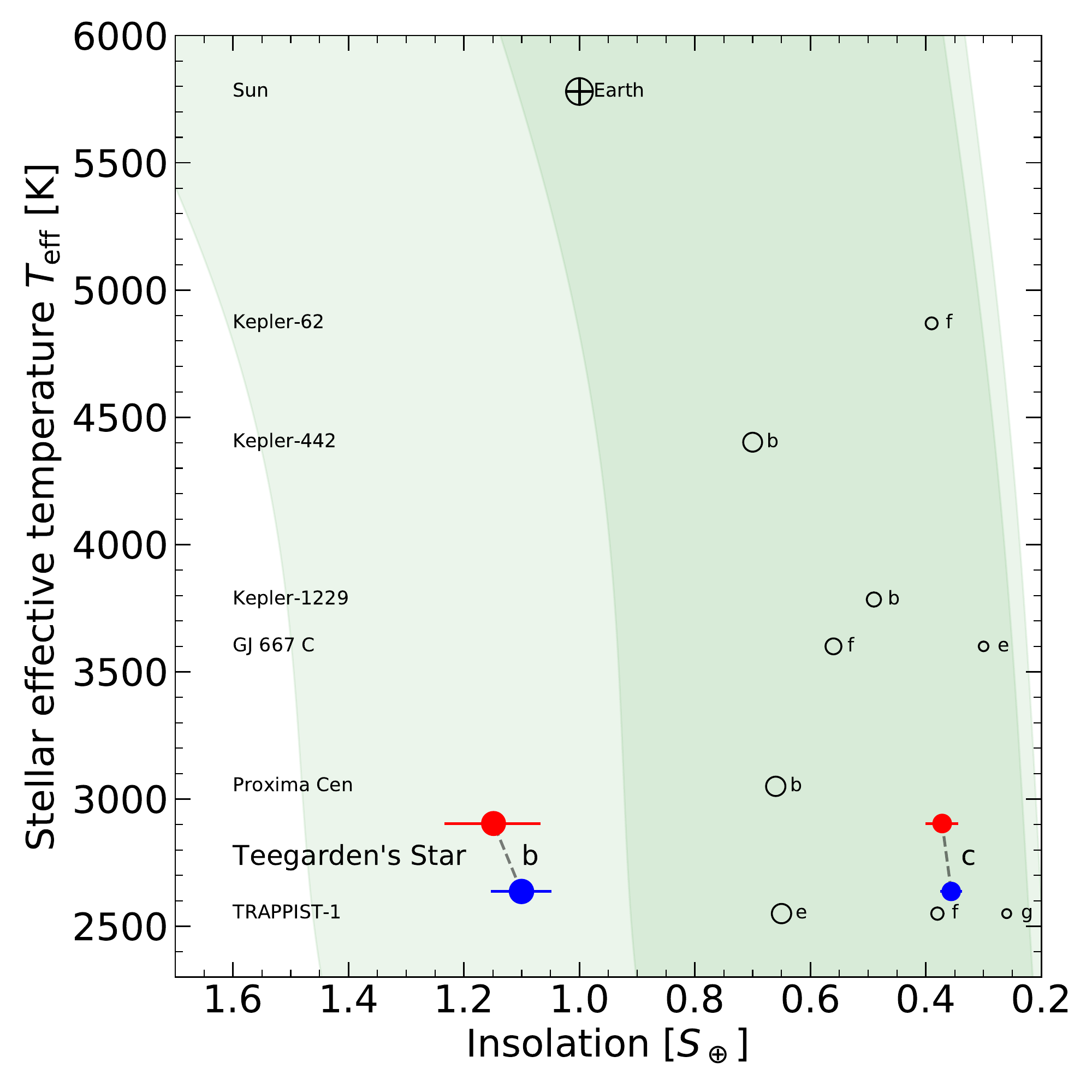}
    \caption{\label{fig:HZ} Optimistic (light green) and conservative (dark green) habitable zone for $1\MEarth$ according to \citet{Kopparapu2014ApJ...787L..29K}. \teegarden planets b and c are shown for two stellar parameter sets (red dots: \citealt{Schweitzer2019A&A...625A..68S}; blue dots: \citealt{Rojas2012ApJ...748...93R, Dieterich2014AJ....147...94D}). The insolation error bar depicts the effect due to intrinsic uncertainty in $T_\mathrm{eff}$. For comparison, we plot nine planets (including Proxima b and TRAPPIST-1 e, f, and g) from the Habitable Exoplanet Catalog\fnref{fn:hec} (black circles) and the Earth (black circled plus) at the top. All symbol sizes scale with the fourth power of the Earth Similarity Index (ESI) assuming a rocky composition.}
\end{figure}

From the analysis we conclude that the two RV signals indicate the presence of two exoplanets. Despite the low amplitudes, the RV signals are significantly detected and have no counterpart, neither in the spectroscopic activity indicators nor in the extensive photometry available to us. Moreover, we have shown that the 4.9\,d signal is stable over the three years of observations in period and amplitude.

Despite intensive monitoring, we cannot derive a rotation period. However, the low photometric variability, the moderate H$\alpha$ emission, and the dLW and CRX indicators suggest a long rotation period. This is in contrast to many other late-M dwarfs, which are very active and rotate rapidly. We therefore conclude that \teegarden is old and that the two signals at 4.91 and 11.4\,d are best explained by the Keplerian motion of exoplanets.

As a note of caution, the late spectral type of \teegarden takes RV analyses into a quite unexplored regime where scaling relations between photometric and RV variations are not established. However, even if RVs turn out to be much more sensitive to stellar variability and one period is connected to stellar rotation, then at least the other signal should be a planet.  Period ratios of 2:1 can be caused by the harmonics of stellar rotation (but also eccentric and resonant planets).  However, the period ratio is $P_\mathrm{c}/P_\mathrm{b}=2.32$, which is close to but noticeably different from 2:1.  Such a discrepancy is hard to explain even with differential rotation, while departures from exact 2:1 ratios are often observed in exoplanet systems \citep{Steffen2015MNRAS.448.1956S}.

Despite the low planetary masses, the dynamical interaction between the two planets is noticeable on longer timescales of decades but not on the rather short timescale of our observations. Using the N-body integrator implemented in Systemic \citep{Meschiari2009PASP..121.1016M}, we integrated the orbits for up to 10$^5$\,yr. It should be noted that the eccentricity changes periodically by about $\Delta e/\mathrm{median}(e)\approx 30\,\%$ on timescales of about 2000\,yr, while the periods are stable within about $\Delta P/\mathrm{median}(P)<0.0003$. 

Because no transits are detected, we cannot derive the planetary radii. We therefore estimated them from mass-radius relations for various compositions \citep{Zeng2016ApJ...819..127Z}. The planetary radii depend weakly on the planetary mass, but strongly on the bulk composition (Fig.~\ref{fig:MR}). Between the two extreme compositions, namely a pure Fe core on the one side and a cold H+He mini-Neptune on the other, the radii differ by a factor of about three (see Fig.~\ref{fig:MR}). We used these radii as well as other stellar and planetary parameters to derive the Earth Similarity Index (ESI) as defined by \cite{Schulze-Makuch2011AsBio..11.1041S}. The ESI is an indicator that compares key parameters to those of our Earth (ESI$_\oplus = 1$). Here we used the weighted ESI, taking into account the equilibrium temperature, atmospheric escape velocity, bulk density, and radius, with a dominant contribution from the equilibrium temperature. The weights were taken from the Habitable Exoplanet Catalog\footnote{\label{fn:hec}\url{http://phl.upr.edu/projects/habitable-exoplanets-catalog}}. Except for the case of a mini-Neptune composition, the two planets have a high ESI. For a potentially rocky composition, the ESI value is 0.94 and 0.8 for planets b and c, respectively. This makes \teegarden b the planet with currently the highest ESI value. However, the ESI is only an estimate, and different weighting of the parameters may lead to changing ESIs. This ESI definition, for example, does not take into account the stellar spectral energy distribution and the resulting planetary atmospheric composition, which very likely have an effect on habitability.

We displayed the two planets of \teegarden (see Fig.~\ref{fig:HZ}) with nine other Earth-like planets within the conservative habitable zone (HZ) listed in the Habitable Exoplanet Catalog and compared their incident stellar flux with their position relative to the HZ according to \citet{Kopparapu2014ApJ...787L..29K} as a function of stellar effective temperature. We also took the different stellar parameters for \teegarden into account. Planet b is placed just outside the hot end of the conservative HZ (but inside the optimistic one), whereas planet c is comfortably within this zone. In addition, planet b receives nearly the same stellar flux as the Earth and therefore has a nearly identical equilibrium temperature, but is outside the conservative HZ. As discussed by \citet{Kopparapu2014ApJ...787L..29K}, this is caused by a runaway greenhouse effect that is due to water vapour starting at lower insolation for low-mass stars. This effect is neglected in the ESI calculation and leads to the curious fact that planet b has a higher ESI value than planets inside the conservative HZ.   

The planets in this planetary system, especially \teegarden c, could become prime targets for further characterisation with the next generation of extremely large telescopes (ELTs). Some of the planned instruments, such as the Planetary Systems Imager \citep[PSI,][]{Guyon2018SPIE10703E..0ZG} for the Thirty Meter Telescope (TMT), aim to achieve contrast values of $10^{-7}$ at $\sim 0.01$\,arcsec. Assuming Earth-size for \teegarden c, the contrast ratio of the reflected light from the planet is also expected to be of the order of $10^{-7}$ at the projected angle of the semi-major axis of $\sim 0.012$\,arcsec. Thus, high-contrast imagers for the new ELTs, both in the Northern and Southern Hemispheres, now have additional motivation because they can stand a good chance of directly imaging nearby and potentially habitable Earth-mass planets. This will enable a detailed study of their properties.

The exoplanet system around \teegarden is a pioneering discovery by the CARMENES survey and remarkable in several aspects. The planets \teegarden b and c are the first planets detected with the RV method around such an ultra-cool dwarf. Both planets have a minimum mass close to one Earth mass, and given a rocky, partially iron, or water composition, they are expected to have Earth-like radii. Additionally, they are close to or within the conservative HZ, or in other words, they are potentially habitable. Our age estimate of 8\,Gyr implies that these planets are about twice as old as the solar system. Interestingly, our solar system currently is within the transit zone as seen from \teegarden. For any potential Teegardians, the Earth will be observable as a transiting planet from 2044 until 2496.

\begin{acknowledgements}
We thank the referee Rodrigo D\'iaz for a careful review and helpful comments and Ren\'e Heller for useful hints.
M.Z. acknowledges support from the Deutsche Forschungsgemeinschaft under DFG RE 1664/12-1 and Research Unit FOR2544 ``Blue Planets around Red Stars'', project no. RE 1664/14-1.

  CARMENES is an instrument for the Centro Astron\'omico Hispano-Alem\'an de
  Calar Alto (CAHA, Almer\'{\i}a, Spain). 
  CARMENES is funded by the German Max-Planck-Gesellschaft (MPG), 
  the Spanish Consejo Superior de Investigaciones Cient\'{\i}ficas (CSIC),
  the European Union through FEDER/ERF FICTS-2011-02 funds, 
  and the members of the CARMENES Consortium 
  (Max-Planck-Institut f\"ur Astronomie,
  Instituto de Astrof\'{\i}sica de Andaluc\'{\i}a,
  Landessternwarte K\"onigstuhl,
  Institut de Ci\`encies de l'Espai,
  Institut f\"ur Astrophysik G\"ottingen,
  Universidad Complutense de Madrid,
  Th\"uringer Landessternwarte Tautenburg,
  Instituto de Astrof\'{\i}sica de Canarias,
  Hamburger Sternwarte,
  Centro de Astrobiolog\'{\i}a and
  Centro Astron\'omico Hispano-Alem\'an), 
  with additional contributions by the Spanish Ministry of Economy, 
  the German Science Foundation through the Major Research Instrumentation 
    Programme and DFG Research Unit FOR2544 ``Blue Planets around Red Stars'', 
  the Klaus Tschira Stiftung, 
  the states of Baden-W\"urttemberg and Niedersachsen, 
  and by the Junta de Andaluc\'{\i}a.
  Based on data from the CARMENES data archive at CAB (INTA-CSIC).
    
This article is based on observations made with the MuSCAT2 instrument, developed by ABC, at Telescopio Carlos S\'anchez operated on the island of Tenerife by the IAC in the Spanish Observatorio del Teide.
Data were partly collected with the 150-cm and 90-cm telescopes at the Sierra Nevada Observatory (SNO) operated by the Instituto de Astrof\'isica de Andaluc\'ia (IAA-CSIC).
Data were partly obtained with the MONET/South telescope of the MOnitoring NEtwork of Telescopes, funded by the Alfried Krupp von Bohlen und Halbach Foundation, Essen, and operated by the Georg-August-Universit\"at G\"ottingen, the McDonald Observatory of the University of Texas at Austin, and the South African Astronomical Observatory.

  We acknowledge financial support from the Spanish Agencia Estatal de Investigaci\'on
  of the Ministerio de Ciencia, Innovaci\'on y Universidades and the European
  FEDER/ERF funds through projects 
  AYA2015-69350-C3-2-P,     
  AYA2016-79425-C3-1/2/3-P, 
  AYA2018-84089,            
  BES-2017-080769,          
  BES-2017-082610,          
  ESP2015-65712-C5-5-R,     
  ESP2016-80435-C2-1/2-R,   
  ESP2017-87143-R,          
  ESP2017-87676-2-2,        
  ESP2017-87676-C5-1/2/5-R, 
  FPU15/01476,              
  RYC-2012-09913,           
  the Centre of Excellence ''Severo Ochoa'' and ''Mar\'ia de Maeztu'' awards to the Instituto de Astrof\'isica de Canarias (SEV-2015-0548), 
  Instituto de Astrof\'isica de Andaluc\'ia (SEV-2017-0709), 
  and Centro de Astrobiolog\'ia (MDM-2017-0737), 
the Generalitat de Catalunya through CERCA programme'',
the Deutsches Zentrum f\"ur Luft- und Raumfahrt through grants 50OW0204 and 50OO1501, 
the European Research Council through grant 694513, 
the Italian Ministero dell'instruzione, dell'universit\`a de della ricerca and Universit\`a degli Studi di Roma Tor Vergata through FFABR 2017 and ``Mission: Sustainability 2016'', 
the UK Science and Technology Facilities Council through grant ST/P000592/1,
the Israel Science Foundation through grant 848/16, 
the Chilean CONICYT-FONDECYT through grant 3180405, 
the Mexican CONACYT through grant CVU 448248, 
the JSPS KAKENHI through grants JP18H01265 and 18H05439, 
and the JST PRESTO through grant JPMJPR1775.

\end{acknowledgements}

\bibliographystyle{aa}
\bibliography{teegarden}

\appendix

\section{\label{sec:phot_inst}Photometric facilities}

\begin{table*}
    \caption{\label{tab:photometry} Photometric facilities.}
    \centering
    \begin{tabular}{@{}llccccl@{}}
        \toprule
        Acronym     &  Location & Tel.  & FOV           & CCD   & Scale                & Band(s)\tabularnewline
                    &           & [m]   & [arcmin$^2$]  &       & [arcsec\,pix$^{-1}$] & \tabularnewline
        \midrule
        CAHA 1.23   & Calar Alto Observatory           & 1.23  &  17.9\,$\times$\,17.9     & 2k\,$\times$\,2k & 0.50 & $I$\tabularnewline
        LCO         & Las Cumbres Observatory Global Telescope    & 0.40  &  29.2\,$\times$\,19.5     & 3k\,$\times$\,2k & 0.57 & $i'$\tabularnewline
        MEarth      & Fred Lawrence Whipple Observatory     & 0.40  &  26.0\,$\times$\,26.0    & 2k\,$\times$\,2k & 0.76 & RG715, $I$\tabularnewline
        MONET-S     & South African Astronomical Observatory  & 1.20  &  12.6\,$\times$\,12.6     & 2k\,$\times$\,2k & 0.37 & $R$\tabularnewline
        MuSCAT2     & TCS Teide Observatory   & 1.52  &   7.4\,$\times$\,7.4     & 1k\,$\times$\,1k & 0.44 & $griz_s$\tabularnewline
        SNO T150    & T150 Sierra Nevada Observatory        & 1.50  &   7.9\,$\times$\,7.9     & 2k\,$\times$\,2k & 0.23 & $VRI$\tabularnewline
        SNO T90     & T90 Sierra Nevada Observatory         & 0.90  &  13.2\,$\times$\,13.2     & 2k\,$\times$\,2k & 0.38 & $R$\tabularnewline
        TJO         & TJO Montsec Astronomical Observatory  & 0.80  &  12.3\,$\times$\,12.3     & 2k\,$\times$\,2k & 0.36 & $R$\tabularnewline
        \bottomrule
    \end{tabular}
\end{table*}

\begin{table}
    \caption{\label{tab:photometry_sets}Properties of the photometric data sets\tablefootmark{a\!}.}
    \centering
    \begin{tabular}{@{}lcccc@{}}
        \toprule
        Data set     & Season & $\Delta T$ & $N_\mathrm{obs}$  & rms \\
                     &            & [d]       &          & [mag]  \\
        \midrule
        CAHA 1.23    & 2018       &    0.3    &  124     & 0.003 \\
        LCO 17       & 2017       &  421      &  116     & 0.012 \\
        LCO 18       & 2018       &  431      &  148     & 0.014 \\
        ME T1 08-1   & 2008--2009 &  351      &  145     & 0.007 \\
        ME T1 08-2   & 2009--2010 &  377      &  342     & 0.007 \\
        ME T1 11     & 2010--2011 &  260      &  504     & 0.007 \\
        ME T1 14-1   & 2011--2015 & 1473      &  824     & 0.004 \\
        ME T1 14-2   & 2011--2015 & 1454      &  638     & 0.004 \\
        ME T8 14-1   & 2013--2014 &  139      &  728     & 0.004 \\
        ME T8 14-2   & 2013--2014 &   87      &  429     & 0.004 \\
        ME all       & 2008--2015 & 2577      & 2547     & 0.004 \\
        MONET-S      & 2018       &   22      & 1201     & 0.004 \\
        MuSCAT2 $g$  & 2018       &   18      &  563     & 0.006 \\
        MuSCAT2 $r$  & 2018       &   55      &  918     & 0.004 \\
        MuSCAT2 $i$  & 2018       &   50      &  827     & 0.001 \\
        MuSCAT2 $z$  & 2018       &   28      &  360     & 0.002 \\
        SNO $V$      & 2017--2019 &  540      & 2205     & 0.006 \\
        SNO $R$      & 2017--2019 &  540      & 2340     & 0.005 \\
        SNO $I$      & 2017--2019 &  540      & 2231     & 0.008 \\
        SNO-T90 $R$  & 2018       &    1      &  236     & 0.003 \\
        SNO-T150 $R$ & 2018       &   20      & 1274     & 0.002 \\
        TJO 17       & 2017       &   54      & 2280     & 0.011 \\
        TJO 18       & 2018       &   35      &  475     & 0.002 \\
        \bottomrule
    \end{tabular}
    \tablefoot{
        \tablefoottext{a}{Data set identifier, season, time span, number of observations, and rms in relative flux after removing the strongest signal in each set.}
    }
\end{table}

\paragraph{CAHA 1.23.}
The Centro Astron\'omico Hispano-Alem\'an operates the 1.23\,m telescope for photometric monitoring projects. The CCD imager mounted at the Cassegrain focus covers a squared field of view (FOV) of 17.9\,arcmin and is equipped with a 2k$\times$2k SITE CCD. Observations were made in the $I$ band.

\paragraph{LCO.}
We obtained $i'$-band images using the 40\,cm telescopes of the Las Cumbres Observatory (LCO) network. The telescopes are equipped with a 3k$\times$2k SBIG CCD camera with a pixel scale of 0.571\,arcsec, providing a field of view of 29.2$\times$19.5\,arcmin$^2$. The data were processed using the {\tt Banzai} pipeline \citep{McCully2018SPIE10707E..0KM}. The photometry of the 2017 set was extracted with {\tt IRAF-PHOT}  and the 2018 set with {\tt AstroImageJ} \citep{Collins2017AJ....153...77C}.

\paragraph{MEarth.}
Since 2008, the MEarth project \citep{Berta2012AJ....144..145B} monitors more than 2000 M dwarfs with eight robotic 40\,cm telescopes (f/9 Ritchey-Chr\'etien Cassegrain) at the Fred Lawrence Whipple Observatory (Arizona, USA). Each telescope covers a squared 26\,arcmin field of view with a 2048$\times$2048 CCD (0.76\,arcsec/pix). MEarth generally used an RG715 long-pass filter, except for the 2010-2011 season, when a $I_{715-895}$ interference filter was chosen. Additionally, \teegarden was monitored with a second telecope (\#8) in 2013 and 2014. We made use of the archival data from the seventh MEarth data release, DR7\fnref{fn:DR7}, which provides long-term monitoring of \teegarden over about 10\,yr.

\paragraph{MONET-S.}
The 1.2\,m MONET/South telescope (MOnitoring NEtwork of Telescopes) is located at the South African Astronomical Observatory (Northern Cape, South Africa). It is equipped with a Finger Lakes ProLine 2k$\times$2k e2v CCD and has a $12.6\times12.6$\,arcmin$^2$ field of view. We performed aperture photometry with {\tt AstroImageJ} using eight comparison stars.

\paragraph{MuSCAT2.}
The Multicolor Simultaneous Camera for studying Atmospheres of Transiting exoplanets 2 \citep[MuSCAT2;][]{Narita2019JATIS...5a5001N} is mounted at Telescopio Carlos S\'anchez in Teide observatory (Tenerife, Spain). MuSCAT2 observes simultaneously in the $g$, $r$, $i$, and $z_s$ bands using a set of dichroics to split the light between four separate cameras with a field of view of 7.4$\times$7.4\,arcmin$^2$ (0.44\,arcsec/pix). MuSCAT2 is designed to be especially efficient for science related to transiting exoplanets and objects varying on short timescales around cool stellar types. Aperture photometry is calculated using a Python-based pipeline especially developed for MuSCAT2 \citep[see][for details]{Narita2019JATIS...5a5001N}.

\paragraph{SNO.}
The T150 telescope at Sierra Nevada Observatory (Granada, Spain) is a 1.5\,m Ritchey-Chr\'etien telescope equipped with a CCD camera VersArray 2k$\times$2k, FOV 7.9$\times$7.9\,arcmin$^2$ \citep{Rodriguez2010MNRAS.408.2149R}. Two sets of observations were collected in Johnson $V$, $R$, and $I$ filters: one set consisted of 54 epochs obtained during the period July 2017 to January 2018, while the other set consisted of 53 epochs collected between July 2018 and January 2019. Each epoch typically consisted of 20 observations per night in each filter of 100, 50, and 20\,s. In addition, the T150 was also used for transit search (SNO T150-$R$ in Table~\ref{tab:photometry_sets}) during two nights in the $R$ filter. The T90 telescope at SNO (SNO T90-$R$ in Table~\ref{tab:photometry_sets}) and $R$ filter were also used for transit search during one night. 

\paragraph{TJO.}
The Telescopi Joan Or\'o is an 80\,cm telescope located in the Montsec Astronomical Observatory (Lleida, Spain). Photometry in Johnson $R$ filter was obtained with the MEIA2 instrument, a 2k$\times$2k Andor CCD camera, with a pixel scale of 0.36\,arcsec and a squared field of view of 12.3\,arcmin. The images were processed with the ICAT pipeline \citep{Colome2006IAUSS...6E..11C} and {\tt AstroImageJ}.

\section{\label{sec:Activity GLS}GLS periodograms for activity indicators}
\begin{figure}
    \centering
    \includegraphics[width=1.\linewidth,trim=0cm 1.4cm 0cm 0cm, clip]{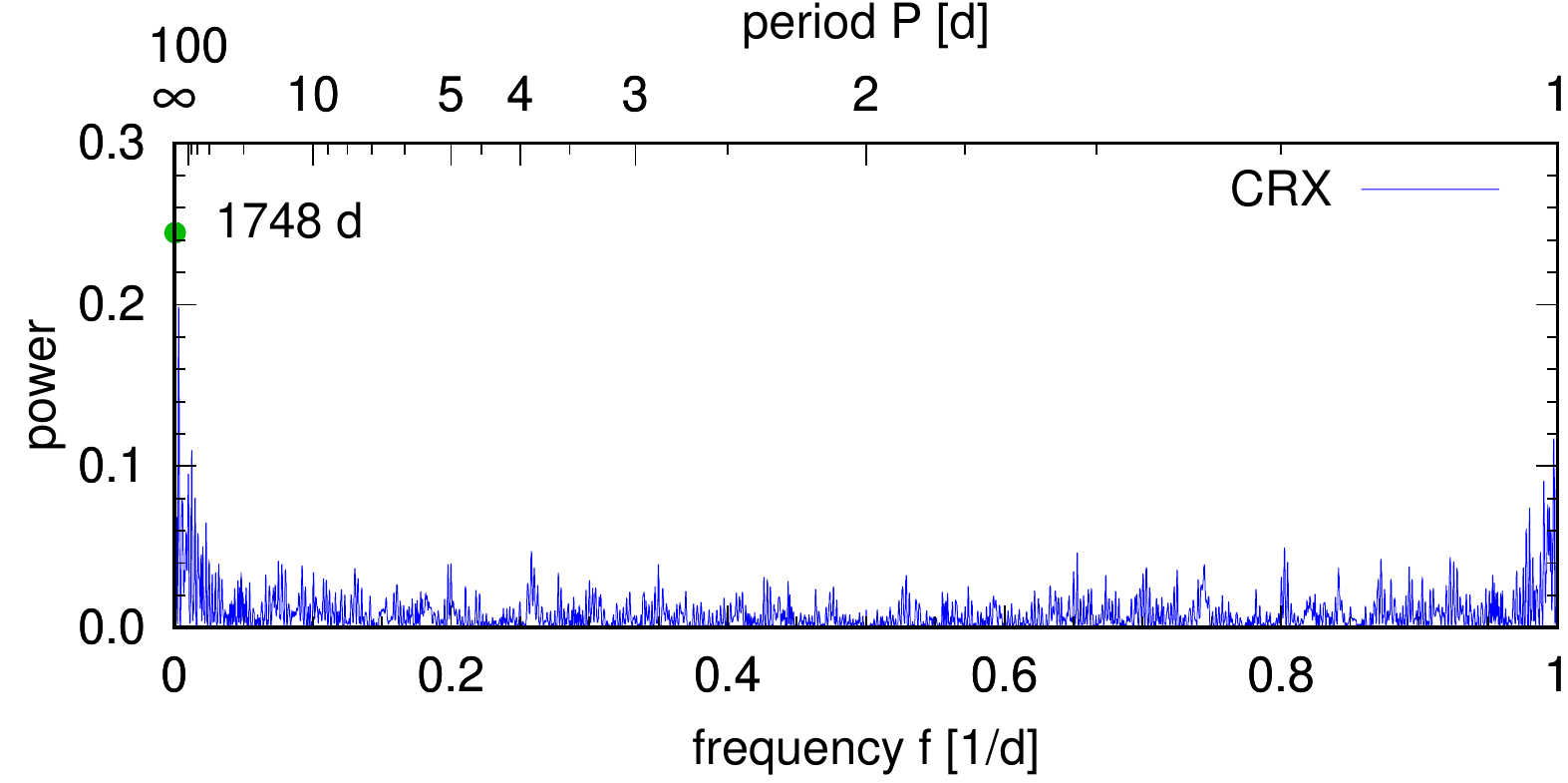}
    \includegraphics[width=1.\linewidth,trim=0cm 1.4cm 0cm 1.3cm, clip]{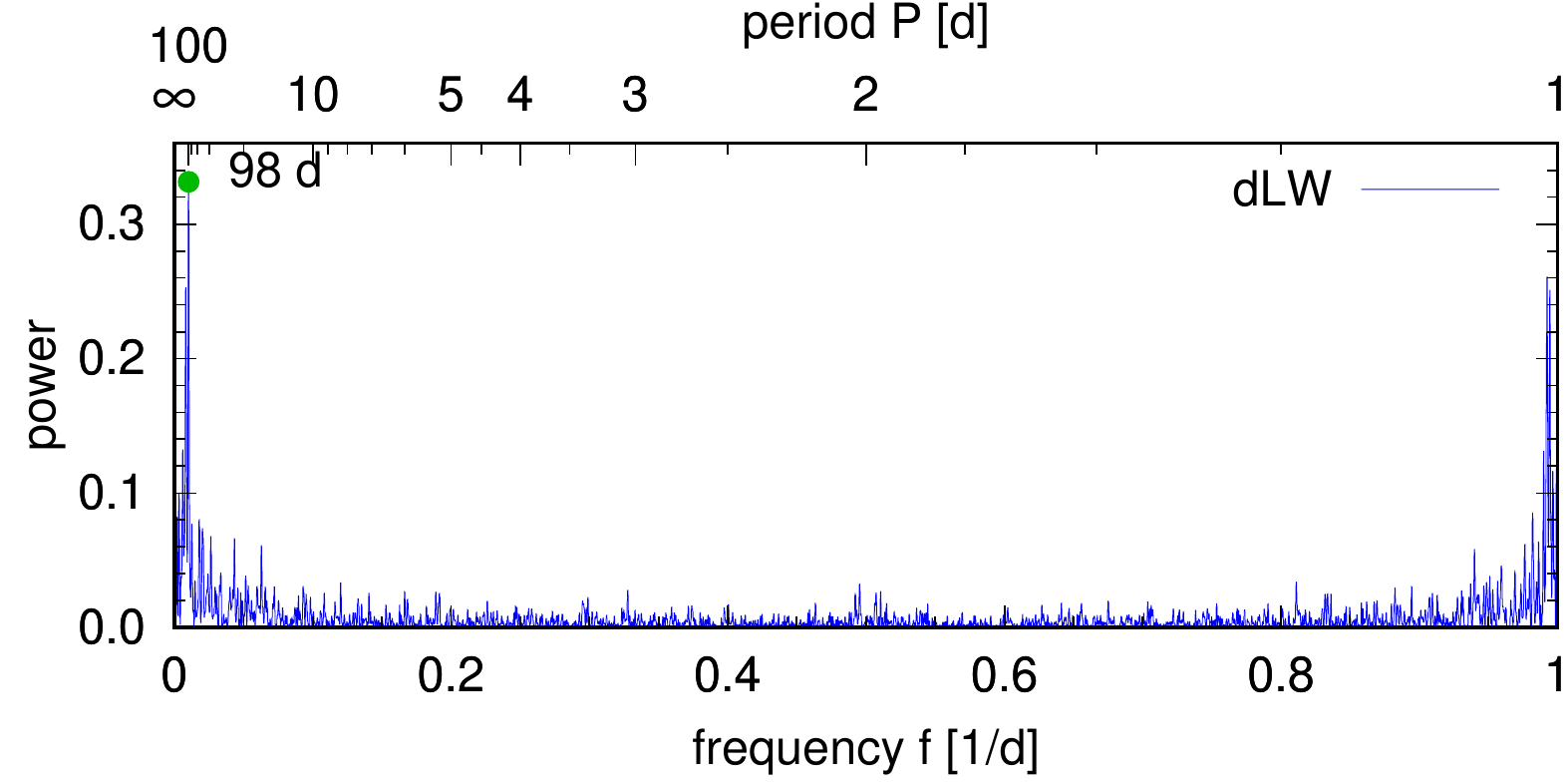}
    \includegraphics[width=1.\linewidth,trim=0cm 0cm 0cm 1.3cm, clip]{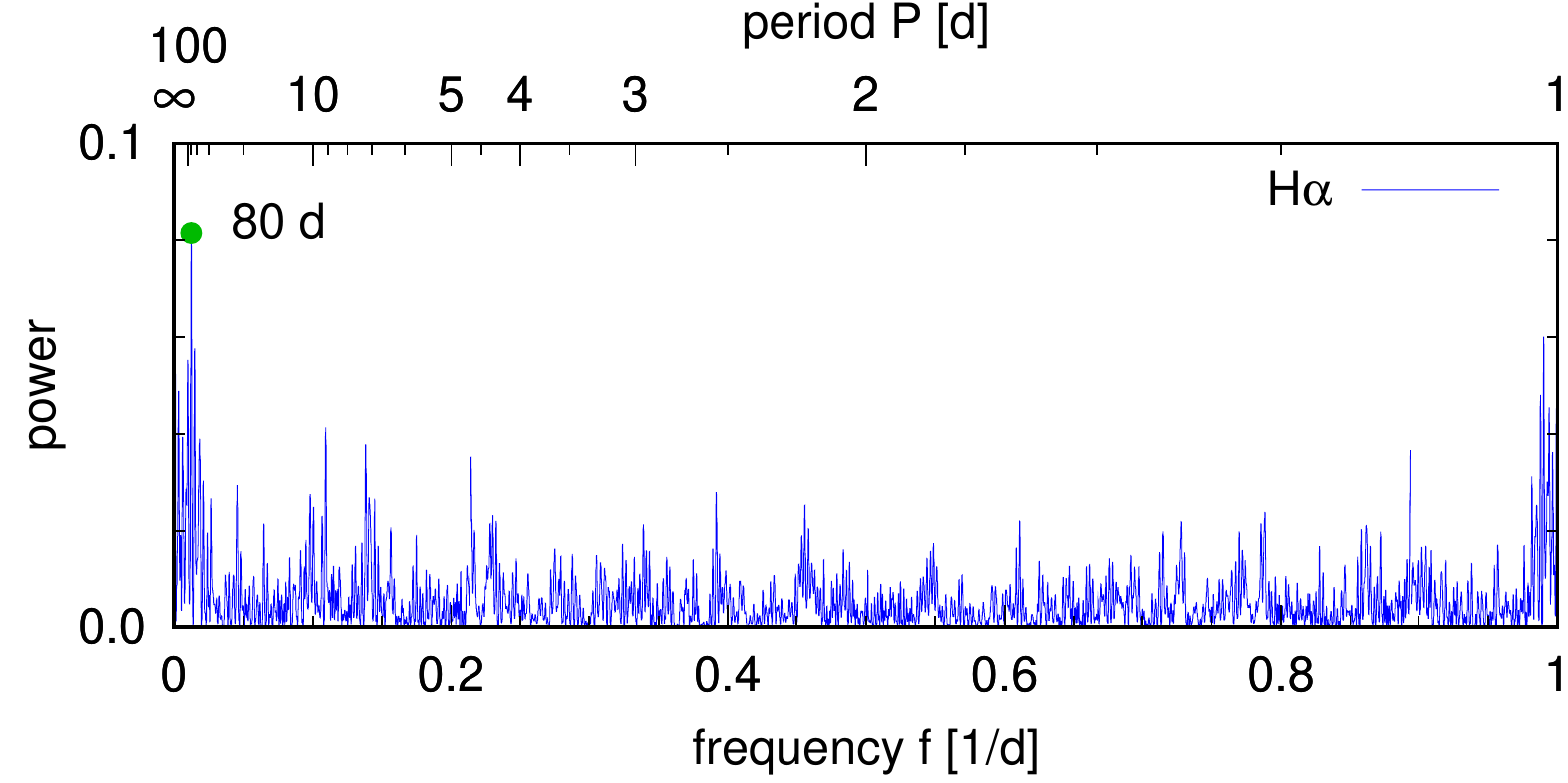}
    \caption{\label{fig:GLS_ActInd}GLS periodograms for CRX, dLW, and H$\alpha$ index.}
\end{figure}

\section{\label{sec:Corner}MCMC corner plots}

\begin{figure*}[!b]
    \centering
    \includegraphics[width=0.5\linewidth]{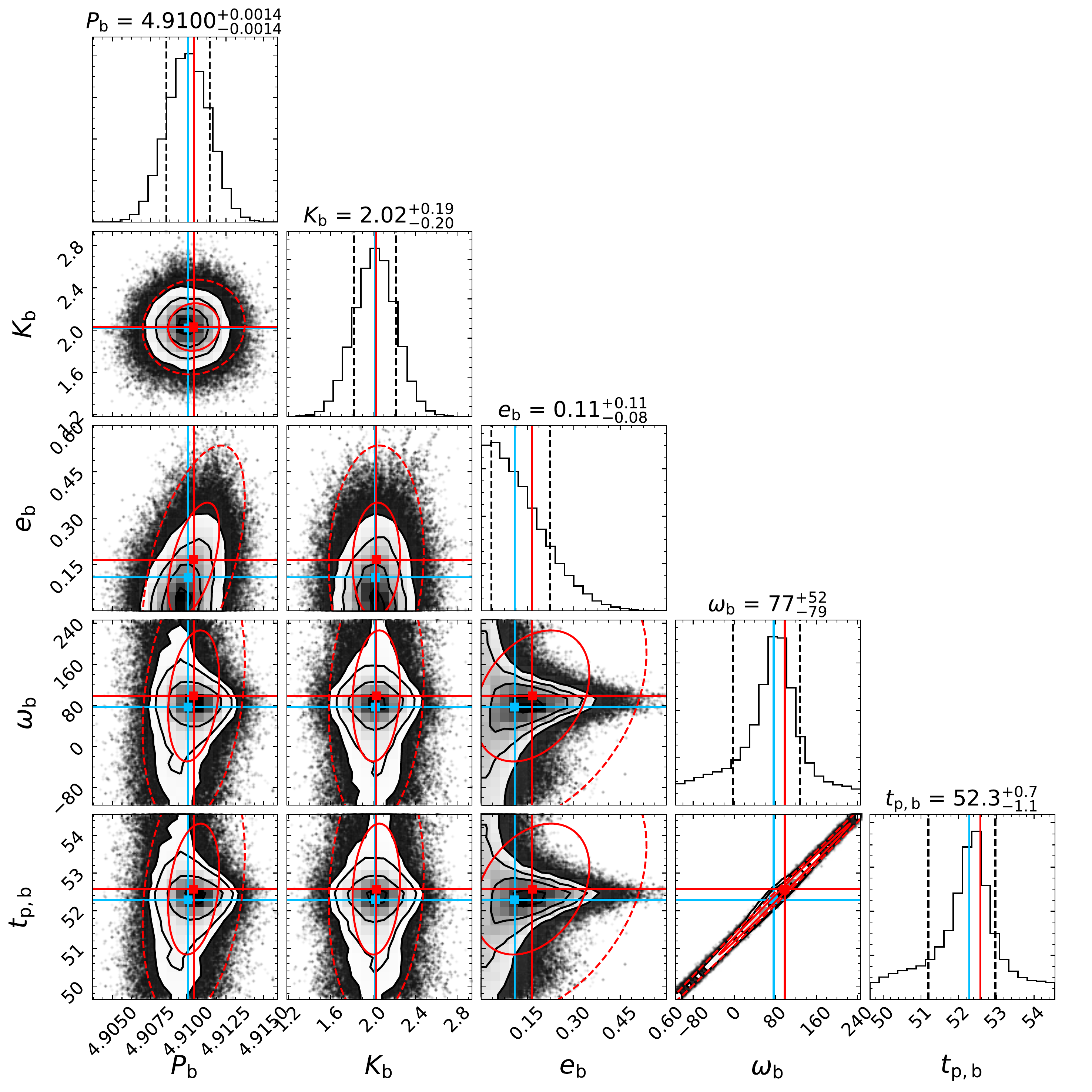}\includegraphics[width=0.5\linewidth]{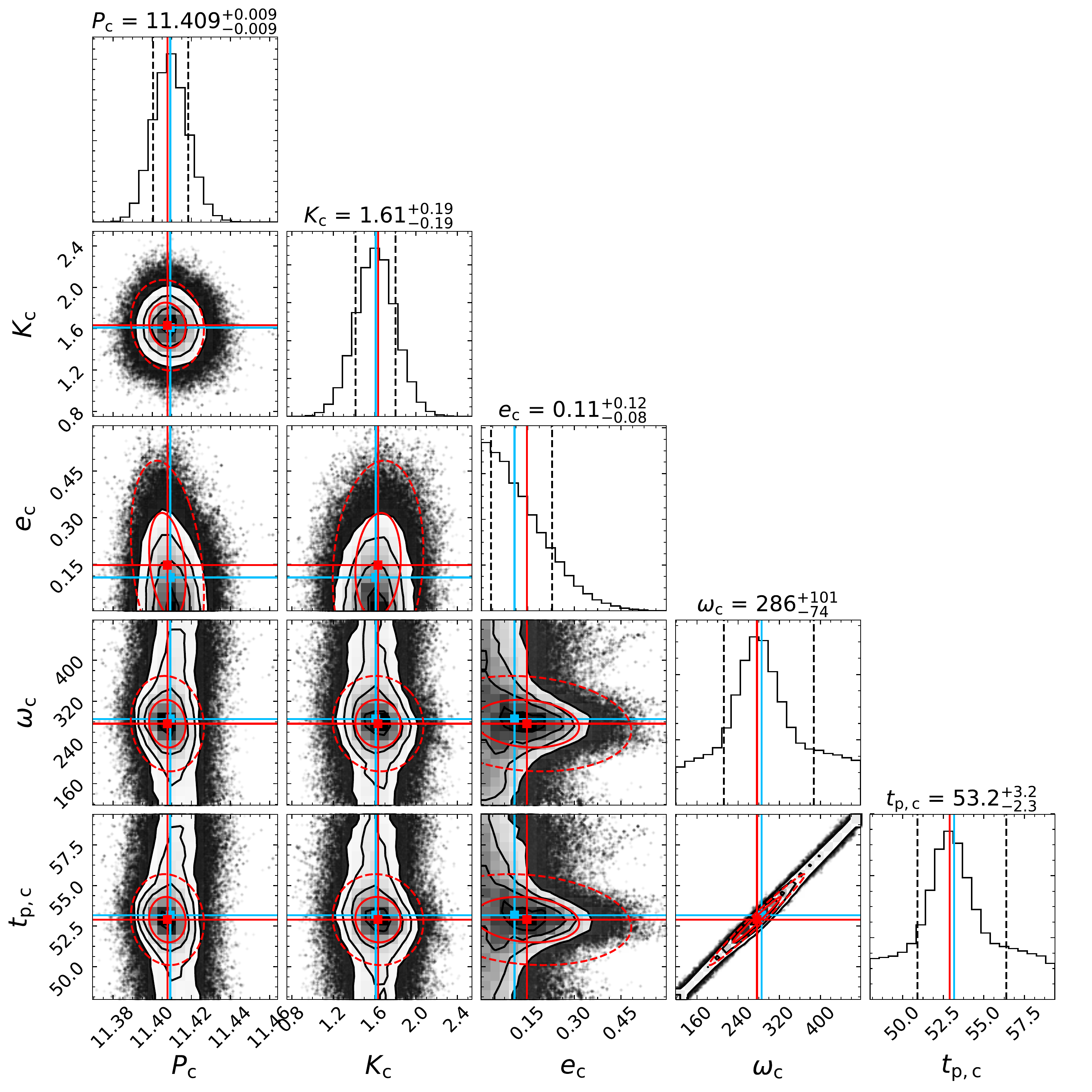}
    \caption{\label{fig:MCMC_bc}MCMC posterior distribution for Keplerian parameters of planets  b ({\em left}) and c ({\em right}). The median values and 15.87 and 84.13\,\% are indicated in light blue and black dashed lines, while best-fit values and $1\sigma$ and $2\sigma$ levels from covariance matrix are overplotted in red.}
\end{figure*}

\begin{figure*}[!b]
    \centering
    \includegraphics[width=0.5\linewidth]{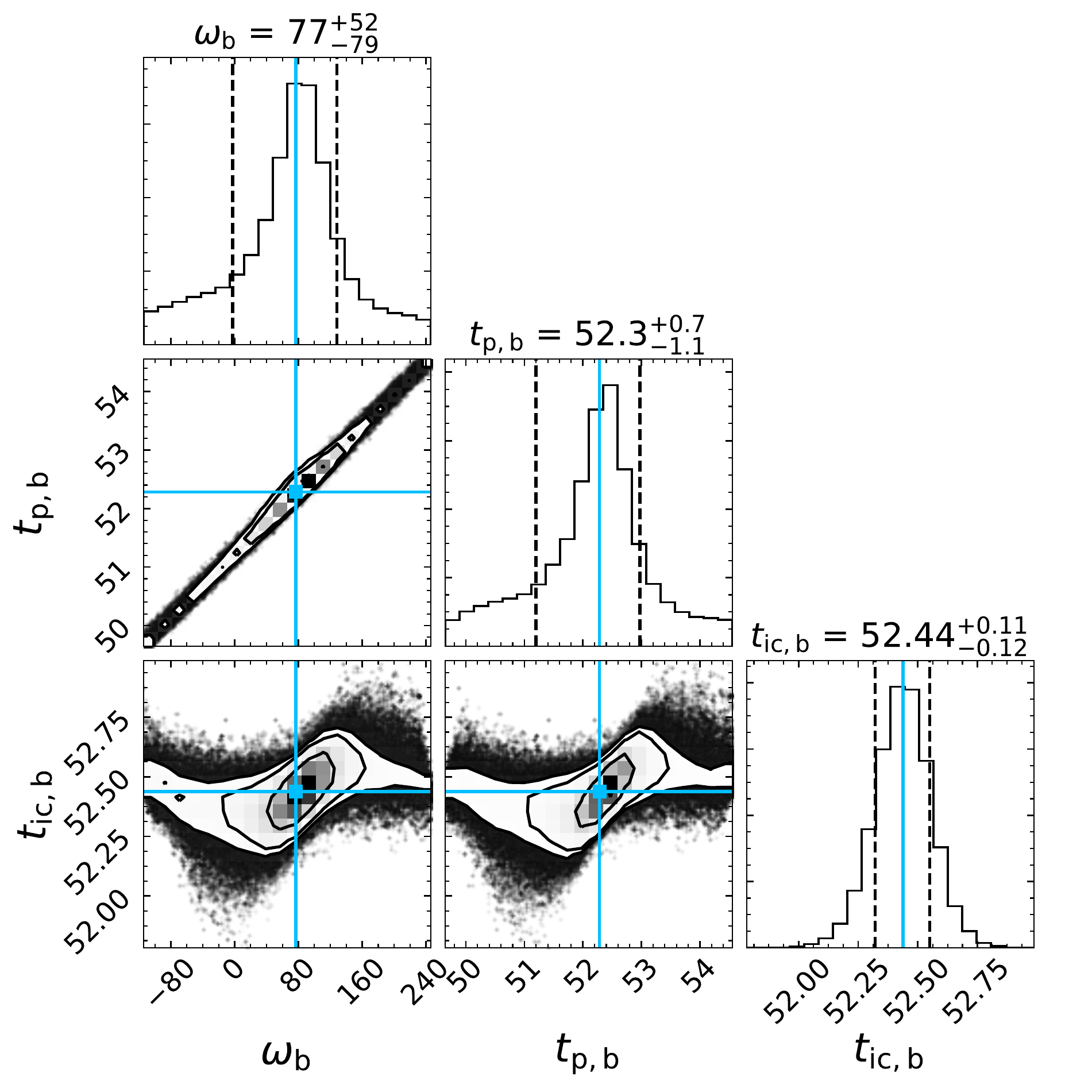}\includegraphics[width=0.5\linewidth]{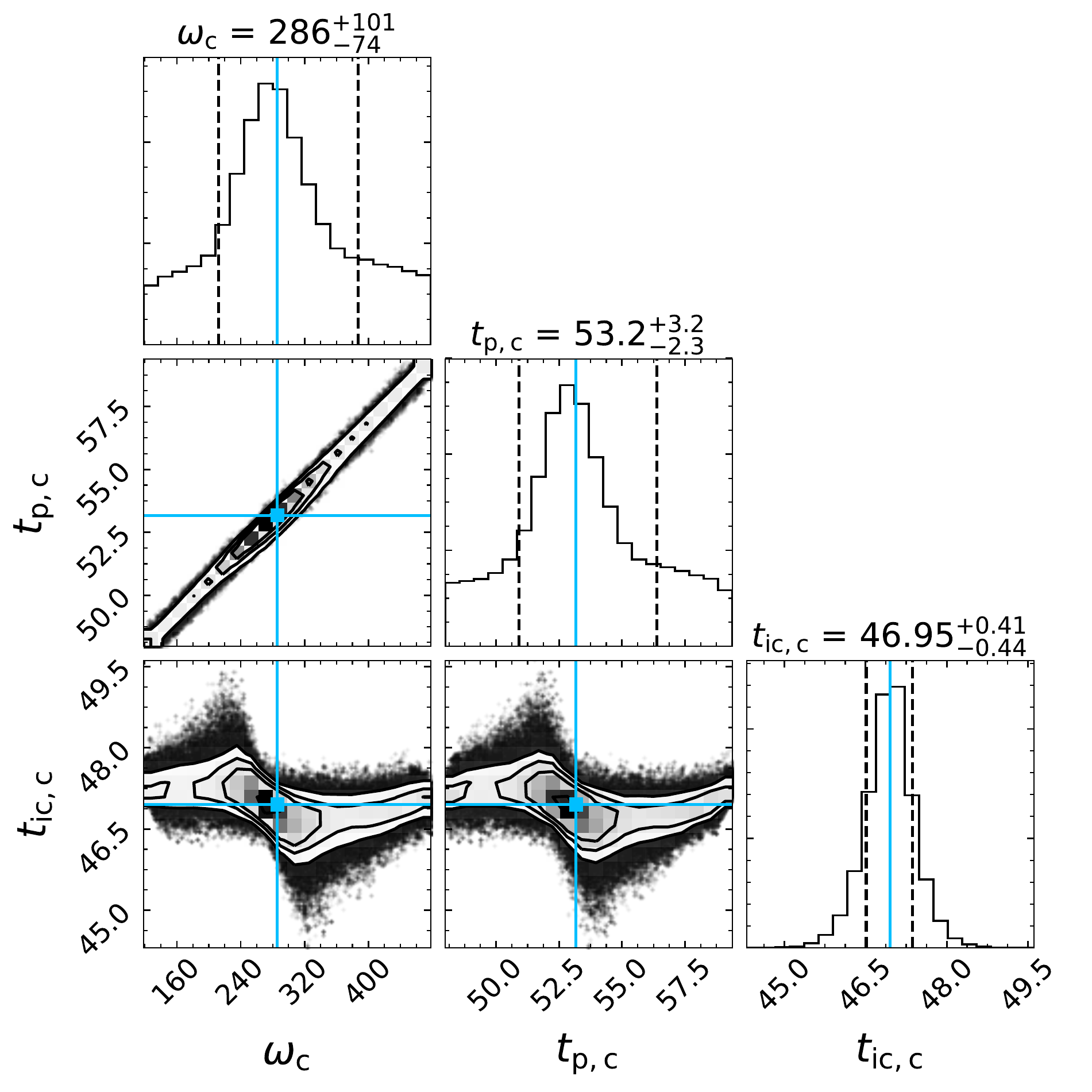}
    \caption{\label{fig:MCMC_tr}MCMC posterior distribution for inferior conjunction times of planet b ({\em left}) and c ({\em right}).}
\end{figure*}

\begin{figure}
    \centering
    \includegraphics[width=1\linewidth]{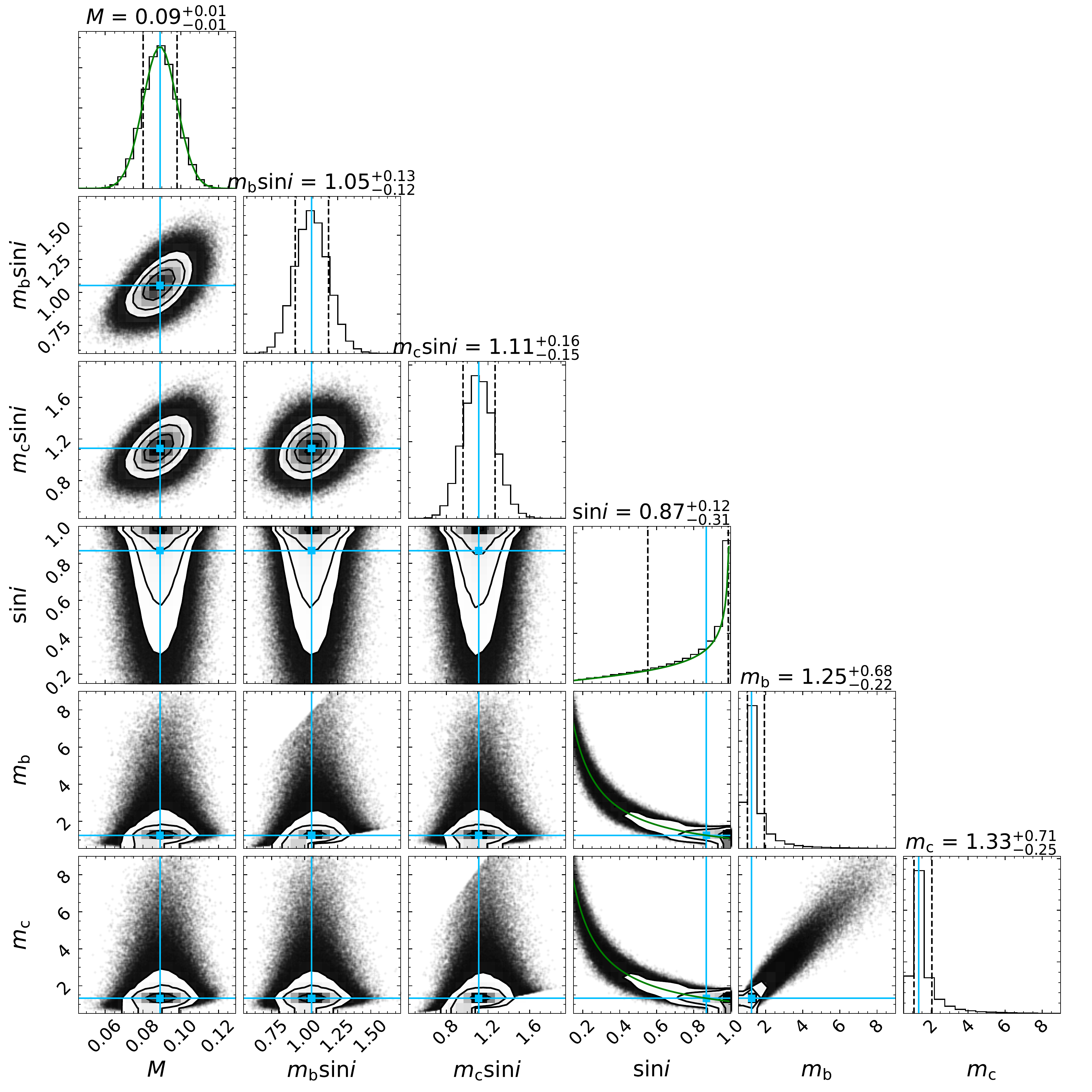}
    \caption{\label{fig:MCMC_mass}MCMC posterior distribution for derived masses of planets b and c. The priors for stellar mass $M$ and inclination $\sin i$ are overplotted in green.}
\end{figure}

\section{Radial velocities}

\longtab[1]{
    \begin{longtable}{@{}crrrrrrrrc@{}}
        \caption{\label{tab:RV_VIS} CARMENES VIS radial velocities and activity indicators. Flag 1 marks RVs without drift correction.}\\
        \hline
        \hline
        BJD & RV & $\sigma_{\rm RV}$ & CRX & $\sigma_{\rm CRX}$ & dLW &  $\sigma_{\rm dLW}$ & H$\alpha$ &  $\sigma_{\rm H\alpha}$ & Flag\\
            & [m/s] & [m/s] & [m/s/Np] & [m/s/Np] & [${\rm m^2/s^2}$] & [${\rm m^2/s^2}$] &  &  & \\
        \hline
        \endfirsthead
        \caption{VIS RVs (cont.)}\\
        \hline
        \hline
        BJD & RV & $\sigma_{\rm RV}$ & CRX & $\sigma_{\rm CRX}$ & dLW &  $\sigma_{\rm dLW}$ & H$\alpha$ &  $\sigma_{\rm H\alpha}$ & Flag\\
            & [m/s] & [m/s] & [m/s/Np] & [m/s/Np] & [${\rm m^2/s^2}$] & [${\rm m^2/s^2}$] &  &  & \\
        \hline
        \endhead
        \hline
        \endfoot
        2457419.28354 &   -3.48 &  1.67 &   -8.92 &   16.09 &  -11.95 &    2.26 &   1.177 &   0.007 &  0\\
2457421.41749 &   -3.88 &  1.22 &   15.95 &   11.47 &  -20.61 &    1.43 &   1.334 &   0.008 &  0\\
2457595.63597 &   -3.51 &  1.89 &   10.18 &   18.24 &   -7.84 &    2.13 &   1.362 &   0.010 &  0\\
2457596.64927 &   -4.83 &  1.70 &   23.92 &   16.45 &   33.64 &    6.28 &   1.718 &   0.010 &  0\\
2457611.63441 &   -0.56 &  2.24 &   45.48 &   20.28 &  -23.07 &    2.73 &   1.145 &   0.008 &  0\\
2457612.66146 &   -0.31 &  1.96 &   12.45 &   16.49 &  -24.16 &    2.18 &   1.190 &   0.008 &  0\\
2457613.64561 &    4.06 &  1.79 &  -22.30 &   13.67 &  -26.48 &    2.38 &   1.193 &   0.008 &  0\\
2457615.68063 &   -4.35 &  2.49 &    9.93 &   19.06 &  -22.24 &    3.40 &   1.127 &   0.010 &  1\\
2457619.65129 &    1.72 &  1.81 &   37.12 &   15.46 &  -24.87 &    3.23 &   1.070 &   0.006 &  0\\
2457621.61312 &    0.55 &  1.66 &   20.62 &   13.78 &  -25.82 &    3.76 &   1.073 &   0.006 &  0\\
2457622.61332 &    6.90 &  1.71 &   33.46 &   15.64 &  -30.82 &    4.65 &   1.508 &   0.011 &  0\\
2457623.68473 &    5.44 &  1.83 &   33.91 &   15.41 &  -31.63 &    3.76 &   1.388 &   0.008 &  0\\
2457625.65334 &   -2.50 &  1.38 &   20.67 &   11.27 &  -30.49 &    3.81 &   1.180 &   0.006 &  0\\
2457628.63497 &    1.78 &  2.08 &   23.58 &   19.39 &  -36.70 &    3.37 &   1.341 &   0.010 &  0\\
2457629.62641 &    2.72 &  1.98 &   24.55 &   17.41 &  -39.78 &    3.66 &   1.421 &   0.010 &  0\\
2457630.64336 &   -0.47 &  2.03 &   62.27 &   19.22 &  -39.87 &    3.75 &   1.125 &   0.009 &  0\\
2457631.61818 &    3.13 &  1.82 &   18.76 &   14.56 &  -36.65 &    3.76 &   1.187 &   0.008 &  0\\
2457632.67351 &    0.47 &  1.85 &   59.42 &   10.89 &  -33.44 &    3.25 &   1.153 &   0.006 &  0\\
2457633.66438 &    3.45 &  1.97 &   37.71 &   20.04 &  -38.09 &    4.14 &   1.101 &   0.009 &  0\\
2457634.59101 &    4.67 &  1.52 &   17.23 &   14.63 &  -35.46 &    3.77 &   1.392 &   0.009 &  0\\
2457635.59839 &    1.72 &  1.78 &   33.14 &   16.98 &  -42.96 &    3.49 &   1.801 &   0.011 &  0\\
2457636.58484 &   -0.84 &  1.92 &   11.14 &   20.54 &  -41.24 &    3.42 &   1.366 &   0.011 &  0\\
2457641.58846 &    2.72 &  2.78 &   30.65 &   27.29 &  -32.23 &    3.38 &   1.144 &   0.010 &  0\\
2457642.66018 &    4.90 &  1.64 &   15.98 &   12.00 &  -30.03 &    6.82 &   1.151 &   0.006 &  0\\
2457643.65701 &    8.40 &  1.92 &   20.82 &   18.41 &  -34.98 &    3.45 &   2.329 &   0.013 &  0\\
2457644.64701 &    5.87 &  1.84 &   29.45 &   18.98 &  -32.06 &    3.52 &   1.044 &   0.010 &  0\\
2457646.56945 &   -2.90 &  2.30 &   41.78 &   24.94 &  -25.05 &    4.39 &   1.082 &   0.015 &  0\\
2457647.58116 &   -0.07 &  1.61 &   14.65 &   14.60 &  -16.68 &    3.16 &   1.161 &   0.008 &  0\\
2457648.66569 &    2.43 &  2.03 &   31.41 &   14.53 &  -19.64 &    3.20 &   1.312 &   0.009 &  0\\
2457649.67788 &    1.02 &  1.69 &   27.30 &   10.99 &  -18.77 &    2.91 &   1.413 &   0.007 &  0\\
2457650.60235 &   -1.08 &  1.60 &   17.68 &   14.40 &  -11.42 &    2.84 &   1.132 &   0.007 &  0\\
2457652.67401 &    2.03 &  1.71 &   -2.91 &   16.10 &  -16.65 &    2.52 &   1.647 &   0.009 &  0\\
2457654.62373 &    2.45 &  1.64 &  -12.86 &   16.50 &  -18.50 &    2.22 &   1.086 &   0.009 &  0\\
2457655.56814 &    0.58 &  1.69 &   -3.43 &   15.94 &  -19.18 &    2.15 &   1.270 &   0.009 &  0\\
2457656.62039 &   -2.73 &  1.65 &    8.81 &   16.07 &  -17.78 &    2.00 &   1.143 &   0.008 &  0\\
2457665.66350 &    0.32 &  1.61 &   -5.53 &   13.89 &   -6.86 &    1.57 &   1.128 &   0.006 &  0\\
2457672.57897 &    2.14 &  1.39 &   30.31 &   11.37 &    0.03 &    1.28 &   1.203 &   0.007 &  0\\
2457673.61955 &    0.28 &  1.58 &  -10.18 &   15.59 &   -6.57 &    2.23 &   1.454 &   0.009 &  0\\
2457673.65319 &    1.00 &  1.14 &   17.73 &    8.69 &   -6.55 &    1.82 &   1.377 &   0.008 &  0\\
2457678.57080 &    4.11 &  1.81 &   39.00 &   17.16 &  124.66 &   19.10 &   1.447 &   0.008 &  0\\
2457689.52936 &    1.35 &  1.51 &   13.40 &   14.60 &  -12.17 &    1.82 &   1.332 &   0.007 &  0\\
2457690.69038 &   -0.47 &  1.69 &    6.85 &   15.67 &   -5.91 &    1.61 &   1.128 &   0.008 &  0\\
2457691.49893 &    0.33 &  1.16 &   18.26 &   10.19 &  -13.29 &    7.07 &   1.447 &   0.008 &  0\\
2457693.44771 &    4.06 &  1.17 &   18.68 &    9.95 &  -15.86 &    6.58 &   1.073 &   0.007 &  0\\
2457694.49906 &   -0.28 &  1.26 &   13.14 &   10.89 &  -15.20 &    7.33 &   1.130 &   0.007 &  0\\
2457695.51622 &   -1.26 &  1.24 &   23.17 &   10.27 &   -7.88 &    1.68 &   1.109 &   0.007 &  0\\
2457709.45391 &   -3.05 &  1.29 &   19.77 &   10.49 &  -12.45 &    2.41 &   1.109 &   0.007 &  0\\
2457949.65499 &   -2.41 &  2.21 &    6.09 &   21.89 &   30.57 &    6.77 &   1.131 &   0.010 &  0\\
2457950.65011 &   -2.92 &  2.02 &  -13.37 &   20.40 &   35.44 &    5.99 &   1.523 &   0.012 &  0\\
2457953.64371 &    2.29 &  2.29 &    2.56 &   19.20 &   26.98 &    4.68 &   1.464 &   0.011 &  0\\
2457959.63636 &   -6.03 &  1.60 &  -20.91 &   12.56 &    2.94 &    1.64 &   1.250 &   0.007 &  0\\
2457960.65963 &   -2.49 &  1.50 &  -20.77 &   13.57 &    4.68 &    2.04 &   1.227 &   0.008 &  0\\
2457961.62634 &   -1.04 &  1.55 &    7.91 &   14.87 &    0.35 &    1.78 &   1.243 &   0.008 &  0\\
2457962.63047 &   -0.65 &  1.41 &   -8.82 &   12.74 &    1.53 &    1.43 &   1.235 &   0.007 &  0\\
2457963.63224 &    0.11 &  1.40 &    6.48 &    9.39 &    2.18 &    1.59 &   1.274 &   0.007 &  0\\
2457964.65801 &   -1.80 &  1.41 &   -0.81 &   12.29 &    2.65 &    1.39 &   1.572 &   0.007 &  0\\
2457969.65929 &   -3.93 &  1.32 &    4.11 &    9.63 &    1.74 &    1.31 &   1.188 &   0.005 &  0\\
2457970.61360 &   -2.99 &  1.49 &   -1.19 &   12.10 &   -1.78 &    1.54 &   1.153 &   0.006 &  0\\
2457971.60575 &   -1.19 &  1.60 &    1.23 &   14.55 &    0.66 &    2.24 &   1.238 &   0.008 &  0\\
2457972.66490 &    8.30 &  3.08 &   23.63 &   28.53 &    2.40 &    4.51 &   1.278 &   0.017 &  0\\
2457975.64930 &    1.31 &  1.65 &    2.69 &   11.84 &   -5.10 &    1.91 &   1.154 &   0.007 &  0\\
2457976.66526 &    0.93 &  1.83 &   19.62 &   10.36 &   -2.37 &    1.91 &   1.344 &   0.008 &  0\\
2457982.66080 &    1.95 &  1.35 &   15.69 &    9.54 &  -10.27 &    1.96 &   1.554 &   0.006 &  0\\
2457986.57142 &    3.93 &  1.66 &    4.23 &   14.56 &  -16.34 &    2.46 &   1.480 &   0.008 &  0\\
2457987.55319 &    2.05 &  1.47 &   15.18 &   12.95 &  -20.28 &    2.57 &   1.665 &   0.009 &  0\\
2457988.64327 &    2.49 &  2.73 &   -3.33 &   12.06 &  -15.85 &    2.58 &   1.244 &   0.006 &  0\\
2457990.56053 &   -4.11 &  1.64 &   19.84 &   14.51 &  -20.72 &    2.85 &   1.113 &   0.007 &  0\\
2457997.65668 &    3.46 &  1.22 &   12.93 &    8.53 &  -13.78 &    2.36 &   1.293 &   0.005 &  0\\
2458002.63169 &    1.81 &  1.37 &   10.76 &    8.78 &   -7.58 &    2.25 &   1.093 &   0.005 &  0\\
2458005.65294 &    0.95 &  1.34 &  -12.28 &   11.27 &   27.07 &   10.43 &   1.152 &   0.007 &  0\\
2458009.64887 &   -1.14 &  1.49 &   -9.49 &   13.62 &  -24.19 &    2.69 &   1.624 &   0.010 &  0\\
2458010.65579 &    2.38 &  1.38 &   11.92 &   11.84 &  -25.78 &    2.16 &   1.386 &   0.008 &  0\\
2458018.64275 &   -3.07 &  1.72 &   20.20 &   12.73 &  -22.24 &    2.29 &   1.113 &   0.006 &  0\\
2458019.63503 &   -2.20 &  1.57 &    0.48 &   11.57 &  -22.64 &    2.35 &   1.352 &   0.006 &  0\\
2458020.62737 &   -0.19 &  1.77 &   -1.13 &   11.33 &  -24.44 &    2.65 &   1.184 &   0.006 &  0\\
2458021.62206 &    2.69 &  1.31 &   14.08 &    9.14 &  -21.98 &    2.20 &   1.126 &   0.005 &  0\\
2458022.66124 &    3.64 &  1.62 &   31.19 &   11.23 &  -22.51 &    2.41 &   1.111 &   0.006 &  0\\
2458023.62656 &    2.76 &  1.65 &    8.12 &   13.67 &  -26.27 &    2.63 &   1.130 &   0.008 &  0\\
2458024.64620 &    1.68 &  5.13 &   40.84 &   40.36 &  -74.79 &   10.26 &   1.117 &   0.042 &  0\\
2458025.62806 &   -0.58 &  1.70 &   -0.75 &    9.79 &  -21.45 &    2.14 &   1.122 &   0.006 &  0\\
2458026.60479 &    2.70 &  1.72 &   -2.95 &   12.37 &  -27.07 &    2.66 &   1.527 &   0.009 &  0\\
2458027.60878 &    0.33 &  1.27 &    9.09 &    9.79 &  -22.77 &    2.79 &   1.146 &   0.005 &  0\\
2458028.61592 &   -4.88 &  1.66 &   -3.56 &   13.02 &  -19.79 &    2.53 &   1.269 &   0.007 &  0\\
2458029.62376 &    1.47 &  1.90 &   39.97 &   10.73 &  -16.68 &    2.40 &   1.146 &   0.005 &  0\\
2458030.69588 &    4.44 &  2.17 &   13.03 &   16.80 &  -26.64 &    2.08 &   1.405 &   0.008 &  0\\
2458031.59252 &    7.49 &  1.87 &   18.86 &   13.48 &  -16.52 &    2.54 &   1.207 &   0.007 &  0\\
2458032.59559 &    7.50 &  1.77 &   39.27 &   11.76 &   -4.93 &    3.06 &   1.065 &   0.006 &  0\\
2458033.65479 &    1.73 &  1.63 &   18.17 &    7.88 &    0.65 &    2.92 &   1.089 &   0.005 &  0\\
2458040.53208 &    2.18 &  2.16 &   21.50 &   11.53 &  -10.10 &    1.49 &   1.164 &   0.005 &  0\\
2458040.66282 &    4.31 &  2.21 &   32.16 &   11.50 &   -8.41 &    2.03 &   1.158 &   0.005 &  0\\
2458041.50453 &   12.68 &  4.61 &   17.65 &    9.80 &   -3.41 &    1.29 &   1.187 &   0.005 &  0\\
2458041.71563 &    8.15 &  4.50 &    9.31 &    9.29 &   -6.04 &    1.73 &   1.236 &   0.007 &  1\\
2458042.49881 &    0.53 &  5.06 &   13.82 &   24.15 &  -31.40 &    3.82 &   1.173 &   0.021 &  1\\
2458043.47476 &  -11.52 &  4.30 &   16.06 &    9.03 &   -2.04 &    1.68 &   1.403 &   0.006 &  1\\
2458047.52314 &    0.13 &  1.77 &  -17.87 &   11.15 &    9.76 &    1.77 &   1.376 &   0.006 &  0\\
2458048.50987 &   -1.54 &  1.42 &   -3.00 &    9.99 &   13.70 &    1.71 &   1.317 &   0.005 &  0\\
2458049.44609 &   -2.75 &  2.20 &  -26.61 &   12.35 &   14.50 &    2.17 &   1.299 &   0.006 &  0\\
2458050.46523 &    1.33 &  1.82 &  -24.55 &   13.73 &   13.54 &    2.83 &   1.208 &   0.007 &  0\\
2458050.67639 &    1.75 &  2.24 &  -29.67 &   19.91 &    9.91 &    2.43 &   2.106 &   0.013 &  0\\
2458051.52644 &   -0.71 &  1.69 &  -14.99 &   11.05 &   15.94 &    2.45 &   1.328 &   0.006 &  0\\
2458051.65272 &   -0.94 &  1.70 &   -6.79 &   10.92 &   16.32 &    2.36 &   1.357 &   0.005 &  0\\
2458052.41748 &   -1.70 &  1.37 &    3.82 &    9.68 &   19.04 &    2.43 &   1.562 &   0.006 &  0\\
2458052.64685 &   -1.72 &  1.35 &    7.08 &    9.88 &   12.77 &    2.69 &   2.046 &   0.007 &  0\\
2458053.47107 &   -0.13 &  1.61 &   -6.99 &   10.40 &   15.76 &    2.76 &   1.290 &   0.006 &  0\\
2458053.64067 &   -0.75 &  1.82 &  -23.06 &   14.05 &   13.52 &    2.83 &   1.215 &   0.007 &  0\\
2458054.42297 &    0.39 &  1.83 &   -9.01 &   11.15 &   19.12 &    2.80 &   1.279 &   0.006 &  0\\
2458054.56012 &    1.12 &  1.76 &  -10.64 &    9.16 &   22.55 &    2.62 &   1.337 &   0.006 &  0\\
2458055.44945 &    4.77 &  1.76 &    4.11 &   15.32 &   18.84 &    2.83 &   1.374 &   0.007 &  0\\
2458056.50768 &    3.33 &  1.42 &   15.33 &    9.00 &   19.62 &    2.98 &   1.399 &   0.006 &  0\\
2458056.60640 &    4.01 &  1.48 &   -1.73 &   10.53 &   18.38 &    4.21 &   4.263 &   0.012 &  0\\
2458057.46819 &    2.24 &  1.38 &   12.09 &   11.35 &   23.39 &    3.08 &   1.882 &   0.007 &  0\\
2458058.37691 &   -2.11 &  1.84 &   -0.22 &   12.25 &   29.72 &    3.30 &   1.349 &   0.006 &  0\\
2458059.45851 &   -2.95 &  1.56 &  -27.37 &   11.97 &   32.56 &    3.49 &   2.390 &   0.009 &  0\\
2458060.35331 &   -3.08 &  1.91 &   -7.48 &   19.65 &  123.66 &   15.49 &   1.317 &   0.008 &  0\\
2458060.49914 &   -7.11 &  2.77 &  -67.54 &   29.35 &  289.93 &   32.45 &   1.170 &   0.009 &  0\\
2458064.53012 &   -3.44 &  3.80 &  -34.78 &   41.51 &   32.36 &    5.89 &   2.791 &   0.020 &  0\\
2458064.66617 &    0.28 &  3.61 &  -54.96 &   38.49 &   45.16 &    6.08 &   2.084 &   0.016 &  0\\
2458065.47674 &    2.62 &  1.48 &  -31.74 &   11.15 &   37.40 &    3.51 &   5.850 &   0.015 &  0\\
2458066.37640 &    2.24 &  2.07 &  -25.52 &   19.92 &   25.66 &    3.92 &   1.240 &   0.015 &  0\\
2458066.50967 &   -0.49 &  3.60 &  -10.17 &   40.70 &   43.06 &    5.68 &   1.527 &   0.020 &  0\\
2458074.39260 &   -3.05 &  1.46 &  -29.66 &    9.57 &   24.70 &    3.32 &   1.571 &   0.006 &  0\\
2458074.56213 &   -4.71 &  1.40 &  -18.63 &    9.28 &   24.14 &    3.41 &   1.957 &   0.007 &  0\\
2458078.32077 &   -1.30 &  1.77 &    2.22 &   13.38 &   20.60 &    2.84 &   1.333 &   0.007 &  0\\
2458078.41927 &   -1.83 &  1.86 &  -18.02 &   14.11 &   16.60 &    2.98 &   1.225 &   0.006 &  0\\
2458079.32107 &    1.85 &  1.65 &   -6.33 &   10.91 &   20.51 &    2.55 &   1.356 &   0.006 &  0\\
2458080.30958 &    0.48 &  1.66 &  -20.68 &   11.63 &  -47.57 &   15.03 &   1.562 &   0.008 &  0\\
2458081.31937 &   -1.28 &  2.42 &  -25.50 &   24.73 &   12.55 &    3.80 &   1.258 &   0.013 &  0\\
2458081.44998 &   -2.29 &  1.76 &  -31.55 &   13.03 &    8.19 &    2.32 &   1.381 &   0.007 &  0\\
2458082.30055 &   -2.22 &  1.71 &   -6.59 &   11.76 &   11.68 &    2.03 &   1.391 &   0.007 &  0\\
2458082.42621 &   -3.53 &  1.63 &    1.86 &   10.58 &    8.21 &    2.37 &   1.303 &   0.006 &  0\\
2458084.36909 &   -3.45 &  1.97 &  -27.02 &   11.71 &    9.23 &    1.70 &   1.381 &   0.008 &  0\\
2458085.55401 &   -1.09 &  2.37 &  -19.69 &   18.63 &   -2.80 &    2.08 &   2.449 &   0.015 &  0\\
2458088.55202 &   -1.24 &  5.79 & -118.31 &   65.01 &  131.21 &   22.78 &   1.181 &   0.026 &  0\\
2458091.48048 &   -2.24 &  1.49 &  -23.28 &    8.74 &    7.00 &    2.38 &   1.173 &   0.005 &  0\\
2458092.50560 &   -4.10 &  1.38 &   -3.18 &   10.35 &   -0.86 &    1.87 &   1.321 &   0.006 &  0\\
2458093.31238 &   -2.53 &  1.14 &   -4.47 &    8.61 &   -7.04 &    1.93 &   1.369 &   0.006 &  0\\
2458094.33746 &   -4.37 &  1.35 &   -8.74 &   10.39 &   -7.34 &    2.29 &   1.288 &   0.005 &  0\\
2458094.47703 &   -1.16 &  1.26 &    3.58 &    9.23 &   -8.44 &    2.07 &   1.613 &   0.006 &  0\\
2458095.38053 &    3.03 &  1.27 &   18.17 &    8.11 &  -22.63 &    4.21 &   1.306 &   0.006 &  0\\
2458099.34850 &   -2.62 &  3.10 &  -28.44 &   35.32 &  -52.33 &   10.83 &   1.309 &   0.024 &  0\\
2458104.35948 &   -0.58 &  2.97 &  -77.71 &   28.59 &  -56.79 &   11.07 &   1.304 &   0.024 &  0\\
2458105.39761 &   -2.53 &  1.71 &   -0.64 &   16.65 &  -36.42 &    5.73 &   1.194 &   0.009 &  0\\
2458106.41814 &    2.11 &  4.91 &  -70.04 &   59.09 &  -71.70 &   10.37 &   1.199 &   0.054 &  0\\
2458109.33039 &    2.42 &  1.76 &   22.83 &   14.94 &  -36.10 &    5.01 &   1.308 &   0.008 &  0\\
2458110.33109 &    2.40 &  1.60 &    1.53 &   13.96 &  -34.56 &    4.47 &   1.405 &   0.006 &  0\\
2458111.29685 &    2.82 &  2.49 &   -2.66 &   24.29 &  -42.87 &    6.19 &   1.073 &   0.012 &  0\\
2458112.31135 &    0.81 &  1.35 &   -3.33 &   11.06 &  -27.38 &    4.43 &   1.150 &   0.005 &  0\\
2458114.27031 &    0.76 &  2.43 &    0.76 &   20.71 &  -47.15 &    7.59 &   1.081 &   0.020 &  0\\
2458118.36114 &   -5.05 &  1.55 &   -0.93 &   12.42 &   -7.84 &    5.12 &   1.674 &   0.007 &  0\\
2458120.38765 &    0.31 &  2.24 &    7.75 &   23.85 &  -24.12 &    4.65 &   1.068 &   0.013 &  0\\
2458121.36531 &   -0.49 &  1.46 &    1.61 &   11.02 &  -19.78 &    4.05 &   2.579 &   0.009 &  0\\
2458122.35020 &   -0.98 &  1.70 &   22.30 &    9.87 &  -24.19 &    3.18 &   1.353 &   0.005 &  0\\
2458123.34833 &   -0.22 &  1.47 &   -0.66 &   12.46 &  -29.27 &    3.30 &   1.179 &   0.006 &  0\\
2458124.34296 &    1.36 &  2.04 &  -21.87 &   12.42 &  -42.72 &    6.25 &   1.081 &   0.009 &  0\\
2458132.35549 &    3.74 &  4.58 &   61.42 &   52.90 &  -26.51 &    5.89 &   1.131 &   0.033 &  0\\
2458134.36060 &    5.10 &  1.76 &    5.89 &   16.48 &  -20.21 &    2.55 &   1.069 &   0.009 &  0\\
2458135.33013 &    1.70 &  2.01 &  -22.74 &   15.76 &  -30.95 &    2.68 &   1.090 &   0.010 &  0\\
2458136.33323 &    0.73 &  1.42 &  -19.82 &    9.10 &  -21.78 &    2.26 &   1.480 &   0.007 &  0\\
2458138.34596 &    0.67 &  2.11 &   19.13 &   18.23 &  -15.99 &    1.70 &   1.394 &   0.011 &  0\\
2458139.34575 &    0.99 &  1.57 &    8.85 &   11.66 &  -11.44 &    1.69 &   1.321 &   0.007 &  0\\
2458140.35047 &   -0.67 &  1.49 &   -2.40 &   11.92 &  -12.52 &    1.54 &   1.790 &   0.008 &  0\\
2458141.41398 &   -3.41 &  1.43 &    7.00 &   12.36 &  -12.63 &    1.66 &   1.389 &   0.008 &  0\\
2458143.31165 &   -3.86 &  1.50 &  -11.07 &    8.19 &    6.55 &    1.17 &   1.290 &   0.005 &  0\\
2458149.31804 &   -0.36 &  1.27 &    7.35 &    9.51 &   21.31 &    2.07 &   1.490 &   0.009 &  0\\
2458159.31587 &    4.07 &  2.76 &  -34.47 &   26.55 &   23.34 &    4.35 &   1.649 &   0.017 &  0\\
2458161.32510 &   -1.00 &  2.47 &   -5.59 &   24.18 &    8.89 &    7.65 &   1.534 &   0.014 &  0\\
2458164.31711 &    0.09 &  2.46 &   -9.59 &   14.19 &   24.61 &    3.57 &   1.542 &   0.008 &  0\\
2458165.29737 &   -0.57 &  2.30 &  -18.11 &   14.86 &   22.86 &    4.27 &   3.584 &   0.016 &  0\\
2458166.32428 &   -3.24 &  2.09 &   -8.53 &   21.37 &   10.55 &    8.12 &   1.553 &   0.014 &  0\\
2458167.29587 &   -6.50 &  2.37 &  -15.39 &   16.82 &   22.89 &    3.93 &   2.004 &   0.012 &  0\\
2458172.30158 &    1.48 &  2.18 &  -22.21 &   18.67 &   15.49 &    6.02 &   2.499 &   0.020 &  0\\
2458173.34394 &    2.23 &  2.22 &   -6.43 &   19.95 &   30.04 &    3.84 &   1.475 &   0.011 &  0\\
2458174.30897 &    2.47 &  2.54 &  -35.41 &   20.71 &   27.88 &    4.04 &   1.954 &   0.015 &  0\\
2458175.32244 &   -4.99 &  1.55 &  -29.69 &   11.86 &   27.00 &    2.79 &   1.344 &   0.007 &  0\\
2458182.32948 &    1.95 &  3.52 &  -19.86 &   35.73 &   -0.62 &    3.90 &   1.807 &   0.024 &  0\\
2458317.64488 &    1.54 &  2.31 &    3.45 &   23.01 &  -16.45 &    2.60 &   1.267 &   0.015 &  0\\
2458329.62479 &    0.09 &  1.31 &   14.97 &   10.02 &    3.21 &    1.33 &   1.413 &   0.007 &  0\\
2458338.66943 &   -1.99 &  1.74 &   33.94 &   12.80 &    0.81 &    1.35 &   1.279 &   0.007 &  0\\
2458339.67027 &    1.82 &  1.73 &  -15.29 &   14.43 &  -12.52 &    5.59 &   1.552 &   0.010 &  0\\
2458340.67229 &    1.86 &  1.32 &  -11.87 &   10.19 &    4.89 &    1.51 &   1.642 &   0.007 &  0\\
2458346.67216 &    3.08 &  1.23 &    7.50 &    8.28 &   -0.69 &    1.55 &   1.335 &   0.007 &  0\\
2458348.66311 &   -0.61 &  1.37 &  -12.34 &   10.89 &    4.39 &    1.35 &   1.714 &   0.007 &  0\\
2458349.67117 &    2.49 &  1.34 &    5.19 &   10.00 &    3.53 &    1.81 &   1.293 &   0.005 &  0\\
2458350.68042 &    3.00 &  1.52 &   -6.49 &   12.34 &    2.09 &    2.01 &   1.693 &   0.008 &  0\\
2458352.68322 &   -1.74 &  1.26 &    4.31 &    8.56 &    6.88 &    1.30 &   1.185 &   0.005 &  0\\
2458356.68144 &   -1.65 &  1.34 &    8.15 &    9.53 &   14.01 &    1.44 &   1.409 &   0.006 &  0\\
2458382.64315 &   -4.78 &  1.46 &  -22.79 &   11.05 &   16.26 &    2.53 &   2.131 &   0.007 &  0\\
2458383.61404 &   -1.09 &  1.38 &  -13.66 &   11.07 &   15.31 &    2.47 &   4.039 &   0.011 &  0\\
2458384.47003 &   -0.39 &  1.49 &  -18.87 &   11.39 &   25.05 &    2.77 &   1.924 &   0.007 &  0\\
2458385.62078 &    1.68 &  1.41 &    2.24 &   10.77 &   20.19 &    2.20 &   1.456 &   0.006 &  0\\
2458386.46614 &   -0.46 &  1.59 &   -5.80 &   13.89 &   52.77 &    7.35 &   1.817 &   0.008 &  0\\
2458391.61215 &   -2.66 &  1.51 &  -21.67 &   13.54 &   61.10 &    6.25 &   1.428 &   0.007 &  0\\
2458393.46595 &   -1.61 &  1.70 &  -25.54 &   15.91 &   12.36 &    2.58 &   2.294 &   0.014 &  0\\
2458396.60058 &   -0.27 &  1.47 &  -22.49 &   11.92 &   18.49 &    2.37 &   1.535 &   0.007 &  0\\
2458399.46780 &    1.90 &  1.61 &  -26.39 &   14.21 &   15.00 &    2.37 &   1.591 &   0.009 &  0\\
2458399.68186 &    1.10 &  1.71 &  -28.74 &   15.44 &   15.43 &    2.52 &   1.535 &   0.010 &  0\\
2458405.65993 &    1.78 &  4.36 &   -9.32 &   43.86 &  -43.90 &   12.93 &   1.404 &   0.046 &  0\\
2458410.63180 &    0.74 &  5.17 &  -94.22 &   38.37 &    1.33 &    4.01 &   1.430 &   0.023 &  0\\
2458413.59966 &   -0.78 &  1.53 &   -9.27 &   12.55 &   17.26 &    1.90 &   1.269 &   0.006 &  0\\
2458421.52970 &    3.93 &  3.92 &   28.16 &   40.01 &    3.71 &    3.88 &   1.486 &   0.029 &  0\\
2458427.37889 &   -2.52 &  2.10 &  -28.36 &   18.95 &   -7.53 &    2.62 &   1.569 &   0.013 &  0\\
2458433.52158 &   -1.46 &  1.61 &   -0.17 &    8.74 &    4.32 &    2.00 &   1.284 &   0.007 &  0\\
2458434.40546 &   -2.26 &  2.39 &  -28.72 &   24.77 &   -4.87 &    2.86 &   1.990 &   0.016 &  0\\
2458449.61992 &    5.35 &  4.88 &  -35.95 &   17.02 &   17.87 &    2.66 &   1.273 &   0.011 &  0\\
2458450.45767 &   -4.09 &  1.49 &   -8.29 &    9.77 &   13.26 &    1.95 &   1.326 &   0.006 &  0\\
2458451.48968 &   -2.93 &  1.55 &  -29.66 &   13.90 &    8.91 &    1.74 &   1.487 &   0.008 &  0\\
2458454.39081 &   -0.80 &  1.52 &  -24.36 &   13.31 &    4.36 &    1.98 &   1.213 &   0.007 &  0\\
2458454.61157 &    2.95 &  2.31 &  -37.17 &   25.52 &  -12.98 &    6.84 &   1.415 &   0.015 &  0\\
2458470.39976 &   -2.97 &  1.97 &   13.03 &   12.17 &   -1.64 &    1.39 &   1.650 &   0.007 &  0\\
2458474.40430 &    0.81 &  1.50 &   -1.90 &    7.90 &    5.20 &    1.67 &   1.196 &   0.005 &  0\\
2458475.37777 &   -3.10 &  1.77 &   16.64 &   18.67 &   51.70 &   11.61 &   1.674 &   0.010 &  1\\
2458476.38504 &   -0.63 &  1.40 &    9.49 &   10.10 &   -0.41 &    1.33 &   1.621 &   0.006 &  0\\
2458477.39026 &    1.76 &  1.22 &   -1.36 &   10.23 &    9.98 &    1.98 &   1.219 &   0.008 &  0\\
2458478.36957 &    0.76 &  1.28 &   -4.59 &    9.80 &   -9.76 &    3.20 &   1.354 &   0.006 &  0\\
2458479.33417 &    5.64 &  2.79 &  -41.50 &   29.74 &  -13.37 &    3.38 &   1.422 &   0.018 &  0\\
2458480.51080 &   -1.84 &  1.63 &    5.28 &   10.21 &   -2.84 &    1.33 &   1.424 &   0.006 &  0\\
2458481.36311 &   -2.38 &  2.35 &  -28.55 &   17.08 &  -20.07 &    6.28 &   1.561 &   0.013 &  0\\
2458483.45312 &   -1.33 &  2.00 &  -28.36 &   15.80 &  -19.30 &    5.40 &   1.337 &   0.010 &  0\\
2458484.35153 &    3.73 &  1.37 &   -8.42 &    9.75 &   -5.30 &    1.91 &   1.184 &   0.006 &  1\\
2458485.35456 &   -4.54 &  1.45 &  -17.24 &   11.07 &   -2.44 &    1.63 &   1.455 &   0.006 &  0\\
2458486.36388 &   -3.06 &  1.36 &   -2.22 &   11.28 &   -4.86 &    1.78 &   1.647 &   0.007 &  0\\
2458487.39957 &    3.61 &  1.76 &  -24.35 &   14.78 &  -15.72 &    5.07 &   1.396 &   0.010 &  0\\
2458488.34965 &    0.63 &  1.46 &  -29.62 &   12.88 &   -7.10 &    3.86 &   1.297 &   0.007 &  0\\
2458490.34181 &   -2.44 &  1.42 &  -26.58 &   10.33 &   -5.86 &    3.53 &   1.231 &   0.006 &  0\\
2458491.47626 &   -0.33 &  1.46 &  -12.04 &   12.11 &   -3.01 &    2.33 &   1.589 &   0.009 &  0\\
2458492.34631 &    7.27 &  3.40 &  -87.32 &   37.87 &  -27.45 &    5.68 &   1.136 &   0.025 &  1\\
2458493.48969 &    2.90 &  1.67 &    3.64 &   17.08 &  -16.89 &    6.37 &   2.824 &   0.020 &  0\\
2458496.34301 &   -2.01 &  1.40 &  -12.20 &   11.02 &    0.46 &    3.50 &   1.228 &   0.006 &  0\\
2458498.46246 &    4.54 &  1.17 &   19.74 &    8.38 &   10.21 &    1.55 &   1.270 &   0.006 &  0\\
2458511.42774 &    4.50 &  4.30 &  -57.77 &   50.84 &  -11.39 &    3.99 &   1.189 &   0.033 &  0\\
2458518.41765 &   -0.06 &  6.92 &  -30.38 &   86.19 &  -17.21 &    6.21 &   1.264 &   0.050 &  0\\
2458525.41433 &    2.93 &  2.36 &  -41.41 &   22.44 &   -5.60 &    2.62 &   1.125 &   0.012 &  0\\
2458528.37952 &   -1.70 &  1.24 &   -4.02 &    9.43 &   -7.21 &    1.45 &   1.313 &   0.006 &  0\\
2458529.38557 &   -3.67 &  1.16 &  -10.71 &    9.56 &   -1.91 &    1.66 &   1.420 &   0.006 &  0\\
2458531.39690 &    1.08 &  2.01 &  -43.60 &   18.83 &   27.79 &    7.39 &   2.006 &   0.015 &  0\\
2458532.38802 &    2.89 &  1.87 &  -31.33 &   18.22 &   13.80 &    4.19 &   1.946 &   0.014 &  0\\
2458534.37035 &   -0.52 &  1.67 &  -23.67 &   16.28 &   31.59 &    6.64 &   2.357 &   0.013 &  0\\
2458535.36254 &    1.07 &  1.68 &  -30.49 &   16.69 &   19.12 &    3.18 &   1.499 &   0.010 &  0\\
2458537.37009 &    2.96 &  1.97 &  -20.13 &   20.76 &    2.81 &    1.93 &   2.050 &   0.015 &  0\\
2458540.34986 &   -5.03 &  1.93 &  -25.91 &   16.48 &   13.50 &    2.28 &   1.687 &   0.010 &  0\\
2458541.34027 &    0.98 &  1.89 &   21.83 &   17.55 &    5.70 &    2.78 &   1.511 &   0.012 &  0\\
2458542.35452 &   -3.39 &  1.65 &    9.50 &   15.78 &    9.12 &    2.35 &   2.179 &   0.011 &  0\\
2458546.34437 &    1.87 &  1.49 &  -25.34 &   11.73 &   17.11 &    2.59 &   1.442 &   0.007 &  0\\
2458554.33741 &   -0.98 &  1.95 &  -37.29 &   18.30 &   48.37 &    5.88 &   1.412 &   0.010 &  0\\
2458555.33436 &    0.46 &  1.94 &   -9.61 &   16.70 &   25.06 &    2.97 &   1.830 &   0.013 &  0\\

        \hline
    \end{longtable}
}

\longtab[2]{
    \begin{longtable}{@{}crrrrrrc@{}}
        \caption{\label{tab:RV_NIR} CARMENES NIR radial velocities and activity indicators. Flag 1 marks RVs without drift correction, flag 2 marks instable instrument (BJD<2457692.0), and flag 4 marks clipped outliers.}\\
        \hline
        \hline
        BJD & RV & $\sigma_{\rm RV}$ & CRX & $\sigma_{\rm CRX}$ & dLW &  $\sigma_{\rm dLW}$ & Flag\\
            & [m/s] & [m/s] & [m/s/Np] & [m/s/Np] & [${\rm m^2/s^2}$] & [${\rm m^2/s^2}$] & \\
        \hline
        \endfirsthead
        \caption{NIR RVs (cont.).}\\
        \hline
        \hline
        BJD & RV & $\sigma_{\rm RV}$ & CRX & $\sigma_{\rm CRX}$ & dLW &  $\sigma_{\rm dLW}$ & Flag\\
            & [m/s] & [m/s] & [m/s/Np] & [m/s/Np] & [${\rm m^2/s^2}$] & [${\rm m^2/s^2}$] & \\
        \hline
        \endhead
        \hline
        \endfoot
        2457595.63569 &    0.04 &  6.89 &   -8.92 &   29.25 &  -41.58 &   28.63 &  2\\
2457596.64878 &   -5.80 &  6.70 &  -52.50 &   30.45 &  -10.59 &   15.20 &  2\\
2457611.63424 &   -5.75 &  6.40 &   27.98 &   23.10 &  -63.44 &   23.16 &  2\\
2457612.66128 &  -12.04 &  6.16 &   10.13 &   23.72 & -105.46 &   40.18 &  2\\
2457613.64612 &    4.83 & 10.89 &  -92.15 &   70.48 & -159.09 &   24.38 &  2\\
2457619.65091 &  -11.10 &  6.67 &   -0.82 &   30.36 & -120.09 &   27.39 &  2\\
2457623.68474 &    2.53 &  9.60 & -128.07 &   61.06 & -179.70 &   22.74 &  2\\
2457625.65310 &   -9.07 &  5.00 &   39.52 &   17.95 & -108.97 &   27.73 &  2\\
2457626.64155 &    0.60 & 11.97 &   12.87 &   59.35 & -113.00 &  101.11 &  2\\
2457628.63520 &   -8.79 &  6.75 &    2.04 &   30.65 & -132.81 &   45.52 &  2\\
2457629.62617 &  -10.06 &  6.19 &   -2.22 &   29.31 & -118.51 &   36.83 &  2\\
2457630.64341 &  -13.13 &  7.06 &  -48.62 &   32.38 & -130.37 &   32.14 &  2\\
2457631.61733 &  -13.25 &  7.22 &  -42.06 &   31.64 & -141.38 &   44.65 &  2\\
2457632.67318 &   -7.50 &  7.57 &   33.40 &   21.85 & -108.00 &   24.36 &  2\\
2457633.66396 &   10.63 & 11.18 & -158.36 &   76.48 & -193.87 &   22.02 &  2\\
2457634.59076 &   -4.71 &  6.28 &  -25.61 &   29.36 & -121.74 &   36.27 &  2\\
2457635.59831 &   -3.84 &  6.07 &  -42.13 &   26.42 & -122.55 &   44.14 &  2\\
2457636.58484 &  -10.95 & 14.36 & -210.37 &   78.31 & -247.27 &   42.88 &  2\\
2457641.58841 &   -5.75 &  8.22 &   -4.37 &   36.46 &  -21.67 &   48.01 &  2\\
2457642.66024 &   -1.18 &  5.91 &   -8.89 &   26.21 &  -87.57 &   26.21 &  2\\
2457643.65577 &   -1.86 &  5.70 &   21.86 &   24.44 &  -87.37 &   20.95 &  2\\
2457644.64792 &   -2.65 & 11.13 & -149.62 &   68.40 & -136.20 &   28.76 &  2\\
2457646.56918 &  -14.30 &  5.01 &  -11.83 &   22.13 &  -80.49 &   33.71 &  2\\
2457647.58109 &   -6.14 &  5.42 &    6.02 &   26.13 &  -76.62 &   24.09 &  2\\
2457648.66607 &    1.33 &  7.02 &   -1.07 &   27.40 &  -30.56 &   33.38 &  2\\
2457649.67827 &    1.43 &  5.95 &   39.37 &   23.90 &  -45.65 &   34.16 &  2\\
2457650.60255 &    3.74 &  5.49 &  -13.46 &   24.28 &  -76.20 &   35.43 &  2\\
2457652.67437 &   -4.44 &  4.75 &   27.20 &   20.46 &  -81.62 &   22.21 &  2\\
2457654.62341 &    6.62 &  5.83 &   -3.95 &   28.85 &  -78.50 &   30.27 &  2\\
2457655.56781 &    6.25 &  5.54 &   21.04 &   24.47 &  -62.90 &   26.70 &  2\\
2457656.62035 &    1.45 &  6.26 &  -14.86 &   28.46 &  -81.43 &   39.65 &  2\\
2457665.66346 &    4.81 &  7.04 &  -20.00 &   28.50 &  -53.37 &   22.29 &  2\\
2457672.57914 &   -4.51 &  4.05 &  -17.07 &   16.55 &    8.05 &    5.38 &  2\\
2457673.61941 &    2.29 &  4.47 &  -33.72 &   17.92 &   -7.72 &    8.61 &  2\\
2457673.65219 &   -3.51 &  4.08 &  -42.37 &   13.01 &    8.73 &    7.13 &  2\\
2457678.57068 &    0.34 &  3.86 &   24.96 &   17.02 &   50.86 &    4.89 &  2\\
2457689.53012 &   12.49 &  9.23 &  -88.66 &   72.72 &  -14.39 &    6.32 &  2\\
2457690.69027 &    3.32 &  4.44 &  -42.32 &   15.65 &  -11.05 &    9.78 &  2\\
2457691.49946 &    0.66 &  3.68 &   10.17 &   16.24 &   32.61 &    7.00 &  2\\
2457693.44773 &   -1.97 &  3.59 &  -33.80 &   14.20 &   27.98 &    5.68 &  0\\
2457694.49916 &    2.94 &  3.72 &  -12.03 &   14.00 &   -0.55 &    3.79 &  0\\
2457695.51670 &    1.52 &  3.54 &  -38.25 &   11.95 &  -14.42 &    6.31 &  0\\
2457699.50289 &   -0.45 &  8.52 &  -23.06 &   42.20 &  -25.87 &   18.81 &  0\\
2457709.45319 &    9.45 &  4.74 &  -76.08 &   13.10 &  -35.52 &    8.91 &  0\\
2457949.65434 &   -8.15 &  7.20 &   11.14 &   30.26 &   17.67 &    9.47 &  0\\
2457950.65026 &   -3.45 &  6.95 &   10.46 &   35.99 &   37.26 &    8.56 &  0\\
2457953.64452 &   -7.89 &  7.38 &   -7.07 &   26.31 &   32.71 &   10.30 &  0\\
2457959.63632 &  -10.02 &  5.36 &   10.58 &   18.49 &   35.87 &    7.69 &  0\\
2457960.66033 &  -12.98 &  4.68 &   29.90 &   21.85 &   35.20 &    8.32 &  0\\
2457961.62662 &   -4.71 &  4.23 &    6.96 &   18.21 &   17.76 &    6.81 &  0\\
2457962.62936 &   -0.38 &  5.03 &   50.46 &   20.13 &   33.41 &    7.93 &  0\\
2457963.63244 &   -2.93 &  4.56 &    2.55 &   19.83 &   23.12 &    7.06 &  0\\
2457964.65834 &   -5.62 &  5.06 &  -31.80 &   18.97 &   13.73 &    6.55 &  0\\
2457969.65934 &  -11.08 &  5.41 &   38.24 &   26.48 &    8.99 &    7.79 &  0\\
2457970.61439 &  -15.14 &  5.30 &   19.29 &   28.56 &   -0.75 &    6.11 &  0\\
2457971.60642 &   -2.82 &  4.85 &   51.34 &   20.50 &   -4.18 &    8.23 &  0\\
2457972.66454 &    0.44 & 15.56 &  -82.04 &   92.86 & -186.84 &   32.30 &  0\\
2457975.64938 &   -6.17 &  5.91 &   24.72 &   17.93 &  -58.57 &    7.56 &  0\\
2457976.66561 &   -1.39 &  5.35 &   13.50 &   24.02 &    1.51 &   10.35 &  0\\
2457982.66120 &  -11.25 &  4.98 &   20.45 &   24.46 &  -39.06 &    6.25 &  0\\
2457986.57169 &  -11.87 &  5.98 &   53.15 &   25.87 &  -56.56 &    8.58 &  0\\
2457987.55358 &   -2.78 &  6.20 &   26.80 &   16.80 &  -57.49 &    9.22 &  0\\
2457988.64394 &    7.93 &  9.19 &   27.39 &   22.75 &  -34.93 &   12.52 &  0\\
2457990.56058 &  -14.73 &  6.24 &    4.60 &   27.51 &  -72.92 &    8.39 &  0\\
2457997.65673 &    3.23 &  7.82 & -108.46 &   34.76 &  -94.35 &   16.35 &  0\\
2458002.63095 &    1.59 &  4.76 &   14.42 &   22.88 &  -47.39 &    9.38 &  0\\
2458003.62928 &    4.75 &  4.33 &   -5.01 &   18.68 &  -37.95 &    7.30 &  1\\
2458005.65418 &   -5.30 &  4.67 &  -25.90 &   21.80 &  -35.67 &    7.67 &  0\\
2458009.64997 &   -0.39 &  4.07 &  -15.16 &   18.15 &  -79.43 &   14.58 &  0\\
2458010.65422 &    3.08 &  3.74 &  -14.05 &   13.87 &  -82.11 &   13.02 &  0\\
2458018.64228 &   -1.12 &  5.35 &   17.35 &   19.49 &  -73.25 &   11.26 &  0\\
2458019.63348 &    7.39 &  4.53 &   11.94 &   17.60 &  -82.04 &    9.40 &  0\\
2458020.62575 &   11.75 &  5.36 &   18.09 &   17.75 &  -71.99 &   10.38 &  0\\
2458021.62104 &   12.74 &  4.23 &  -18.52 &   16.03 &  -68.61 &    9.41 &  0\\
2458022.66319 &   -4.49 &  4.61 &   33.44 &   11.37 &  -82.53 &   11.07 &  0\\
2458023.62793 &   -0.08 &  5.73 &   67.12 &   25.93 &  -83.45 &   12.70 &  0\\
2458024.64498 &    6.05 & 10.39 &   69.69 &   58.78 & -146.98 &   28.17 &  0\\
2458026.60610 &    6.68 &  5.53 &    8.77 &   22.46 &  -88.52 &   11.29 &  0\\
2458027.60825 &    1.84 &  4.36 &   41.82 &   21.61 &  -73.02 &    6.74 &  0\\
2458028.61802 &   -8.76 &  5.00 &   50.62 &   16.46 &  -81.95 &    9.10 &  0\\
2458029.62506 &   -2.78 &  5.36 &   30.77 &   17.95 &  -73.26 &    8.43 &  0\\
2458030.69502 &    9.55 &  4.91 &   28.48 &   22.34 &  -88.29 &   10.34 &  0\\
2458031.59235 &    4.42 &  4.64 &   49.01 &   16.03 &  -65.64 &    8.98 &  0\\
2458032.59794 &   -0.44 &  4.09 &   23.56 &   13.68 &  -65.14 &    8.02 &  0\\
2458033.65481 &    3.46 &  4.31 &   12.06 &   12.37 &  -54.45 &    6.34 &  0\\
2458040.53119 &   -3.41 &  6.10 &    0.80 &   15.61 &  -21.49 &    4.39 &  0\\
2458040.66077 &    1.27 &  6.18 &   -6.29 &   16.75 &  -21.74 &    4.80 &  0\\
2458041.50493 &    5.31 &  4.76 &  -24.63 &   19.06 &  -16.40 &    4.55 &  0\\
2458041.71520 &    0.47 &  4.66 &  -27.89 &   17.70 &  -15.65 &    6.32 &  0\\
2458043.47119 &   13.07 &  4.30 &   10.19 &   11.71 &   -4.85 &    5.11 &  0\\
2458047.52317 &   -2.03 &  6.58 &   34.40 &   13.35 &   30.50 &    6.78 &  0\\
2458048.50962 &   -3.24 &  4.04 &   -7.22 &   14.23 &   41.58 &    7.92 &  0\\
2458049.44657 &    6.23 &  6.39 &   40.39 &   14.68 &   47.82 &    7.43 &  0\\
2458050.46514 &   -8.56 &  4.57 &   40.90 &   15.50 &   42.29 &    8.26 &  0\\
2458050.67294 &   -0.70 & 10.23 &  -12.76 &   81.45 &   -8.30 &    8.80 &  0\\
2458051.52526 &    8.62 &  4.83 &    3.59 &   18.57 &   61.09 &    8.78 &  0\\
2458051.65013 &    8.83 &  4.81 &    5.79 &   19.28 &   71.90 &   10.04 &  0\\
2458052.41852 &   -7.88 &  4.24 &    4.29 &   15.25 &   69.17 &   10.19 &  0\\
2458052.64568 &   -6.51 &  4.36 &   -0.81 &   16.39 &   67.12 &    9.63 &  0\\
2458053.47137 &   -1.34 &  5.32 &  -43.77 &   20.75 &   69.40 &   10.08 &  0\\
2458053.63888 &   -0.78 &  5.31 &  -47.13 &   19.99 &   60.67 &    9.73 &  0\\
2458054.42368 &   -7.29 &  5.00 &   15.87 &   16.14 &   75.49 &   10.73 &  0\\
2458054.56042 &   -5.58 &  4.89 &    4.24 &   15.15 &   79.04 &   12.35 &  0\\
2458055.44974 &   -1.56 &  4.64 &  -52.09 &   17.93 &   69.30 &   12.63 &  0\\
2458056.50724 &    3.17 &  4.62 &  -31.06 &   13.92 &   95.04 &   11.64 &  0\\
2458056.60665 &    5.57 &  4.88 &  -38.56 &   16.53 &   87.12 &   10.60 &  0\\
2458057.46876 &    0.68 &  4.46 &   26.26 &   21.06 &  118.97 &   13.36 &  0\\
2458058.37703 &   -8.58 &  7.05 &   -3.14 &   28.82 &  124.65 &   14.02 &  0\\
2458058.59744 &   14.90 & 10.15 &  -47.47 &   72.23 &   26.62 &   12.26 &  0\\
2458059.45895 &   -9.31 &  6.48 &   25.04 &   26.62 &  129.02 &   14.92 &  0\\
2458060.35395 &    0.36 &  6.07 &   26.36 &   31.82 &  151.39 &   16.42 &  0\\
2458060.49895 &   -4.28 &  7.76 &   65.17 &   37.97 &  210.25 &   21.12 &  0\\
2458060.62450 &  -16.21 & 15.27 &   -1.47 &   78.36 &  210.09 &   50.81 &  0\\
2458064.66624 &   15.59 &  8.66 &  -80.61 &   58.67 &  -28.25 &   21.79 &  0\\
2458065.47672 &   11.70 &  6.60 &   43.93 &   33.09 &  148.60 &   17.07 &  0\\
2458066.37602 &    9.68 &  9.15 &   34.42 &   45.07 &  119.30 &   17.95 &  0\\
2458066.50945 &   13.54 &  8.47 &   28.01 &   38.52 &  125.58 &   19.09 &  0\\
2458074.39111 &    2.54 &  4.49 &   12.12 &   13.73 &   92.63 &   13.84 &  0\\
2458074.56259 &    0.34 &  7.19 &  -69.35 &   50.57 &   37.45 &    8.27 &  0\\
2458078.32070 &    6.47 &  3.93 &   -5.28 &    8.71 &   94.67 &   14.67 &  0\\
2458078.41918 &    0.90 &  4.11 &   10.59 &   10.96 &   77.62 &   12.23 &  0\\
2458079.32152 &    1.72 &  4.71 &  -17.56 &   12.42 &   76.12 &   12.35 &  0\\
2458080.31022 &    2.74 &  5.23 &   30.33 &   17.62 & -153.74 &   35.64 &  0\\
2458081.31976 &    8.08 &  5.47 &   10.49 &   22.57 &   44.53 &   10.58 &  0\\
2458081.44974 &    0.12 &  4.68 &   27.41 &   14.37 &   60.61 &    9.79 &  0\\
2458082.30039 &    0.80 &  5.41 &   27.24 &   17.69 &   56.50 &    9.28 &  0\\
2458082.42428 &   -0.20 &  5.07 &   16.08 &   12.84 &   45.78 &   10.31 &  0\\
2458084.36875 &    2.18 &  4.90 &    9.54 &   21.12 &   44.88 &    7.49 &  0\\
2458085.55441 &    0.78 &  7.89 &   31.11 &   25.02 &   20.19 &    5.73 &  0\\
2458088.55157 &    9.01 &  9.19 &  -29.21 &   37.06 &   -3.10 &   12.96 &  0\\
2458091.48090 &    4.81 &  4.59 &   -9.12 &   13.55 &    0.15 &    5.48 &  0\\
2458092.50612 &    4.03 &  7.25 &  -98.65 &   53.29 & -118.63 &   15.37 &  0\\
2458093.31269 &    2.38 &  3.74 &   22.36 &   13.39 &  -18.97 &    5.80 &  0\\
2458094.33691 &    2.44 &  3.50 &    1.99 &   14.55 &  -20.03 &    6.44 &  0\\
2458094.47679 &   -2.36 &  3.39 &   15.71 &   12.93 &  -24.56 &    6.55 &  0\\
2458095.38127 &   -1.65 &  3.80 &    4.08 &   15.85 &  -27.43 &    6.69 &  0\\
2458097.36992 &  -36.95 & 15.24 &    8.24 &   76.46 & -108.69 &   41.24 &  4\\
2458099.34862 &  -10.34 &  7.56 &   14.26 &   35.22 &  -84.58 &   14.88 &  0\\
2458104.35974 &   -2.20 &  6.61 &  -26.69 &   20.31 &  -71.94 &   17.21 &  0\\
2458105.39903 &   -6.79 &  4.69 &    7.53 &   19.61 &  -76.09 &   13.51 &  0\\
2458106.41826 &    9.59 &  9.82 &   55.71 &   47.06 & -127.83 &   30.40 &  0\\
2458108.46389 &  -53.51 & 21.71 &  258.91 &   92.94 & -195.94 &   56.81 &  4\\
2458109.33128 &    0.68 &  4.08 &    0.06 &   15.21 &  -76.63 &   12.64 &  0\\
2458110.32852 &    5.86 &  3.96 &    7.58 &   15.47 &  -74.48 &   11.13 &  0\\
2458111.29538 &   -0.23 &  4.28 &   37.78 &   12.11 &  -75.71 &   13.74 &  0\\
2458112.31264 &   -3.94 &  3.31 &   26.71 &    9.26 &  -65.47 &    9.94 &  0\\
2458114.27031 &   -2.01 &  6.71 &   56.83 &   26.48 &  -91.88 &   14.92 &  0\\
2458114.44953 &  -13.62 & 16.46 & -201.16 &   67.79 &  -83.10 &   30.75 &  0\\
2458118.36021 &   -0.81 &  3.46 &    2.35 &   13.20 &  -69.51 &   12.23 &  0\\
2458120.38764 &    3.81 &  4.57 &   21.96 &   19.77 &  -87.19 &   17.21 &  0\\
2458121.36520 &    4.47 &  3.63 &   16.72 &   12.87 &  -70.78 &   11.83 &  0\\
2458122.35001 &   -0.52 &  4.00 &  -15.16 &   15.61 &  -77.38 &   12.23 &  0\\
2458123.34725 &    3.85 &  7.25 & -127.90 &   15.15 & -122.11 &   25.28 &  0\\
2458124.34337 &    8.31 &  6.16 &    5.18 &   15.21 &  -67.05 &   12.08 &  0\\
2458132.35562 &   -7.67 &  9.94 &   46.75 &   47.55 & -118.34 &   17.83 &  0\\
2458134.36129 &    4.74 &  3.67 &  -14.43 &   14.83 &  -53.34 &    9.46 &  0\\
2458135.33041 &    3.58 &  5.00 &   18.11 &   20.35 &  -56.82 &    8.80 &  0\\
2458136.33389 &    2.00 &  3.06 &   -1.80 &    5.82 &  -36.30 &    7.23 &  0\\
2458138.34678 &   -4.12 &  5.38 &  -23.70 &   16.08 &  -38.14 &    8.50 &  0\\
2458139.34550 &    1.32 &  3.96 &  -37.43 &   13.10 &  -22.90 &    4.68 &  0\\
2458140.35037 &   -0.62 &  3.91 &  -39.23 &   14.15 &  -25.04 &    5.46 &  0\\
2458141.41381 &   -8.70 &  3.72 &  -17.89 &   13.50 &  -22.60 &    4.71 &  0\\
2458149.31840 &   -1.09 &  3.35 &   22.63 &   12.41 &   50.89 &    9.38 &  0\\
2458159.31625 &   -0.48 &  5.97 &  -34.14 &   20.14 &   63.29 &   15.61 &  0\\
2458161.32497 &    6.35 &  9.19 &  -94.20 &   51.54 &  -75.72 &   33.52 &  0\\
2458164.31702 &   -3.87 &  7.70 &   -1.89 &   15.99 &  101.90 &   14.90 &  0\\
2458165.29676 &   -6.35 &  7.18 &  -40.89 &   23.56 &  105.24 &   16.67 &  0\\
2458166.32434 &   -6.82 &  5.63 &   31.94 &   24.75 &   87.71 &   17.22 &  0\\
2458167.29668 &   -5.95 &  6.86 &   32.51 &   22.80 &   95.64 &   14.09 &  0\\
2458172.30159 &   -9.31 &  4.52 &   -3.30 &   17.61 &   59.82 &   14.56 &  0\\
2458173.34398 &    1.45 &  6.18 &   45.47 &   17.85 &   65.06 &   11.86 &  0\\
2458175.32258 &   -7.01 &  4.28 &   47.50 &   14.30 &   77.42 &   12.80 &  0\\
2458182.32921 &   13.15 & 13.22 & -151.47 &   63.35 & -257.35 &   83.43 &  0\\
2458317.64551 &   -2.21 &  5.98 &   14.87 &   25.93 &  -23.59 &   10.04 &  0\\
2458329.62422 &    0.73 &  4.11 &  -11.33 &   20.69 &    1.61 &    4.04 &  0\\
2458338.66937 &   -4.02 &  4.79 &    8.47 &   14.88 &   17.84 &    6.85 &  0\\
2458339.67110 &   -0.22 &  3.97 &    9.99 &   11.68 &    9.63 &    7.68 &  0\\
2458340.67263 &    4.96 &  4.25 &   57.80 &   13.16 &   21.26 &    6.11 &  0\\
2458346.67070 &    2.49 &  4.03 &   38.38 &   13.73 &   17.87 &    5.81 &  0\\
2458348.66234 &   -3.83 &  3.87 &   24.55 &   15.58 &   31.34 &    6.67 &  0\\
2458349.67104 &    2.54 &  3.56 &    1.21 &   12.37 &   30.78 &    7.24 &  0\\
2458350.67943 &    6.24 &  3.89 &  -28.58 &   13.12 &   28.34 &    6.73 &  0\\
2458352.68312 &   -1.20 &  3.53 &   -9.58 &   11.01 &   27.91 &    7.25 &  0\\
2458356.68142 &   -2.97 &  3.62 &   -3.20 &   11.36 &   41.70 &    8.58 &  0\\
2458382.64379 &   -8.57 &  4.32 &   16.59 &   12.80 &   64.36 &   11.41 &  0\\
2458383.61440 &    1.25 &  3.90 &   24.24 &   14.69 &   66.31 &    9.41 &  0\\
2458384.47047 &    2.00 &  4.30 &   24.78 &   19.98 &   70.71 &    7.99 &  0\\
2458385.61865 &    8.19 &  4.00 &   27.33 &   17.24 &   67.04 &   11.98 &  0\\
2458386.46617 &    6.73 &  4.51 &   22.63 &   19.74 &   80.05 &   10.58 &  0\\
2458391.61619 &   -8.60 &  7.75 & -107.49 &   56.15 &  -28.89 &   13.93 &  0\\
2458393.46570 &   12.59 & 10.46 & -135.04 &   61.80 & -105.05 &   23.70 &  0\\
2458396.60119 &    2.44 &  4.11 &   26.74 &   15.62 &   60.55 &   11.81 &  0\\
2458399.46753 &   -0.34 &  5.08 &   48.30 &   20.80 &   53.68 &   10.47 &  0\\
2458399.68120 &    0.41 &  4.81 &   20.70 &   20.78 &   48.21 &    9.33 &  0\\
2458405.65981 &  -14.61 &  9.72 &   20.25 &   43.98 &   -2.17 &   17.65 &  0\\
2458410.63176 &    6.30 & 15.21 &   33.35 &   26.62 &   10.51 &   10.51 &  0\\
2458413.59889 &    2.48 &  4.17 &   10.53 &   16.39 &   43.27 &    6.46 &  0\\
2458421.52968 &    1.06 & 10.70 &   58.30 &   33.34 &  -30.59 &   15.76 &  0\\
2458427.37966 &    8.54 &  8.77 &  -63.18 &   53.90 & -121.35 &   26.46 &  0\\
2458433.52034 &    5.39 &  4.39 &    5.47 &   18.15 &   44.93 &    6.91 &  0\\
2458434.40555 &    2.13 &  5.16 &   28.94 &   19.99 &   30.55 &    6.09 &  0\\
2458447.47133 &   -8.61 & 23.56 &   15.72 &  115.33 &  -45.79 &   44.20 &  0\\
2458449.61881 &    4.31 &  8.86 &  -35.51 &   22.54 &   42.64 &    9.94 &  0\\
2458450.45685 &    5.62 &  4.85 &   22.52 &   16.22 &   53.92 &    9.54 &  0\\
2458451.48983 &   -0.79 &  3.72 &   26.07 &   14.82 &   43.63 &    8.17 &  0\\
2458454.39053 &    8.97 &  4.21 &  -15.33 &   19.19 &   44.34 &    8.07 &  0\\
2458454.61182 &    2.88 &  5.97 &  -14.84 &   34.87 &   16.17 &    9.27 &  0\\
2458470.39941 &   -4.79 &  4.38 &   45.29 &    5.53 &   18.01 &    3.88 &  0\\
2458474.40310 &    5.50 &  4.32 &   61.24 &   11.96 &   17.37 &    4.92 &  0\\
2458475.37871 &   -2.73 &  4.63 &   79.39 &   10.93 &   26.71 &    4.97 &  0\\
2458476.38550 &    0.14 &  2.94 &   -3.34 &    9.01 &   13.63 &    4.21 &  0\\
2458477.39012 &    7.07 &  3.61 &   10.96 &   15.01 &    3.65 &    4.00 &  0\\
2458478.36685 &   11.85 &  3.34 &   -0.22 &   11.59 &   10.40 &    3.88 &  0\\
2458479.33550 &   -6.04 &  5.45 &   37.81 &   21.48 &    6.70 &   10.03 &  0\\
2458480.51023 &   -5.59 &  3.51 &   12.72 &   11.69 &   10.55 &    3.71 &  0\\
2458481.36265 &   -0.71 &  6.92 &  -11.19 &   16.71 &  -45.38 &    5.27 &  0\\
2458483.45347 &   17.75 & 11.68 & -168.17 &   74.55 & -124.86 &   21.66 &  0\\
2458484.35157 &    2.35 &  4.16 &  -15.80 &   11.60 &   14.69 &    5.13 &  0\\
2458485.35678 &   -8.31 &  3.56 &   14.04 &   12.83 &   13.13 &    4.35 &  0\\
2458486.36431 &   -1.23 &  3.59 &   23.55 &   10.14 &   16.73 &    4.93 &  0\\
2458487.39667 &   10.27 &  3.32 &   -5.98 &   12.17 &    9.00 &    4.32 &  0\\
2458488.34863 &    7.44 &  3.00 &   -6.51 &    9.12 &   15.71 &    4.24 &  0\\
2458490.34335 &   -1.60 &  4.21 &   46.47 &   10.64 &  -75.01 &   14.08 &  0\\
2458491.47599 &    6.20 &  3.75 &   -6.31 &   14.42 &   13.99 &    4.23 &  0\\
2458492.34608 &  -12.53 &  5.19 &   35.04 &   23.61 &  -16.52 &    7.65 &  0\\
2458493.48897 &    5.81 &  4.10 &   -7.01 &   18.12 &    1.53 &    4.64 &  0\\
2458496.34421 &   -6.06 &  4.30 &   58.35 &   12.38 &   26.09 &    7.30 &  0\\
2458498.46151 &    1.90 &  3.43 &  -11.86 &   13.60 &   31.28 &    6.26 &  0\\
2458511.42794 &  -26.83 &  8.33 &  -10.15 &   34.10 &  -20.62 &   14.92 &  4\\
2458518.41796 &  -28.98 & 10.51 &  -10.72 &   50.34 &  -65.43 &   22.28 &  4\\
2458525.41316 &    0.02 &  5.61 &    6.57 &   25.13 &  -47.82 &    6.13 &  0\\
2458528.37840 &   18.55 &  7.75 & -140.96 &   58.28 & -112.45 &   18.21 &  0\\
2458529.38581 &   -6.69 &  3.49 &   30.97 &   16.06 &   -5.50 &    4.13 &  0\\
2458531.39644 &   -7.27 &  5.70 &  -21.47 &   23.96 &    7.80 &    4.51 &  0\\
2458532.38801 &    1.01 &  3.74 &   -0.33 &   13.73 &   -6.69 &    6.81 &  0\\
2458534.37013 &   -5.44 &  4.72 &   40.77 &   21.82 &   12.73 &   21.75 &  0\\
2458535.36152 &    1.27 &  3.68 &   12.63 &   15.71 &   20.38 &    7.45 &  0\\
2458537.37008 &   -5.93 &  4.54 &   39.84 &   18.89 &   24.16 &    7.79 &  0\\
2458540.34713 &   13.02 &  8.09 & -101.75 &   47.60 &  -84.26 &   23.85 &  0\\
2458541.33948 &    3.06 &  5.53 &   76.69 &   19.54 &   41.69 &    7.63 &  0\\
2458542.35411 &    2.18 &  4.66 &   18.93 &   22.87 &   47.11 &   10.36 &  0\\
2458546.34411 &    2.69 &  4.50 &   18.43 &   21.61 &   80.32 &   10.02 &  0\\
2458554.33745 &    3.51 &  4.40 &   -8.81 &   20.55 &   75.54 &    9.53 &  0\\
2458555.33466 &   -1.34 &  6.25 &   38.21 &   25.80 &   78.39 &   13.05 &  0\\

        \hline
    \end{longtable}
}

\end{document}